%% file: main.tex
\documentclass[acmtog]{acmart}

\acmSubmissionID{146}

\usepackage{multirow}
\usepackage{bm}
\usepackage{soul}
\usepackage{caption}
\usepackage{subcaption}
\usepackage[normalem]{ulem}
\usepackage{booktabs} 
\usepackage{csquotes}
\usepackage[linesnumbered,vlined,ruled]{algorithm2e}

\setcopyright{none}
\acmPrice{15.00}

\setcitestyle{square}

\newcommand{\handstate}{\ensuremath{{\bm{s}_\text{hand}}}}

\newcommand{\chopstate}{\ensuremath{{\bm{s}_\text{chop}}}}
\newcommand{\kinechopstate}{\ensuremath{{\tilde{\bm{s}}_\text{chop}}}}
\newcommand{\objstate}{\ensuremath{{\bm{s}_\text{obj}}}}

\setcopyright{acmcopyright}\acmJournal{TOG}
\acmYear{2022}\acmVolume{41}\acmNumber{4}\acmArticle{95}\acmMonth{7} \acmDOI{10.1145/3528223.3530057}

\begin{document}

\title{Learning to Use Chopsticks in Diverse Gripping Styles}


\author{Zeshi Yang}
\affiliation{%
  \institution{Simon Fraser University}
  \city{Vancouver}
  \country{Canada}
  \institution{and CFCS Peking University}
  \city{Beijing}
  \country{China}}
\email{zeshiy@sfu.ca}
\author{Kangkang Yin}
\affiliation{%
  \institution{Simon Fraser University}
  \city{Vancouver}
  \country{Canada}}
\email{kkyin@sfu.ca}
\author{Libin Liu}
\affiliation{%
  \institution{CFCS Peking Univeristy}
  \city{Beijing}
  \country{China}}
\email{libin.liu@pku.edu.cn}
\authornote{corresponding author}

\renewcommand{\shortauthors}{}

\begin{abstract}

{Learning dexterous manipulation skills is a long-standing challenge in computer graphics and robotics, especially when the task involves complex and delicate interactions between the hands, tools and objects. In this paper, we focus on chopsticks-based object relocation tasks, which are common yet demanding. The key to successful chopsticks skills is steady gripping of the sticks that also supports delicate maneuvers. We automatically discover physically valid chopsticks holding poses by Bayesian Optimization (BO) and Deep Reinforcement Learning (DRL), which works for multiple gripping styles and hand morphologies without the need of example data. Given as input the discovered gripping poses and desired objects to be moved, we build physics-based hand controllers to accomplish relocation tasks in two stages. First, kinematic trajectories are synthesized for the chopsticks and hand in a motion planning stage. The key components of our motion planner include a grasping model to select suitable chopsticks configurations for grasping the object, and a trajectory optimization module to generate collision-free chopsticks trajectories. Then we train physics-based hand controllers through DRL again to track the desired kinematic trajectories produced by the motion planner. We demonstrate the capabilities of our framework by relocating objects of various shapes and sizes, in diverse gripping styles and holding positions for multiple hand morphologies. Our system achieves faster learning speed and better control robustness, when compared to vanilla systems that attempt to learn chopstick-based skills without a gripping pose optimization module and/or without a kinematic motion planner. Our code and models are available at \href{https://github.com/chopsticks-research2022/learning2usechopsticks}{this link}.}\footnote{https://github.com/chopsticks-research2022/learning2usechopsticks}
\end{abstract}



\definecolor{tmp}{rgb}{0.6,0.,0.6}
\definecolor{myBlack}{rgb}{0,0.,0}
\definecolor{myRed}{rgb}{1,0.,0.}
\definecolor{myOrange}{rgb}{0.8,0.3,0.}
\definecolor{myGreen}{rgb}{0,0.5,0.27}
\definecolor{myBlue}{rgb}{0,0.0,0.9}
\definecolor{libinBlue}{rgb}{0,0.0,0.5}
\newcommand{\tmp}[1]{\textcolor{black}{#1}}
\newcommand{\zeshi}[1]{\textcolor{tmp}{#1}}
\newcommand{\add}[1]{\textcolor{black}{#1}}
\newcommand{\kk}[1]{\textcolor{myOrange}{#1}}
\newcommand{\kkr}[1]{\textcolor{myBlue}{#1}}
\newcommand{\libin}[1]{\textcolor{libinBlue}{#1}}
\newcommand{\tobedeleted}[1]{{\textcolor{black}{#1}}}
\newcommand{\commentText}[1]{{}}

\input{fig/teaser.tex}
\maketitle
\input{sec/1_introduction.tex}
\input{sec/2_relatedwork.tex}

\input{sec/3_overview.tex}
\input{sec/4_control_RL.tex}

\input{sec/5_holding_pose.tex}

\input{sec/6_motion_planner.tex}
\input{sec/7_results.tex}

\input{sec/8_conclusions.tex}

\input{sec/x_ack.tex}

\bibliographystyle{ACM-Reference-Format}
\bibliography{main}


\input{sec/x_appendix.tex}
\end{document}

%% file: fig/teaser.tex
\begin{teaserfigure}
  \centering
    \begin{subfigure}[b]{0.49\linewidth}
         \centering
         \includegraphics[width=\linewidth]{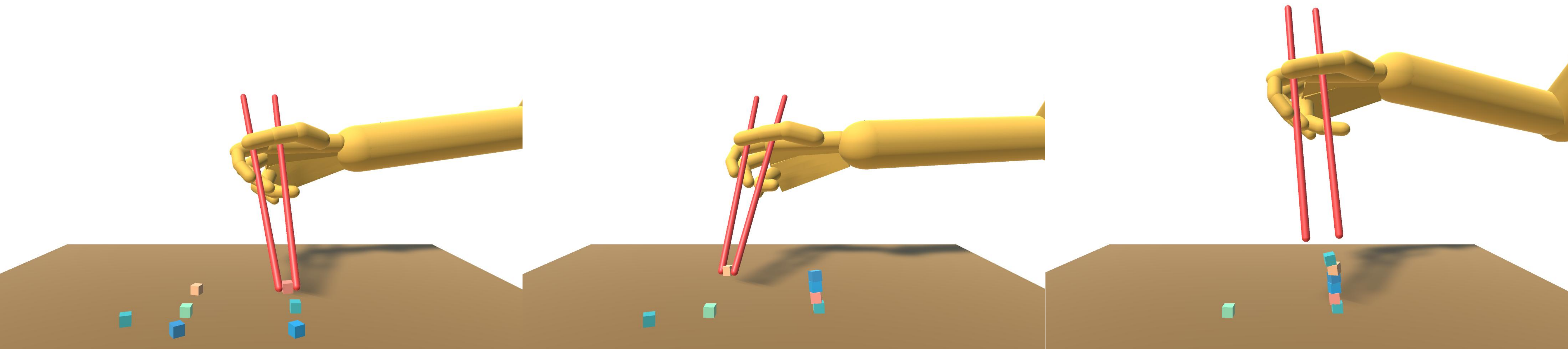}
         \caption{}
         \label{fig:stack}
     \end{subfigure}
     \begin{subfigure}[b]{0.49\linewidth}
         \centering
         \includegraphics[width=\linewidth]{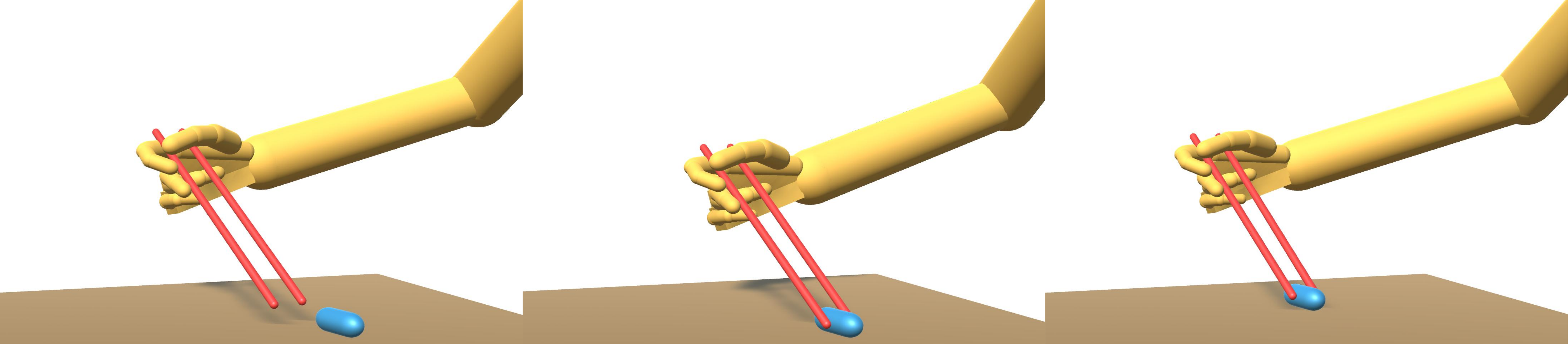}
         \caption{}
     \end{subfigure}
    \begin{subfigure}[c]{0.49\linewidth}
         \centering
         \includegraphics[width=\linewidth]{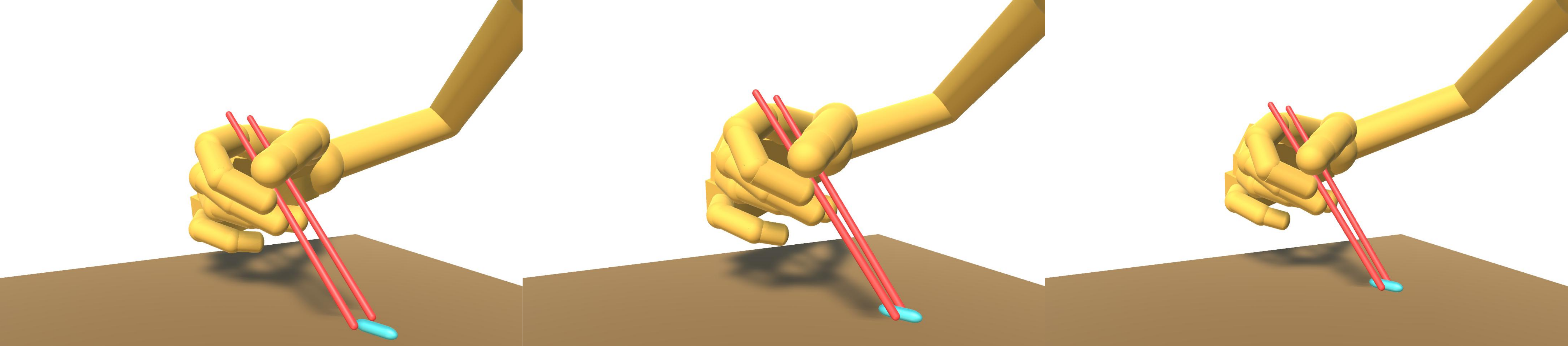}
         \caption{}
         \label{fig:teaserLargehand}
     \end{subfigure}
     \begin{subfigure}[d]{0.49\linewidth}
         \centering
         \includegraphics[width=\linewidth]{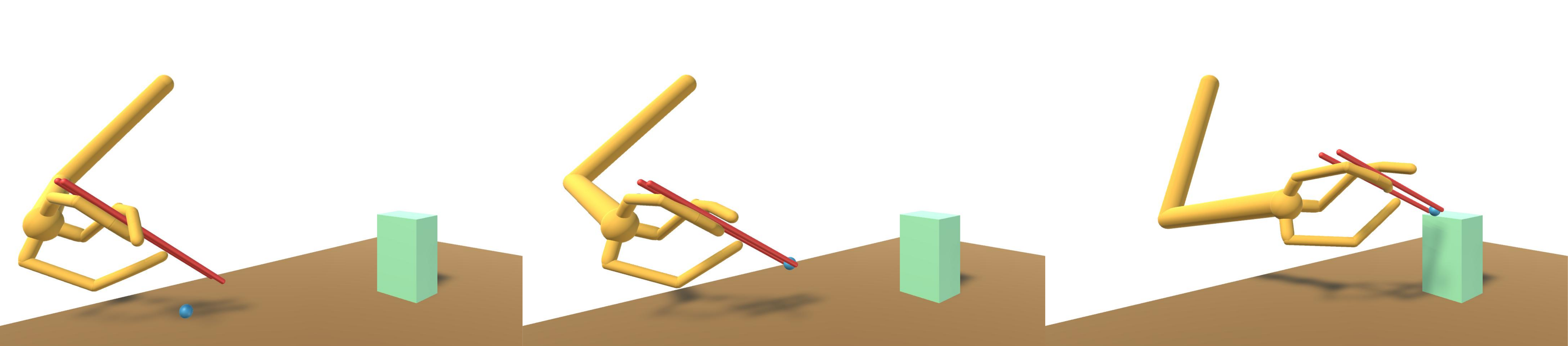}
         \caption{}
         \label{fig:teaserTrifinger}
     \end{subfigure}
  \caption{{Our system learns how to use chopsticks in diverse gripping styles for multiple hand morphologies. The trained physics-based hand controllers can pick up and relocate objects of various shapes and sizes in realtime.}}
\end{teaserfigure}

%% file: sec/1_introduction.tex
\section{Introduction}

{Dexterous manipulation and tool usage has been a long-standing challenge in computer animation and robotics. The main difficulties of tool use include the high degrees of freedom of the hands; the underactuation of the tools; and the complex interplay between the hands, tools and objects. The difficulty level also depends on the type of tools involved. Some tools only need to be grasped firmly in hand, such as hammers. Some tools need to be grasped and manipulated by hand, such as scissors. In this paper, we consider one of the most challenging tools: chopsticks.}

{Chopsticks are pairs of equal-length sticks used in Asian dinning for millenniums. Their simple design, or lack of design, poses several challenges in terms of control. First, the hand needs to grip and manipulate two independent sticks at the same time. Second, there are no obvious holding structures on the chopsticks, such as finger rings for a pair of scissors, to stabilize hand-tool contacts. Lastly, the chopstick-object contacts lie at the tip of the chopsticks, which are usually far away from the chopstick-hand contact points near the rear end of the chopsticks. Children usually need years of practice to master chopstick maneuvers. Even adults may find learning chopsticks challenging, if they did not grow up using them. The steep learning curve associated with chopsticks usage has spurred many video tutorials on YouTube, and the invention of training chopsticks that provide position retainer loops to stabilize finger-stick contacts.}

{The simple form of chopsticks, however, does enable their popularity and versatility. An estimated 33\% of the world’s population use chopsticks on a daily basis. Chopsticks are available everywhere, on dining tables as well as in the woods. They can pick up and move all kinds of foods: rice, meat, or noodles. They can manipulate in many different ways, so that spatulas, whisks, or pasta ladles are not necessary in Asian cooking. In robotics, research has been carried out to adopt chopsticks for eating assistance \cite{chang2007pincer,yamazaki2012autonomous}, micro-manipulation \cite{ramadan2009developmental}, and medical surgery \cite{sakurai2016thin,joseph2010chopstick,ragupathi2010robotic}.}

{The practicality and generality of chopsticks come at the cost of control complexity. In robotics, the chopsticks are usually rigidly attached to robot arms with reduced Degrees of Freedom (DoFs). In graphics, research on using chopsticks is nonexistent so far, to the best of our knowledge. Chopsticks usage is emblematic of a wider category of difficult-to-solve multi-contact manipulation-and-control problems. We focus on solving the problem of using truly underactuated chopsticks, with reasonable robustness in diverse gripping styles and holding positions for multiple morphologies. Inspired by how parents teach children chopsticks skills, we tackle this challenging control problem by decomposing it into two sub-problems: how to hold chopsticks properly? And then how to manipulate objects using chopsticks?}

{To use chopsticks effectively, users need to firstly hold them firmly. Although there is a consensus on a so-called standard grip being the most efficient way to use chopsticks \cite{Tomoko2010}, many people find their own ways to grip chopsticks during learning \cite{mukai1978study,osera2018relationship,yamakawa2018development}, such as the various grips shown in Figure~\ref{fig:pose_name}. We characterize a gripping style by the contact relationships between each finger and either chopstick. We optimize the gripping pose in a particular style by combining deep reinforcement learning and Bayesian optimization. Such an approach enables automatic discovery of diverse gripping poses for unusual hand morphologies. Using the output grips of our BO optimization, a moving virtual hand can hold the chopsticks firmly in physics simulation, and achieve some basic open-and-close chopsticks maneuvers.}

{To use chopsticks proficiently, users also need to control finger movements precisely in order to relocate objects via the tips of the chopsticks. Such fine motor controls are most likely impossible to design manually, which were possible for locomotion tasks. We design a two-level control system that first plans the chopsticks movements kinematically, and then trains physics-based hand controllers via model-free deep reinforcement learning. The high-level kinematic motion planner consists of a grasping model to select the best chopsticks configuration for grasping the objects, and then a trajectory generator to optimize for a collision-free chopsticks trajectory based on the start and goal transformations of the object. The hand and arm trajectories are then solved from the desired chopsticks trajectory using inverse kinematics. All the planned reference trajectories along with the optimized gripping pose in a desired style are then passed to the DRL system to train the low-level hand controls using simple tracking rewards. Our learned low-level hand controllers are able to grasp and move or throw objects of various shapes and sizes in realtime, and the high-level motion planner can plan or replan trajectories at interactive rates.}

\input{fig/sixStylePhotos.tex}

{In summary, the contributions of this work are mainly twofold:}
\begin{enumerate}
\item {We present a learning and control framework for object relocation using chopsticks. The high-level motion planner synthesizes collision-free kinematic trajectories at interactive rates, and the low-level physics-based hand controllers track the planned trajectories in realtime once trained. No sophisticated reward tuning or motion capture of human demonstrations are needed.} 
    
\item {We use Bayesian optimization combined with deep reinforcement learning to discover physically valid gripping poses in multiple styles. The optimized grips correspond well to human experiences and no manual specification is needed. Such an imitation-free method can generalize easily to different hand morphologies and is thus applicable to a broad range of graphics and robotics applications.}
\end{enumerate}

%% file: fig/sixStylePhotos.tex
\begin{figure}[t]
  \centering
   \begin{subfigure}[b]{0.3\linewidth}
         \centering
         \includegraphics[width=\linewidth]{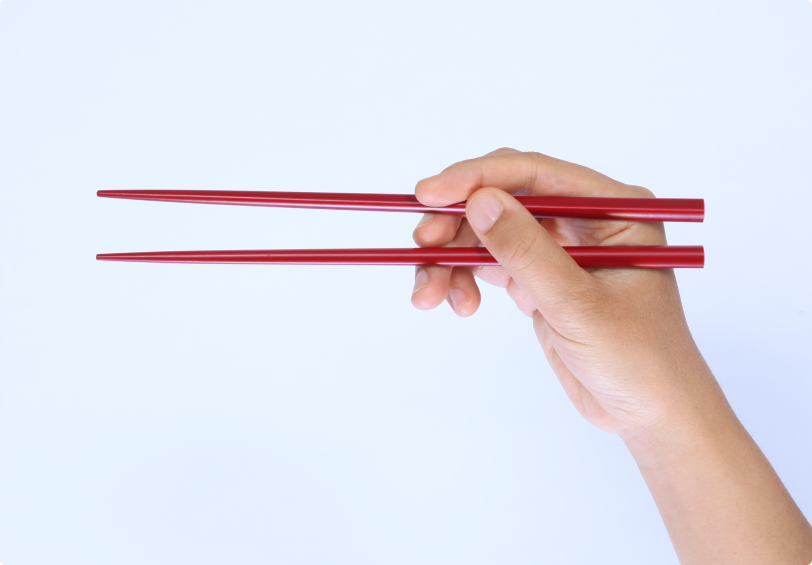}
         \caption{Standard}
         \label{fig:standard_pic}
     \end{subfigure}
     \begin{subfigure}[b]{0.3\linewidth}
         \centering
         \includegraphics[width=\linewidth]{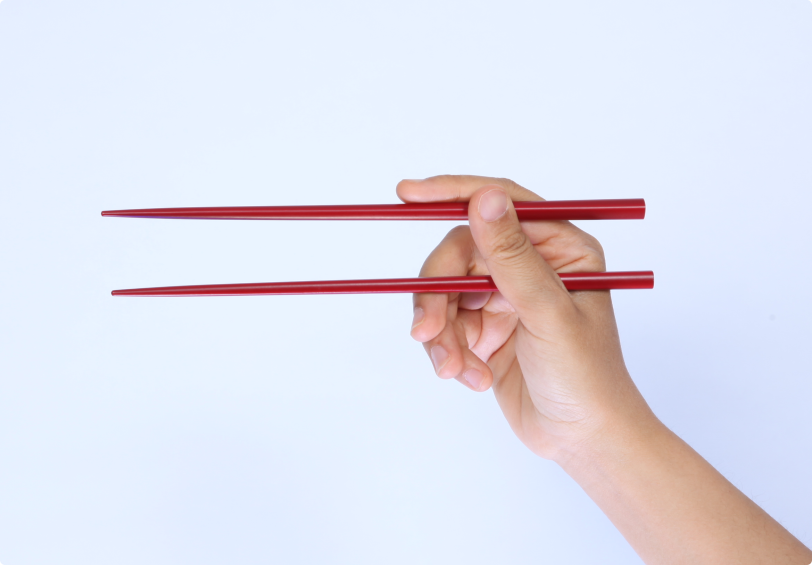}
         \caption{Right-hand rule}
         \label{fig:righthand_pic}
    \end{subfigure}
          \begin{subfigure}[b]{0.3\linewidth}
         \centering
         \includegraphics[width=\linewidth]{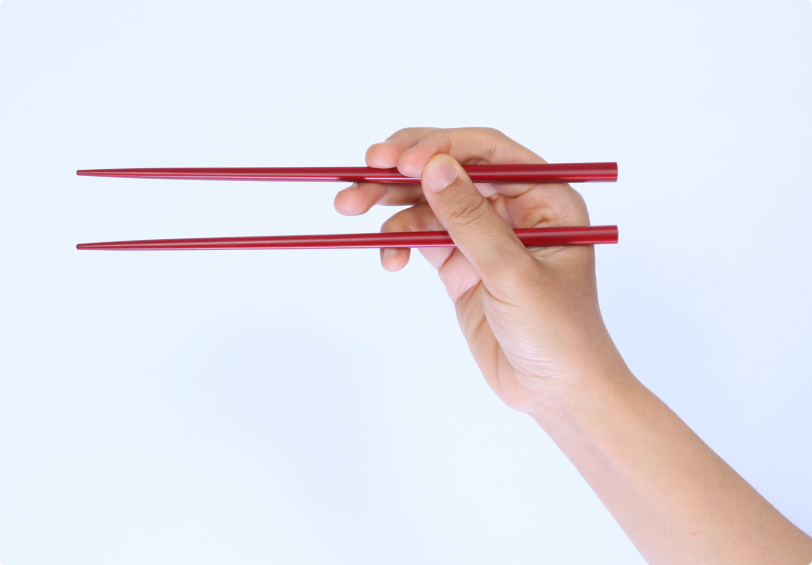}
         \caption{Forsaken pinky}
         \label{fig:forsaken_pic}
     \end{subfigure}
     \begin{subfigure}[b]{0.3\linewidth}
         \centering
         \includegraphics[width=\linewidth]{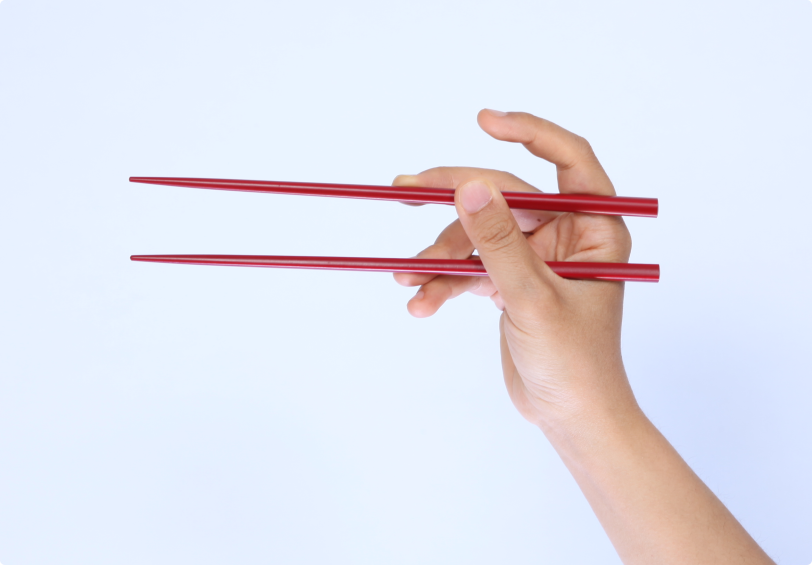}
         \caption{Dino claws}
         \label{fig:dino_pic}
    \end{subfigure}
     \begin{subfigure}[b]{0.3\linewidth}
         \centering
         \includegraphics[width=\linewidth]{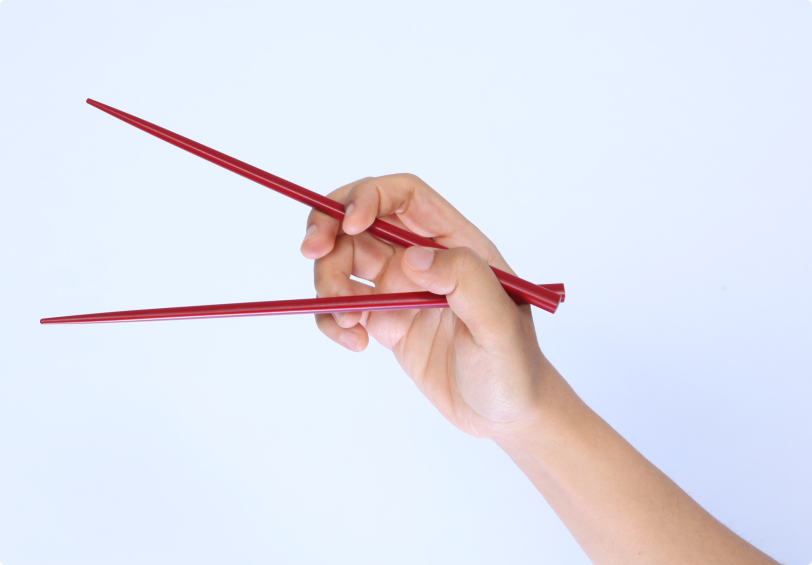}
         \caption{Dangling stick}
         \label{fig:dangling_pic}
    \end{subfigure}
     \begin{subfigure}[b]{0.3\linewidth}
         \centering
         \includegraphics[width=\linewidth]{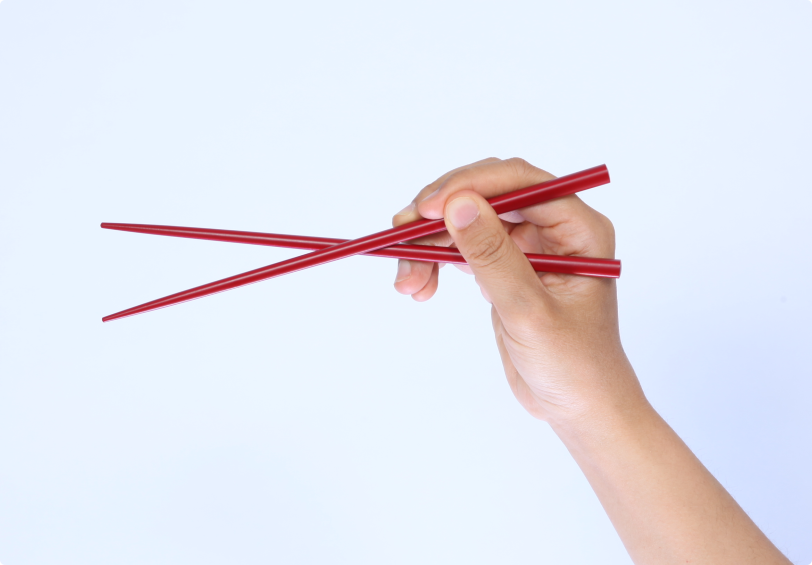}
         \caption{Italian}
         \label{fig:dangling_pic}
    \end{subfigure}
  \caption{Multiple ways of holding chopsticks. Our system can discover similar gripping poses for seventeen styles, many of which correspond to the commonly used chopstick grips given in \cite{macro2021}.}
  \label{fig:pose_name}
\end{figure}

%% file: sec/2_relatedwork.tex
\section{Related Work}

{Human hands and tool usage are the special anatomy and function that helped driving the evolution of human brain, which ultimately differentiated humans from the rest of the animal kingdom. Significant amount of research endeavors have been invested into tackling the challenging problem of synthesizing dexterous manipulation in simulation or on robots. We classify the most relevant recent works into two categories: hand manipulation and tool usage. Hand manipulation is manipulation of objects directly by fingers and the palm, such as grasping and relocation of objects \cite{zhao2013robust,kry2006interaction,liu2009dextrous}, and in-hand manipulation \cite{zhang2021manipnet}. Tool usage is manipulation of objects by tools that are operated by hands or robot arms. We refer interested readers to other orthogonal dimensions of research in hand and finger animation to the excellent survey provided in \cite{wheatland2015state}.}
 
\subsection{Hand Manipulation}

\subsubsection{Kinematic Methods}
{Hand manipulations such as grasping or playing musical instruments can be synthesized through traditional inverse kinematic methods \cite{huang1995multi,elkoura2003handrix,kim2000neural,koga1994planning,aydin1999database}. More recently, deep neural networks have also been utilized to synthesize hand manipulations \cite{GRAB:2020,karunratanakul2020grasping}. For example, \cite{GRAB:2020} constructed a hand-object interaction database through motion capture, and trained a neural network to predict object grasping poses for human hands. \cite{zhang2021manipnet} showcased  impressive hand-object manipulation results such as turning a torus in hands. Kinematic methods, however, cannot generate motions responsive to a dynamic environment. The synthesized manipulations could also display artifacts such as penetrations into the objects. Moreover, data-driven kinematic methods usually require high-fidelity hand manipulation capture, which is hard and expensive in most cases.}

\subsubsection{Physics-based Methods}

{Physics-based control methods leverage physics simulation to generate motions with physical realism and environmental interactions. The key challenge, however, is to design or learn robust controllers to drive the simulated characters or hands. High-fidelity motion capture data has been utilized to help with synthesizing physically realistic hand manipulations \cite{zhao2013robust,kry2006interaction}. In the absence of motion capture data, trajectory optimization provides a viable approach to synthesizing physics-based dexterous manipulations \cite{liu2008synthesis,liu2009dextrous,ye2012synthesis,mordatch2012contact,wang2013video}. These methods usually need to model the dynamics to a great extent, such as incorporating friction constraints into the objective functions to avoid undesirable movements between hands and objects during grasping. Therefore, such methods usually simplify the dynamics to a certain degree to reduce the control complexity. For our case of dealing with five fingers and two chopsticks, the contact and friction dynamics will probably be too overwhelming to handle, even with simplified physics.}

{We therefore opt for a model-free approach using deep reinforcement learning. DRL has been widely used in computer graphics and robotics to learn diverse motion skills, such as locomotion \cite{peng2017deeploco,peng2018deepmimic,bergamin2019drecon,park2019learning, tan2018sim,peng2021amp}, athletic skills \cite{yin2021strategy, liu2017learning}, and manipulations \cite{rajeswaran2018learning,popov2017data,nagabandi2020deep,chen2021simple,andrews2012policies,garcia2020physics}. Challenging manipulation tasks, such as solving a Rubik's cube with a robot hand, have been demonstrated using DRL-based control learning methods \cite{akkaya2019solving}. It remains an open problem how to design suitable DRL reward functions to learn natural-looking skills for complex tasks, however. One idea is to leverage human demonstrations as reference skills for physical agents to imitate, which has been proven to improve both the learning efficiency and control robustness \cite{rajeswaran2018learning}. Multiple reference skills can be combined to help solve more challenging tasks using hierarchical deep reinforcement learning, such as dribbling a soccer ball or carrying objects to target locations \cite{peng2019mcp,merel2020catch}. More recently, adversarial imitation learning DRL system has been quite successful at learning motion priors from large datasets of unstructured motion clips \cite{peng2021amp}. The motion priors obviate the need for manually designed imitation objectives or a high-level motion planner.}

{Capturing chopsticks skills, however, may be impractical as severe occlusions and subtle movements are involved. We also wish our solution to generalize well to graphics applications with monster-hand morphologies, and robotics applications with non human-hand-like manipulators. We thus rely on Bayesian optimization coupled with DRL to discover diverse and physically valid chopsticks gripping poses. We then use a motion planner to generate chopsticks configurations and trajectories to satisfy kinematic task objectives. The discovered gripping poses and synthesized trajectories are then passed to our DRL-based training system to learn chopsticks skills using simple tracking rewards. Therefore, the advantages of our system include its simplicity in terms of system setup and tuning, and diversity in terms of gripping styles and hand morphologies.}

\subsection{Tool Usage}
{Tool usage has not been explored too much in the graphics community. The most relevant work is \cite{zhang2020learning}, which employed DRL to learn control policies to manipulate amorphous materials, such as gathering rice with scrapers or flipping pancakes with pans. Their tools are driven by a virtual proportional derivative controller. Then hand motions are reconstructed via inverse kinematics. There are more research activities on tool usage in the robotics community \cite{wu2019learning,fang2020learning,toussaint2018differentiable,ke2020telemanipulation,ke2021grasping,kim2021integrated}, although most of them turn it into a simpler problem by attaching the tools directly onto the robot arm. For example, chopsticks are attached to a robot arm to grasp objects in \cite{ke2021grasping}, where human demonstrations were also used to help the learning of grasping policies. We study the problem of controlling underactuated chopsticks by hands, meaning that the chopsticks can actually move inside and fall out of the hand. We have not been able to find any prior work on exactly the same problem in the literature.}

\subsection{Bayesian Optimization}
{Bayesian Optimization (BO) is a class of optimization methods for expensive black-box function optimization. The function is optimized purely through evaluations as no gradient information is readily available. In Bayesian optimization, a Bayesian statistical model, such as a Gaussian Process \cite{rasmussen2003gaussian}, is maintained to predict the values and uncertainties of the objective function. An acquisition function is applied to query the most promising and informative regions based on current estimations. Recently BO has seen more and more adoptions in robotics and computer graphics for various applications \cite{brochu2007preference,brochu2010bayesian,koyama2020sequential,hu2021neural}. Most relevant to our work, BO has been employed to tune parameters of a bipedal locomotion controller \cite{rai2018bayesian}, optimize hyperparameters of physics-based character animation systems \cite{yang2021CMFBO}, and discover diverse athletic jumping strategies \cite{yin2021strategy}. For our problem, the chopsticks gripping pose can be viewed as a hyperparameter of the hand control policies trained through DRL, and therefore optimized by BO. There are multiple BO algorithms, such as Gaussian-Upper Confidence Bound (GP-UCB) \cite{srinivas2010gaussian} and Entropy Search \cite{hennig2012entropy}. Our gripping pose optimization is based on the GP-UCB implemented with the Gaussian process framework~\cite{gpy2014}. Our results show that the optimized poses correspond well to commonly used chopstick grips by humans.}

%% file: sec/3_overview.tex
\begin{figure}[t]
    \centering
    \begin{subfigure}[b]{1\linewidth}
         \centering
         \includegraphics[width=\linewidth]{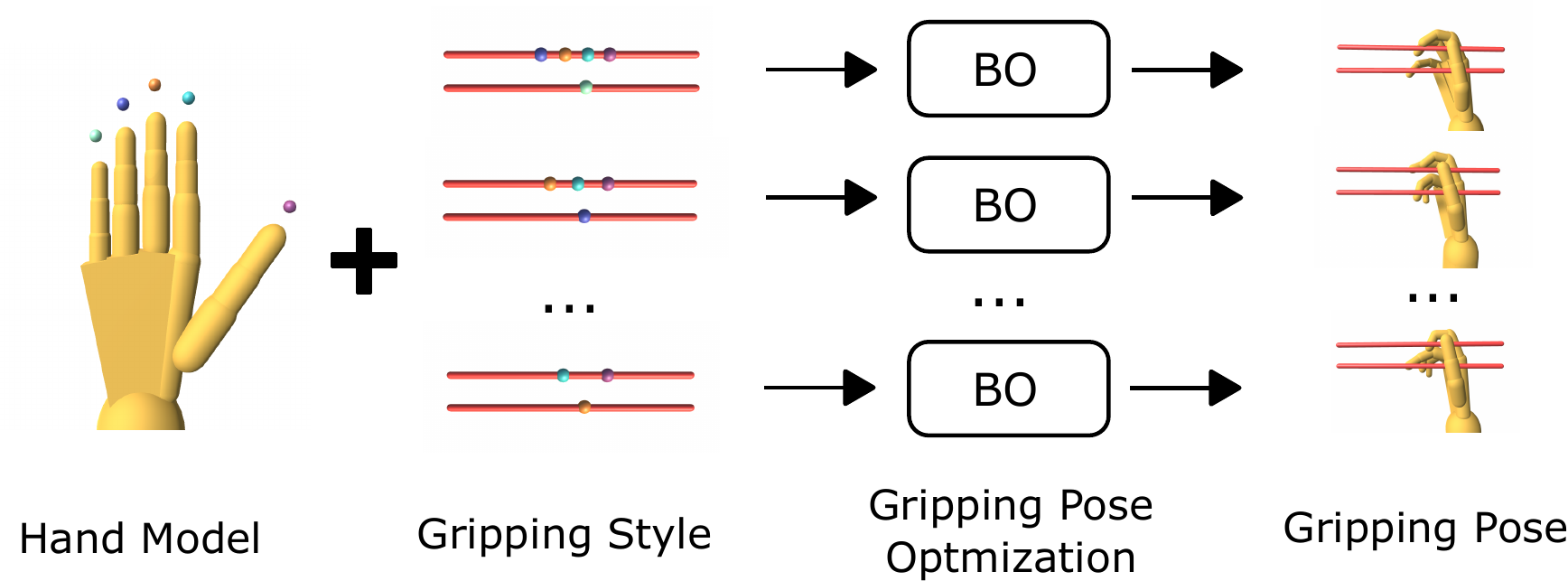}
         \caption{Gripping pose optimization (Section~\ref{sec:poseOptimization}).}
         \label{fig:bo_pose}
     \end{subfigure}
    \begin{subfigure}[b]{1\linewidth}
         \centering
         \includegraphics[width=\linewidth]{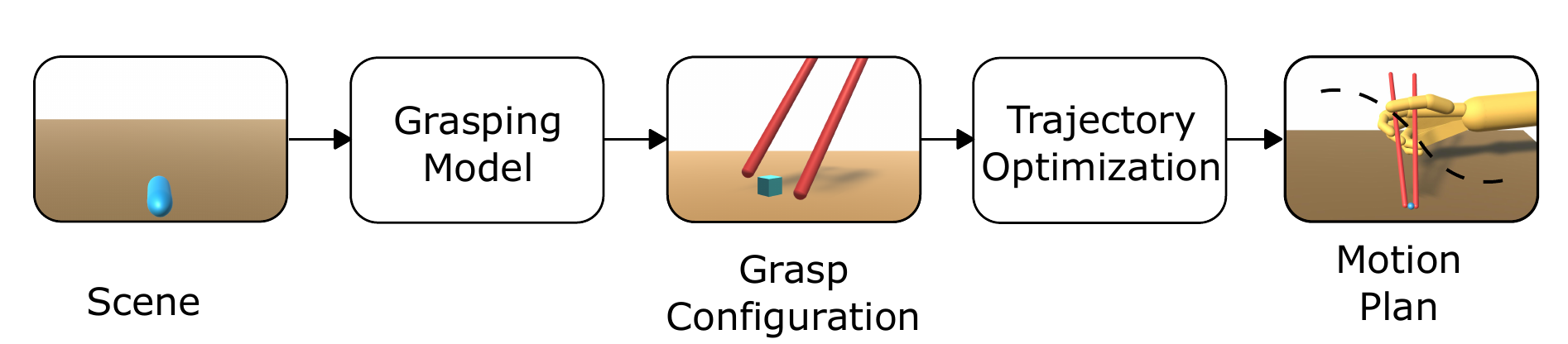}
         \caption{High-level motion planning (Section~\ref{sec:motionplanner}).}
         \label{fig:highplan}
    \end{subfigure}
    \begin{subfigure}[b]{1\linewidth}
         \centering
         \includegraphics[width=\linewidth]{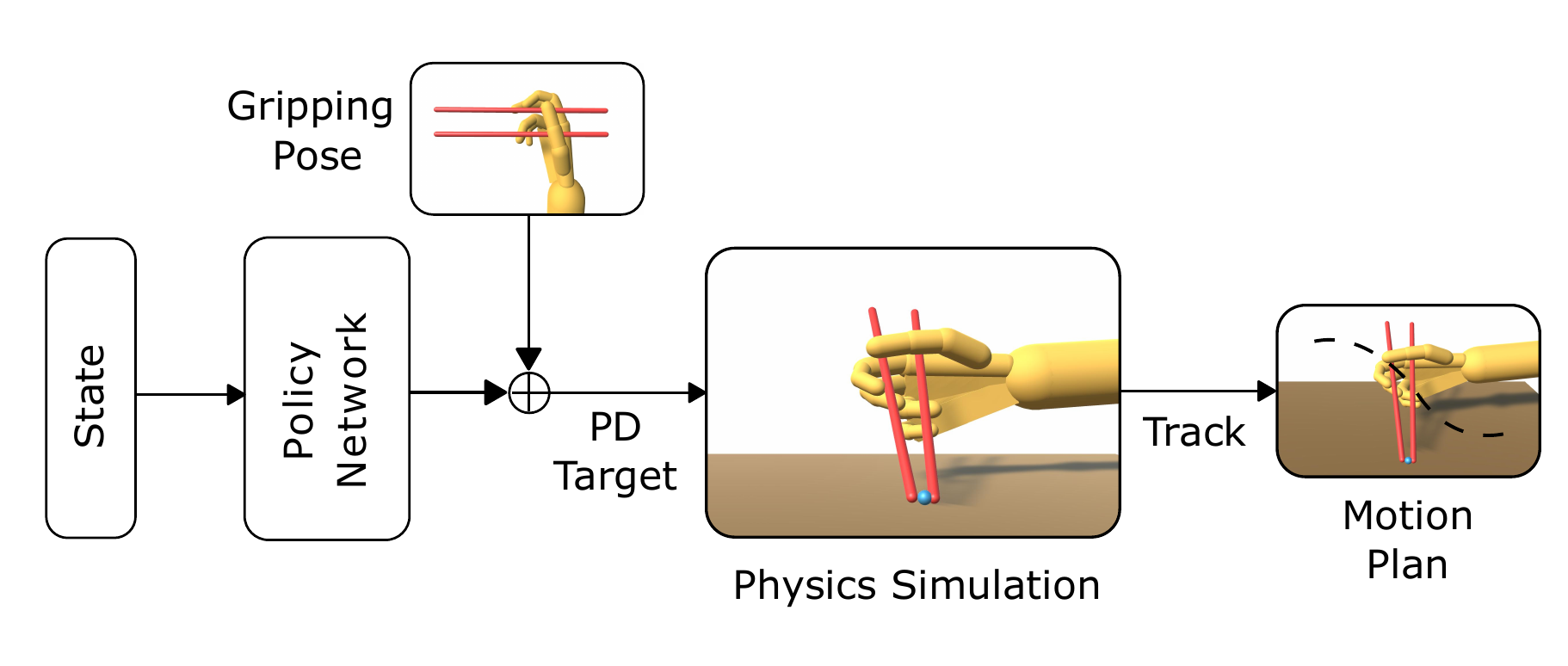}
         \caption{Low-level hand control (Section~\ref{sec:trackingcontroller}).}
         \label{fig:lowhand}
    \end{subfigure}
    \caption{The three main components of our learning and control framework. (a) For a desired gripping style, we employ BO and DRL to optimize for a physically valid gripping pose of the hand. (b) To achieve an object relocation task, the motion planner first selects a suitable chopsticks configuration for grasping, and then synthesizes collision-free trajectories for the chopsticks and hand. (c) Then we train policy networks using DRL to track the planned trajectories for a chosen gripping pose.}
    \label{fig:pipeline}
\end{figure}

\section{Overview}
\label{sec:overview}

{Figure~\ref{fig:pipeline} provides an overview of our learning and control framework. Taken the models of one hand and two chopsticks as input, our goal is to learn robust hand controls to use the chopsticks for a variety of object grasping and relocation tasks. We achieve this goal by solving two sub-problems: first finding physically valid gripping poses in multiple styles in a \emph{gripping pose optimization} step, and then learning \emph{hand control policies} to hold the chopsticks in a specific gripping pose to pick up and relocate objects in simulation.}

{In the gripping pose optimization step as illustrated in Figure~\ref{fig:bo_pose}, we employ Bayesian Optimization (BO) to find the optimal gripping pose for a desired style. A gripping style is characterized by a set of finger-chopstick contact relations, or specifically, which finger should be in contact with which stick. Gripping styles can be specified either manually or automatically as will be described later. Given a desired gripping style, our BO algorithm iteratively proposes a set of finger-stick contact positions, which are then converted into a hand pose using Inverse Kinematics (IK). We then employ DRL to evaluate the quality of the candidate gripping pose by training policies to perform simple open-and-close chopsticks maneuvers. The performance of the trained policy, characterized by its average reward in simulation, is then fed back to the BO to come up with the next proposal for contact positions.} 

{After good gripping poses are found, we build hand controllers to relocate objects with chopsticks for each gripping pose. To this end, we design a two-level learning and control framework. The high-level kinematic motion planner as shown in Figure~\ref{fig:highplan} first selects the best chopsticks configuration in order to grasp the object of interest. We develop a neural-network based grasping model to recommend such chopsticks configurations based on the shape and transformation of the object. The planner then optimizes for collision-free trajectories for the chopsticks and hand, based on the relocation task required. Next the low-level hand control policy as shown in Figure~\ref{fig:lowhand} is trained through model-free DRL to track the generated motion plan. Once trained, the control policies can track novel reference trajectories generated by the planner in realtime to relocate objects of various shapes and sizes.}

%% file: sec/4_control_RL.tex
 \section{Physics-based Hand Tracking Control}
\label{sec:trackingcontroller}

{We train tracking controllers for the hand to follow given trajectories of the object, the chopsticks, the hand, and the arm, while holding the chopsticks in a specific gripping pose. The trajectories to track are generated by a high-level motion planner in our system, but motion capture examples or manually defined keyframe animations could be used here if so wished. We use Deep Reinforcement Learning (DRL) to train these tracking controllers. The same DRL components are used for both the gripping pose optimization and the final control policy learning for actual object manipulations, although their reward terms are slightly different.}

\subsection{Deep Reinforcement Learning}
\label{sec:DRL}
{Our policy learning is formulated as a standard Reinforcement Learning (RL) problem, where an agent interacts with the environment and learns from its experience for the optimal control policy that maximizes the total reward. We denote the control policy as $\pi_{\theta}(\bm{a}_{t}|\bm{s}_{t})$, which models the conditional distribution of the action $\bm{a}_{t}$ given the current state $\bm{s}_{t}$, and $\theta$ represents the learnable parameters. At each time step $t$, the agent takes an action $\bm{a}_{t}$ according to $\pi_\theta$, gets a reward $r_{t}$ from the environment, and then transits to another state $\bm{s}_{t+1}$ according to a probabilistic dynamics model $p(\bm{s}_{t+1}|\bm{s}_{t}, \bm{a}_{t})$. Starting from an initial state $\bm{s}_0$, this procedure will generate a trajectory $\tau=\{\bm{s}_0, \bm{a}_0, \bm{s}_1, \bm{a}_1, ...\}$. We calculate the cumulative reward of $\tau$ with a discount factor $\gamma$ as $R(\tau)=\sum_t\gamma^{t}r_{t}$ . Then the expected return $J(\theta)$ is computed over all possible trajectories induced by policy $\pi_\theta$ and the dynamics $p$ as:
\begin{equation}
    J(\theta) = \mathbb{E}_{\tau\sim{}p_{\theta(\tau)}}\left[ R(\tau)\right]
\end{equation}
where $p_{\theta(\tau)}=p(\bm{s}_0)\prod_t{\pi_{\theta}(\bm{a}_{t}|\bm{s}_{t})p(\bm{s}_{t+1}|\bm{s}_{t},\bm{a}_{t})}$, and $p(\bm{s}_0)$ is the initial state distribution.}

{We train our policies using the Proximal Policy Optimization (PPO) algorithm \cite{schulman2017proximal}, which has been frequently adopted in physics-based character animation and robotics for its robustness and simplicity. PPO is often implemented within an actor-critic framework, where a critic network is trained to estimate the value of states, and an actor network is trained to output the control policy $\pi$. We employ multi-step returns $TD(\lambda)$ and the generalized advantage estimator $GAE(\lambda)$ \cite{schulman2015trust} to facilitate training of the neural networks, similar to the PPO implementation in \cite{peng2018deepmimic} where interested readers can find more details.}

\input{sec/4.2_simulation.tex}

\subsection{Learning of Tracking Control}
\label{sec:tracking}

{We model our hand tracking control policy $\pi$ as a fully-connected neural network with two $256$-unit hidden layers and $\textrm{ReLU}$ activations. We train these controllers using the PPO algorithm as described in Section~\ref{sec:DRL}, where the value network shares the same structure as the policy network except that its last layer is a single linear unit. At runtime, the controller $\pi$ is responsible for computing an action $\bm{a}$ given the current state $\bm{s}$ of the system.}

\subsubsection{States and Actions}
{The state $\bm{s}$ is the input to the hand controllers. In our system, $\bm{s}$ consists of the simulation states of the chopsticks, the hand, and the object being manipulated, as well as the contact information between them. We compute the state of the chopsticks and the object with respect to the local coordinate frame of the palm, to facilitate learning and improve robustness. In addition, a short segment of the desired motion trajectory is also included as part of $\bm{s}$ to facilitate tracking. More specifically, the desired kinematic states of the hand, chopsticks, and objects of the next six frames from the planned trajectories are included in $\bm{s}$. A complete list of all the components of $\bm{s}$ is provided in \appendixname~\ref{sec:app:states}}.

{The action $\bm{a}$ is the output of the hand controllers. In our system, $\bm{a}$ represents a corrective offset pose $\delta\bm{q}$ that will be added to a chosen gripping pose $\bm{q}^*$ to compose the final target pose $\hat{\bm{q}}$ for the PD-servos as in Equation~\ref{eqn:pdcontrol}}. 

\subsubsection{Rewards}
{The reward function is designed to encourage the hand to hold the chopsticks firmly in a chosen style and move the object following a desired trajectory. More specifically, it consists of four reward terms:
\begin{equation}
    r = e^{r_{\textrm{hand}} + r_{\textrm{chop}} + r_{\textrm{object}} + r_{\textrm{contact}}}
\end{equation}
The hand control term $r_{\textrm{hand}}$ encourages the hand and arm to match their planned trajectories:
\begin{equation}
    r_{\textrm{hand}} = -10 \left\Vert\bm{q}_{\textrm{hand}} - \tilde{\bm{q}}_{\textrm{hand}}\right\Vert
\end{equation}
where $\bm{q}_{\textrm{hand}}$ is the simulated hand pose, and $\tilde{\bm{q}}_{\textrm{hand}}$ is the desired hand pose in the planned trajectory.}

{Similarly, the chopsticks term $r_{\textrm{chop}}$ and the object term $r_{\textrm{obj}}$ are defined as:
\begin{align}
    r_{\textrm{chop}} &= -40\sum_{i\in\{1,2\}}\left\Vert\bm{p}_i - \tilde{\bm{p}}_i\right\Vert - 10\sum_{i\in\{1,2\}}\Theta(\bm{o}_i, \tilde{\bm{o}}_i) \\
     r_{\textrm{obj}} &= -40\left\Vert\bm{p}_{\textrm{obj}} - \tilde{\bm{p}}_{\textrm{obj}}\right\Vert - 10 \cdot \Theta(\bm{o}_{\textrm{obj}}, \tilde{\bm{o}}_{\textrm{obj}}) \label{eqn:objloss}
\end{align}
where $\bm{p}$ and $\bm{o}$ represent positions and orientations respectively. The scalar function $\Theta(\bm{o}_1,\bm{o}_2)$ computes the absolute angle between two quaternions $\bm{o}_1$ and $\bm{o}_2$.}

{Lastly, the term $r_{\textrm{contact}}$ prevents fingertips from leaving or slipping on the chopsticks:
\begin{equation}
    r_{\textrm{contact}} = -10 \cdot \sum_{i=1}^{i=N}d_{i}
\end{equation}
where $d_{i}$ is the minimal distance between the fingertip $i$ (the part called distal phalanx in hand anatomy) and its desired contact position on the chopsticks as determined by the gripping style. $N$ is the number of fingers that should remain in contact with the chopsticks in the corresponding gripping style, as will be described next.}

%% file: sec/4.2_simulation.tex
\subsection{Simulation Setup}
\label{sec:simulation}
{We simulate our hand models of potentially different morphologies together with a common two-link arm model as an articulated rigid-body system, allowing a moderate task space for object relocation tasks. The shoulder joint is fixed in position and only has three rotational Degrees of Freedom (DoFs). We note that hereafter whenever we refer to the state or pose of the hand, we really mean the state or pose of the whole hand-and-arm structure. We parameterize the state of the hand in generalized coordinates as $\handstate=(\bm{q},\dot{\bm{q}})$, where $\bm{q}$ represents the joint angles and $\dot{\bm{q}}$ the rotational speeds. All joints are actuated using PD-servos, and the positional PD targets $\hat{\bm{q}}$ are computed by the control policy $\pi$. The joint torques $\bm{\tau}$ are then computed as
\begin{equation}
    \bm{\tau} = k_p(\hat{\bm{q}} - \bm{q}) - k_d\dot{\bm{q}}
    \label{eqn:pdcontrol}
\end{equation}
where $k_p$ and $k_d$ are the stiffness and damping parameters of the joint-level PD-servos.}

{We model chopsticks as a pair of long rigid capsules of the same length and radius. The two sticks can move independently so the total DoFs of a pair of chopsticks is twelve. We denote the state of the chopsticks as $\chopstate=(\bm{p}_i,\bm{o}_i,\bm{v}_i,\bm{\omega}_i), i=1,2$, where $\bm{p}_i$ is the position of the Center of Mass (CoM) of Chopstick $i$, $\bm{o}_i$, $\bm{v}_i$ and $\bm{\omega}_i$ are the orientation, linear velocity and angular velocity of Chopstick $i$. Unless otherwise noted, the rotations are parameterized in quaternions in our system. The upper stick is indexed as Chopstick 1 and the lower stick is Chopstick 2, as shown in Figure~\ref{fig:pose_pregrasping}.}

{The objects being manipulated by the chopsticks are modeled as rigid bodies from a predefined set of shapes and range of sizes as shown in Table~\ref{tab:geom_range}. We use a tuple $\bm{t}_{\text{obj}}$ to indicate the shape and size of the corresponding object. The state of an object is denoted by a tuple $\objstate=(\bm{p}_\text{obj},\bm{o}_\text{obj},\bm{v}_\text{obj},\bm{\omega}_\text{obj})$, which consists of the object's position, orientation, and linear and angular velocity.}

%% file: sec/5_holding_pose.tex
\section{Gripping Pose Optimization}
\label{sec:poseOptimization}

{A good chopstick gripping pose is a strong determinant of successful manipulations using chopsticks. Some gripping poses make it easy and efficient to use chopsticks, while others may feel awkward or even make it infeasible to manipulate with chopsticks. The goal of the gripping pose optimization component is to find the optimal pose with which the hand can hold the chopsticks steadily while still can move in a coordinated and flexible fashion to accomplish object manipulation using the tips of the chopsticks. Such an objective is difficult to formulate in closed form, thus we opt for a learning-based evaluation scheme coupled with Bayesian optimization.}

{More specifically, we learn a control policy using reinforcement learning to track basic chopsticks maneuvers for a candidate gripping pose. The performance of the learned policy is then used as an assessment of the gripping pose. As the DRL-based evaluation is relatively expensive and not differentiable, we employ the sample-efficient and gradient-free Bayesian optimization to suggest candidate poses. Our gripping pose optimization algorithm is summarized in Algorithm~\ref{alg:BayesianOptimization}. Note that the arm is controlled to maintain a static pose at this stage, and dynamic arm movements will be synthesized later in Section \ref{sec:motionplanner}.}

\begin{algorithm}[t]
    \DontPrintSemicolon{}
    \SetAlgoLined{}
    \KwIn{A gripping style $\bm{c}$; maximal BO iterations \it{maxIter}. }
    \KwOut{The optimal gripping pose $\bm{q}^*$ for $\bm{c}$.}
    $i=0$; $r^*=-\inf$ \;
    \While(//BO iterations){\textrm{i} $<$ \textrm{maxIter}}{
        $\bm{x} \leftarrow $ contact position proposal by BO from $\bm{c}$; \;
        $\bm{q} \leftarrow$ solving gripping pose by IK from $\bm{x}$; \;
        $r \leftarrow$ average reward of DRL-trained policy for $\bm{q}$;  \;
        update BO record with $\bm{x}, r$; \;
        update Gaussian surrogate model; \;
        \lIf{$r > r^*$}{$(\bm{q}^*, r^*) \leftarrow (\bm{q}, r)$;}
        $i$++; \;
    }
    \Return $\bm{q^*}$
    \caption{Gripping Pose Optimization with BO and DRL\label{alg:BayesianOptimization}}
\end{algorithm}

\subsection{Bayesian Optimization}
{Bayesian Optimization (BO) is one of the ideal choices for optimizing expensive black-box functions with no gradients. It is designed to minimize the number of function evaluations by querying the most promising and informative data points.
For a given objective function, BO searches for its optimum through a series of evaluations, where the history of those evaluations are recorded to fit an \textit{acquisition function}, which will determine the next candidate data point.
Our gripping pose optimization is based on the GP-UCB algorithm \cite{srinivas2010gaussian} implemented with the Gaussian process framework~\cite{gpy2014}. Interested readers can refer \cite{srinivas2010gaussian} for more details. We set the maximal allowed number of function evaluations to ten in our implementation.}

\subsection{Chopsticks Gripping Style}
{In our system, a chopsticks gripping style is characterized by the contact relationships between each finger and either chopstick. For a human hand with five fingers, namely the thumb, index, middle, ring, and little finger, we can represent a gripping style with a 5-tuple $\bm{c}=(c_1, c_2, c_3, c_4, c_5)$, where $c_i=j, j\in\{0,1,2\}$ indicates that finger $i$ should be in contact with chopstick $j$. $0$ denotes no contact. Following this notation, the standard gripping style shown in Figure~\ref{fig:pose_name} can be represented by the tuple $(1,1,1,2,0)$, for example. In general, for a hand model with $N$ fingers, its chopsticks gripping style can be defined with an $N$-tuple.}

{For hands with a small number of fingers, we could simply enumerate all values of the gripping style tuple and optimize a gripping pose for each of them. A human hand, for example, has potentially a maximal of $3^{5}=243$ gripping styles. However, many of them may be infeasible, inefficient, or unnatural.  We thus employ two heuristics to prune the style space to eliminate bad styles, as well as to reduce the optimization workload. First, the thumb plays a crucial role in performing precise tool use, and therefore we do not allow $c_1$ to be zero. We also observe that the thumb is usually used to support the upper Chopstick 1 as shown in Figure~\ref{fig:pose_name}. This is because the lower stick has some default support from the valley between the thumb and the index finger, while the upper stick needs the thumb more for support and movement. Consequently, we set $c_1=1$ and shrink the style space by $2/3$. Second, finger crossing usually leads to awkward or infeasible grasping poses, thus the gripping styles with finger crossings are excluded from further pose optimization. After applying these two heuristics, there are seventeen gripping styles left, for each of which we run BO to obtain an optimized gripping pose.}

\subsection{Gripping Pose Generation}
{When given a chopsticks gripping style as input, represented by an $N$-tuple, we use BO to search for the optimal gripping pose for that style. To reduce the degrees of freedom of the problem, we first optimize the finger-stick contact positions according to the contact patterns specified in the style tuple. Specifically, we parameterize each contact position using a single scalar $x$ representing the contact location along the chopstick. Then the BO algorithm just needs to optimize a vector $\bm{x}$ up to $N$ dimensions. Once the contact positions are determined, the gripping pose can be obtained by an Inverse Kinematics (IK) solver. The quality of the gripping pose is then evaluated by the performance of its trained policy using deep reinforcement learning as described in Section~\ref{sec:tracking}.}

\begin{figure}[t]
    \centering
    \includegraphics[width=0.9\linewidth]{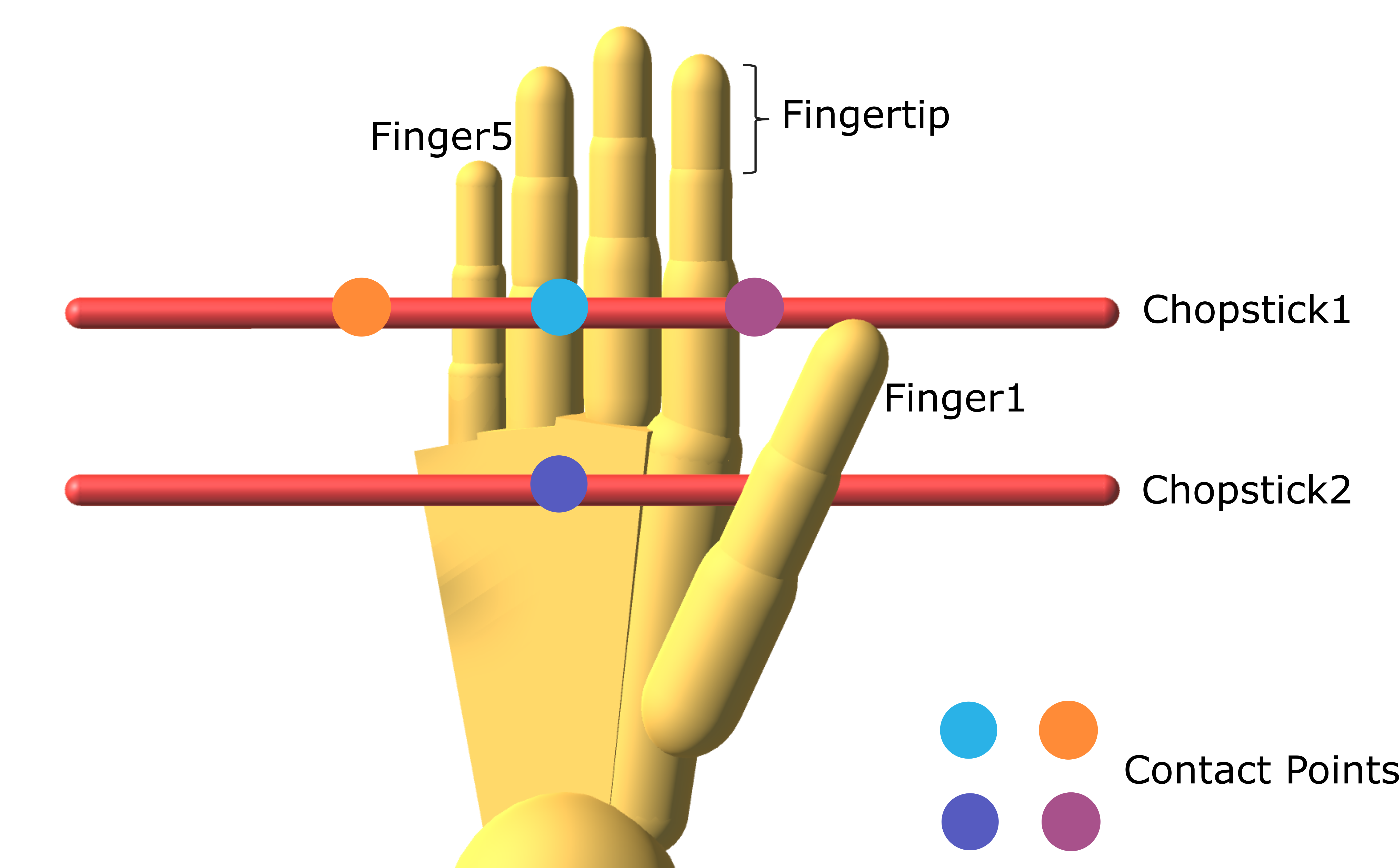}
    \Description{}
    \caption{The default T-Pose used by our IK solver. The upper stick is indexed as Chopstick 1 and the lower stick is Chopstick 2. Fingers are indexed from the thumb to pinky as Finger 1 to 5. A fingertip is the first segment of a finger.}
    \label{fig:pose_pregrasping}
\end{figure}

\subsubsection{Inverse kinematics}
We implement an optimization-based IK solver. The objective function is formulated for a position vector $\bm{x}$ as follows:
\begin{equation}
    \label{eqa:ik_loss}
     \min_{\bm{q}}\sum_{i=1}^{N}\Vert{}f_{i}(\bm{q}) - \bm{p}_i(\bm{x})\Vert_2^2 + \textrm{clog}(\delta_{i}, 0.001)
\end{equation}
where $\bm{p}_i(\bm{x})$ represents the 3D position of the contact point on the chopstick that finger $i$ should touch, $f_{i}(\bm{q})$ calculates the 3D position of the point on fingertip $i$ that is closest to $\bm{p}_i(\bm{x})$, and $\delta_{i}$ represents the depth of penetration between fingertip $i$ and its contacting chopstick. $\textrm{clog}(\cdot)$ is a modified clamped log-barrier function defined as:
\begin{equation}
    \textrm{clog}(z, z_0)=
    \begin{cases}
        -\frac{(z-z_0)^{2}}{z}\textrm{ln}(\frac{z}{z_0}), & 0<z<z_0,\\
        0, & z\ge z_0
    \end{cases}
    \label{eqn:barrier}
\end{equation}
The first term in Equation~\ref{eqa:ik_loss} encourages fingers to touch chopsticks at the positions proposed by BO, and the second term penalizes potential penetrations between the fingers and the chopsticks. Note that we only explicitly control contacts between the chopsticks and the fingertips. Chopsticks contacts on other parts of the fingers or the hand, such as contacts with the valley between the index finger and the thumb, emerge naturally during our DRL policy learning.

{We solve this IK problem using L-BFGS. As the objective function in Equation~\ref{eqa:ik_loss} is highly non-convex, we use the default pose shown in Figure~\ref{fig:pose_pregrasping} to initialize the IK solver. This effectively eliminates poor local minima, such as fingers touching the wrong side of the chopsticks.}

\subsection{Gripping Pose Evaluation}
\label{sec:grip-evaluation}

\begin{figure}[t]
    \centering
    \includegraphics[width=\linewidth]{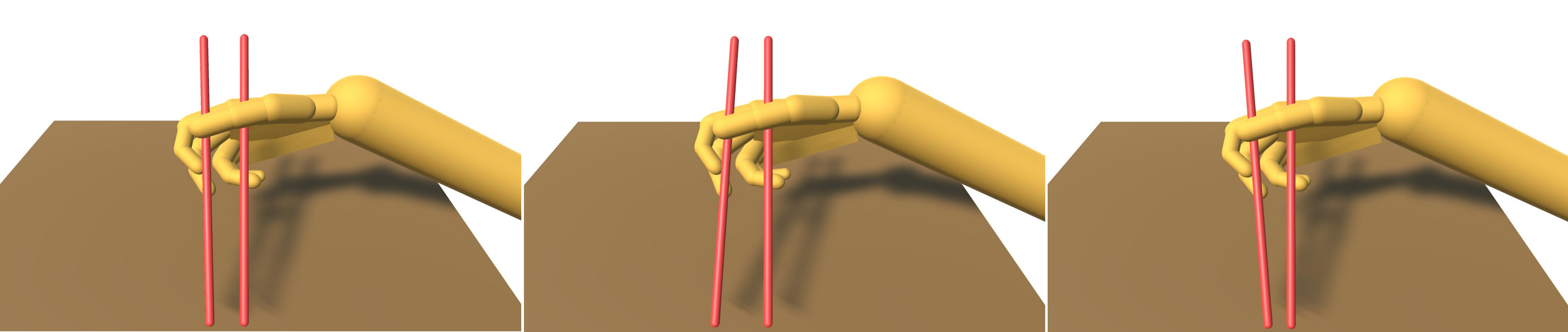}
    \includegraphics[width=\linewidth]{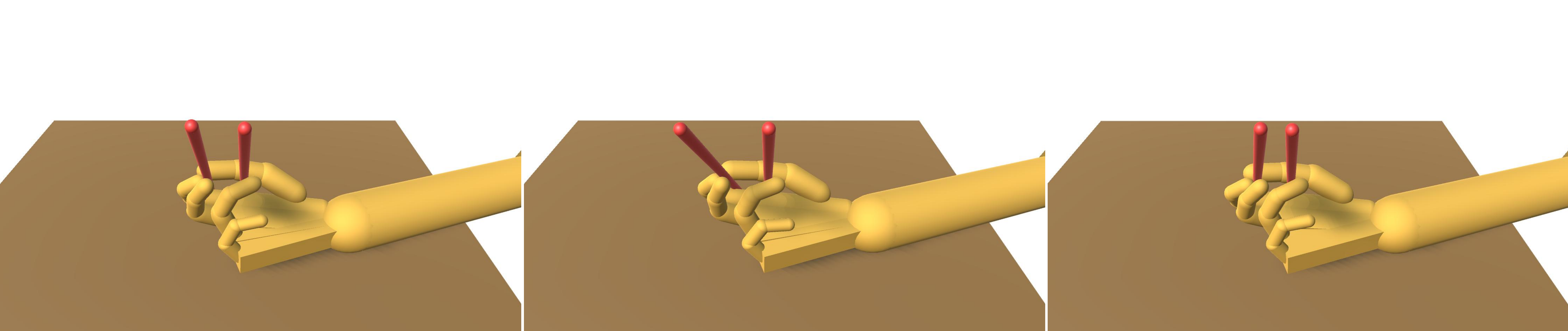}
    \includegraphics[width=\linewidth]{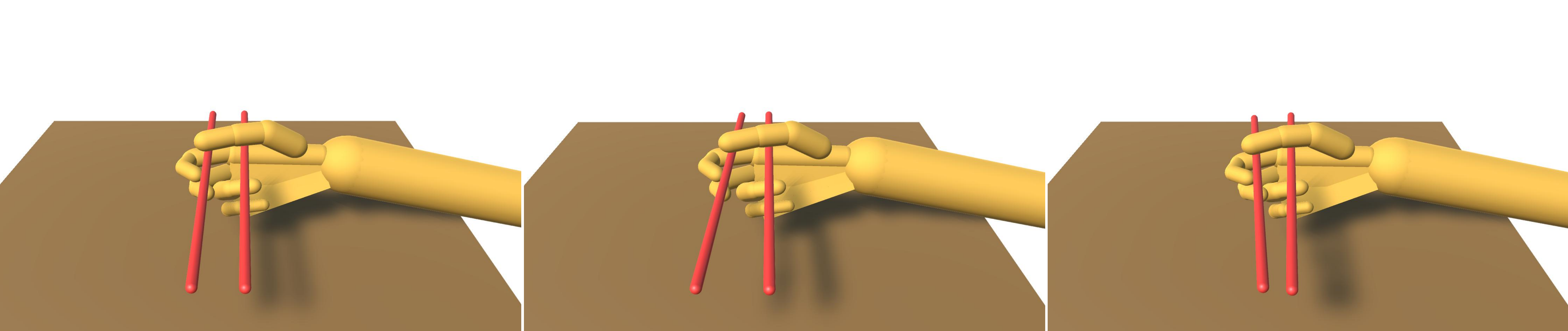}
    \caption{For gripping pose evaluation, three one-second long motions are used to train the hand controller to open and close chopsticks while pointing to different directions.}
    \label{fig:basicskill}
  \end{figure}

We evaluate each candidate gripping pose proposed by BO via deep reinforcement learning of a tracking control policy to accomplish basic chopsticks maneuvers without manipulating any object. The details of the DRL training can be found in Section~\ref{sec:trackingcontroller}. Since no objects are involved in these maneuvers, the corresponding reward term of Equation~\ref{eqn:objloss} is excluded from the reward function. Three one-second long motions as shown in Figure~\ref{fig:basicskill} are used for training, where the chopsticks open and close several times while pointing to different directions. We train each tracking policy for $500$ epochs, then run the learned controller to perform all the test motions again. The average reward of the simulated motions, i.e., the undiscounted cumulative reward divided by the episode length, is sent back to BO as the quality score of the input candidate pose. More complicated maneuvers could be used here for gripping pose evaluation, but the cost of computation will go up as well.

%% file: sec/6_motion_planner.tex
\section{High-level Motion Planning}
\label{sec:motionplanner}

{Given a relocation task characterized by an object to be moved and its target location, we propose a hierarchical control framework where a high-level motion planner is responsible for generating feasible kinematic motion trajectories for the hand, the chopsticks, and the object to accomplish the task. A low-level hand tracking controller, as described in Section~\ref{sec:trackingcontroller}, then tries to follow these trajectories to drive the simulated hand to move towards the object, pick it up and then drop it at the target location using the chopsticks.}

{The motion planner is queried once for each object to be relocated. It synthesizes feasible trajectories in two steps. First, it proposes a chopsticks configuration from a grasping model to ensure quality of the grasp, as will be detailed shortly in Section~\ref{sec:graspingModel}. The grasping model consists of pretrained neural network based models independent of the gripping styles and hand morphologies. Then a trajectory generation module computes the actual trajectories for the hand and arm, through trajectory optimization and inverse kinematics as will be described in Section\ref{sec:trajectoryGeneration}. The trajectory generation algorithm takes as input the chopsticks configuration proposed by the grasping model, the object start and goal locations, as well as the hand morphology and desired gripping style.}

{The trajectories generated by the high-level motion planner are then passed to the low-level hand control policy for tracking. We train one hand controller for each chopstick gripping pose in the desired style by tracking a large set of trajectories created by the motion planner for moving objects between random start and goal locations. Once trained, the policy is robust enough to generalize to new trajectories generated by the same motion planner for new tasks not present in the training set.}

\begin{figure}[t]
    \centering
        \includegraphics[width=0.9\linewidth]{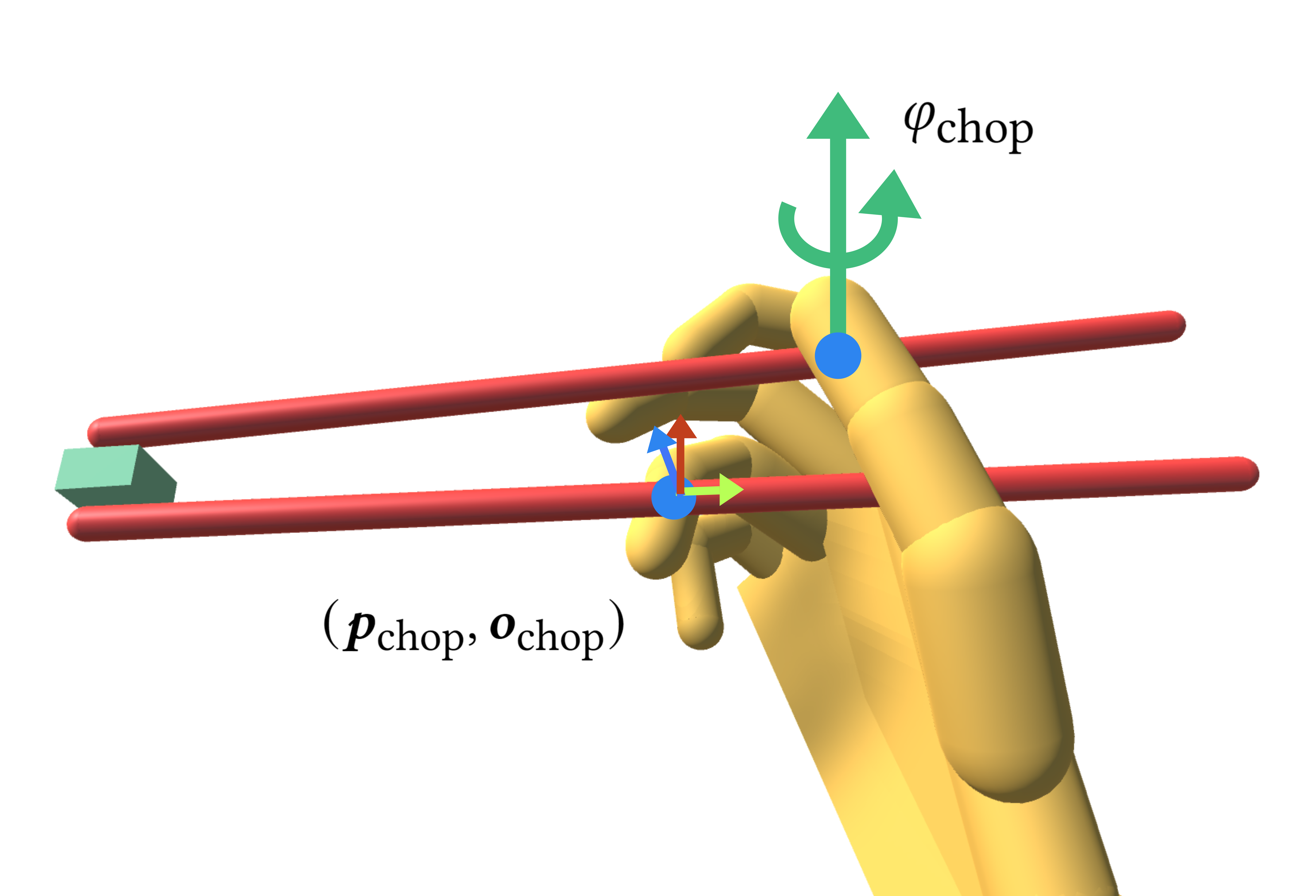}
    \caption{The $7$-DoF chopsticks model used in motion planning. The hand controllers still use the 12-DoF chopsticks model in simulation.}
    \Description{}
    \label{fig:7dofchop}
  \end{figure}
  
\subsection{Grasping Model}  
\label{sec:graspingModel}
  
{Determining how and where to grasp an object using chopsticks is a core mission of the motion planner, which is particularly challenging when dealing with objects of various shapes. We therefore simplify the planning problem in two ways. First, we reduce the DoFs of the chopsticks from twelve to seven, similar to a parallel gripper as illustrated in Figure~\ref{fig:7dofchop}. This is based on the observation that humans usually hold the lower chopstick firmly and rotate the upper chopstick around the object to prepare for grapsing. The configuration of the 7-DoF chopsticks can thus be described by a tuple $\bm{q}_{\text{chop}}=(\bm{p}_{\text{chop}}, \bm{o}_{\text{chop}}, \varphi_{\text{chop}})$, where $\bm{p}_{\text{chop}}$ and $\bm{o}_{\text{chop}}$ represent the position and orientation of the lower chopstick, and the scalar $\varphi_{\text{chop}}$ represents the relative rotation of the upper chopstick with respect to the lower stick around the axis perpendicular to both sticks. Note that this simplified chopsticks model is only used in motion planning, and the full 12-DoF chopsticks are still used in simulation.}

{We further assume that the line connecting the two tips of the chopsticks go through the CoM of the object when grasped, which further reduces the planning problem to three DoFs, i.e., the orientation of the lower chopstick $\bm{o}_{\text{chop}}$. Once $\bm{o}_{\text{chop}}$ is known, we can compute $\bm{p}_{\text{chop}}$ from the object's position and $\varphi_{\text{chop}}$ from the object's shape information.}

{We develop a neural-network based grasping model to predict the optimal grasping configuration for the chopsticks, from the shape and configuration of the object to be grasped. Our grasping model supports efficient planning and replanning at runtime, and is partially inspired by the GraspNet model proposed in~\cite{fang2020graspnet}. More specifically, our grasping model consists of two neural networks. Given the shape parameter of an input object, the \emph{configuration network} nominates a number of candidate chopsticks configurations for grasping, together with their probabilities of success. The chopsticks configurations are specified in the local coordinate frame of the object in the configuration network. We then transform them into the global coordinate frame and pass them into a \emph{reachability network} to estimate how likely each candidate configuration is within the reachable space of the simulated hand and arm. We also compute a continuity score for a configuration by measuring how close it is to the current chopsticks configuration in simulation. Closer chopstick configurations help produce natural hand and arm motions in sequential relocation tasks. We then multiply the probability from the configuration network, the score from the reachability network, and the continuity score all together as the final quality score of a candidate grasping configuration. The motion planner then selects the candidate configuration with the highest score for further trajectory planning.}

\subsubsection{Configuration network}
{There are usually multiple ways to grasp an object, especially for symmetric objects such as a ball. We thus design the configuration network to evaluate multiple candidate configurations simultaneously. Specifically, we uniformly discretize the three dimensional chopsticks configuration space into a set of $N_c$ candidate configurations denoted as $\mathcal{C}_\text{chop}$. As the configuration space corresponds to the orientation of the lower chopstick, this discretization can be performed easily using Euler angles. Then the configuration network takes the shape parameters of an object as input, and computes an $N_c$-dimensional vector representing the success probability for each configuration.}

{The configuration network is implemented as a fully-connected neural network with two 512-unit hidden layers and tanh as the activation function. A softmax layer is appended after the last linear layer to turn its output into probabilities. The network is trained as a multi-class classification problem, matching a given object to configurations in $\mathcal{C}_\text{chop}$ with estimated success probabilities. The top $n_c$ configurations will be passed to the reachability network for further assessment, as will be described shortly. In our implementation, we choose $N_c=2000$ and $n_c=10$.} 

{We use synthetic data for our supervised learning problem. We first generate 100 objects from three primitive shapes and uniformly sample their sizes in the range as shown in Table~\ref{tab:geom_range}. For each object, we then optimize for the grasping configuration using the particle swarm optimization algorithm with $n_c$ random initial solutions \cite{pyswarmsJOSS2018}. The optimization objective follows the grasp quality metric employed by \cite{zhao2013robust}, which encourages the center of the line connecting the chopsticks tips to stay close to the object CoM, and the direction of the connecting line to align with contact surface normals. Each of the optimized configurations is then mapped to its nearest neighbour in $\mathcal{C}_\text{chop}$, for which the success probability in its corresponding one-hot vector is labeled as one.}

\begin{table}[t]
  \caption{{The range of size of our tested geometry primitives.}}
  \label{tab:geom_range}
  \begin{tabular}{c|c|c|c}
    \hline
    \hline
            & Width/Radius    & Length        & Height        \\ \hline
    Sphere  & {[}0.5cm,1cm{]} &               &               \\ \hline
    Capsule & {[}0.5cm,1cm{]} & {[}2cm,4cm{]} &               \\ \hline
    Box     & {[}1cm,2cm{]}   & {[}1cm,2cm{]} & {[}1cm,2cm{]} \\ \hline
    \hline
  \end{tabular}
\end{table}

\subsubsection{Reachability network}
{The reachability network evaluates each candidate chopsticks configuration proposed by the configuration network, in terms of reachability in the global frame. We implement the reachability network as a fully-connected neural network with two 256-unit hidden layers, tanh activation functions, and a sigmoid output layer. Different from the configuration network, the reachability network is trained as a binary classification problem, which outputs the probability that whether a configuration is reachable by the hand and arm.} 

{We use synthetic data for training as well. We first randomly sample $10000$ chopsticks configurations in an operating cuboid of $0.5m \times 0.5m \times 0.25m$ over the table on which objects are placed. We then solve their corresponding arm poses using the analytical arm IK algorithm proposed in \cite{tolani1996ik}. More specifically, from the chopsticks configuration and the chosen gripping pose and hand morphology, we can compute the desired rigid transformation for the hand, from which the IK algorithm then solves for a pose for the 7-DoF arm. If the hand transformation is reachable, the IK solver returns a solution, otherwise the IK solver returns no solution. For training, chopsticks configurations with an IK solution are labeled with probability one while configurations with no IK solutions are labeled with zero.} 

{The reachability network could simply be replaced by the arm IK solver, which is relatively fast and generates solutions of acceptable quality. We choose the neural network based model for reachability tests mainly for the potential of switching to a more expensive IK algorithm in the future that not only tests the reachability of the desired end-effector transformations, but also assesses the naturalness of the solved arm poses. We note that in robotics, machine learning models such as Gaussian mixture models are also used to rapidly predict a feasible catching configuration for object catching tasks \cite{kim2014catching}.}

\subsubsection{Configuration Continuity}
{For a sequence of object relocation tasks, the hand and arm need to manipulate the chopsticks from one configuration to the next. The closer the chopsticks configurations are, the smoother the hand and arm trajectories will be, which improves the overall continuity and naturalness of the animation for consecutive relocation tasks. We thus calculate a continuity score between the candidate chopstick orientation and the current simulated chopstick orientation $\xi =\exp(-5\cdot\Theta(\bm{o}_{\textrm{chop}}, \bm{o}_{\text{chop}}^{\text{0}}))$, where $\bm{o}_{\text{chop}}^{\text{0}}$ is the simulated lower chopstick orientation when the motion planner is activated for a new object relocation task. We then multiply the continuity score with the grasping success probability and the reachability score as the final quality score estimated for a candidate configuration.}

\subsection{Trajectory Generation}
\label{sec:trajectoryGeneration}
{Once the optimal chopsticks configuration is selected by the grasping model, the motion planner needs to generate trajectories for the chopsticks, the hand and arm, and the object. This is done in three steps. First, a collision-free trajectory for the chopsticks is computed through trajectory optimization. Then the arm trajectory is solved by the arm IK algorithm, which also considers continuity when applied to a sequence of IK problems \cite{tolani1996ik}. Lastly, the object trajectory is easily synthesized from the tips of the chopsticks. As the latter two steps are straightforward, we will only discuss our trajectory optimization method for the chopsticks.}

{Following the conventions in hand manipulation research, we segment an object relocation task into three phases: \emph{approaching} the object, \emph{relocating} the object, and then \emph{releasing} the object. In the \emph{releasing} phase, we simply keep the transformation of the chopsticks fixed and set $\varphi_{\text{chop}}$ to $0$. We utilize trajectory optimization for the first two phases. Two additional control points are inserted between the start and end chopsticks configurations, then the chopsticks positions are computed by a degree-three B\'{e}izer curve from the four control points, and the chopsticks orientations are computed with spherical linear interpolation between two control points. We then solve for the optimal control points that generate the shortest collision-free trajectory, which can be formulated as
\begin{equation}
\min_{\bm{q}_1,\bm{q}_2} \int_t \left\Vert{}\dot{m}(t)\right\Vert dt + \text{clog}\left(\delta({m}(t)\right), 0.001) dt
\label{eqn:traj_optim_loss}
\end{equation}
where ${m}(t)={m}(t;\bm{q}_1,\bm{q}_2)$ represent the trajectory defined by the control points $\bm{q}_1,\bm{q}_2$, and $\delta$ computes the penetration depth between the chopsticks and the environment in trajectory $m(t)$, and $\textrm{clog}(\cdot)$ is the same barrier function as defined in Equation~\ref{eqn:barrier}.
The duration of this trajectory is computed as $d_{\text{chop}}/\tilde{v}_{\text{chop}}$, where $d_{\text{chop}}$ is the length of the displacement vector of the chopsticks in the current phase, and $\tilde{v}_{\text{chop}}$ is a hyperparameter indicating the desired speed of the chopsticks. We use $\tilde{v}_{\text{chop}}=0.25\,m/s$ for our demo results. We use numerical integration with a descretized time step $10ms$ to compute the integral in Equation~\ref{eqn:traj_optim_loss}. We use the L-BFGS algorithm with multi-start points to solve the optimization, similar to \cite{li2011hybrid}.
}

%% file: sec/7_results.tex
\begin{figure}[t]
  \centering
  \caption{Our simulated hand has 30 DoFs in total. The thumb has 6 DoFs, and the other fingers each has $4$ DoFs. The four bones connecting the root of the hand to the base of the fingers, indicated as green arrows, each has $2$ DoFs to model small deformations of the palm.}
  \label{fig:handModel}
\end{figure}

\section{Results}
\label{sec:results}
{We implement our system with Pytorch version 1.9.0 \cite{paszke2017automatic} and the MuJoCo physics simulator version 1.5 \cite{todorov2012mujoco}. We use two $6$-DoF capsules to model the chopsticks. The length and radius of the chopsticks are $26$ cm and $0.4$ cm respectively. Our hand model as shown in  Figure~\ref{fig:handModel} has $30$ DoFs in total, and shares the same joint hierarchy as the hand in \cite{ye2012synthesis}. The only difference is that the palm part connecting the thumb and index finger, the so-called union valley in acupuncture, is modeled with a capsule instead of a box. We found that the chopsticks need stable contacts with the valley to perform well, which aligns with human experiences. We mount the hand on a $7$-DoF robot arm consisting of a shoulder, an elbow, and a wrist joint. We run the simulation and control at the same frequency at $100$ $Hz$. Torque limits for hand joints are taken from values of the shadow hand robot used in \cite{rajeswaran2018learning}.}

\input{fig/fiveStyles.tex}

{For low-level DRL policy training, we set $\lambda$ to $0.95$ for both $\mathop{\mathrm{TD}}(\lambda)$ and $\mathop{\mathrm{GAE}}(\lambda)$. The discount factor $\gamma$ is set to $0.99$ and the PPO policy clip ratio is $0.2$. The learning rates of the actor and critic network are set to $3\times10^{-5}$ and $3\times10^{-4}$, respectively. We sample $1\times10^{4}$ state-action tuples with $32$ parallel simulation environments for each training epoch, as all experiments are performed on a workstation with $32$ Intel-i9-9980XE CPUs. Both networks are updated $10$ times in each epoch with a mini-batch size of $256$. During training the action noise linearly decreases from $0.1$ to $0.01$ in the first $5\times 10^{7}$ simulation steps to encourage exploration in early training and exploitation in later training. For learning basic chopsticks maneuvers of fixed length in the stage of gripping pose optimization, we run PPO for $500$ epochs. For learning object relocation, we gradually increase the episode length to facilitate policy learning similar to \cite{peng2018deepmimic}. More specifically, the episode length is $100$ control steps for the first $5000$ epochs, and then increases by $100$ steps every $500$ epochs. The tracking trajectories are generated by the motion planner for $100$ multi-object relocation tasks and last for about $30$ minutes in duration. Each multi-object relocation task contain eight objects to be moved between random locations. We run PPO for $2\times10^{4}$ epochs for each gripping pose in a particular style, which takes roughly two days on our workstation.} 

{For high-level motion planning, the planner is queried whenever a new object is required to be moved. The planning takes roughly $3 \sim 5$ seconds using a non-optimized single-thread implementation on our workstation. The computation bottleneck is the trajectory optimization component. It is possible to further improve the efficiency of the optimizer, but real-time performance may be hard to achieve using the current algorithms.}

{We show optimized chopsticks gripping poses in different styles in Section~\ref{sec:BO_results}. Various learned chopsticks skills are demonstrated with multiple hand morphologies in Sections \ref{sec:skill_learned}, \ref{sec:morphology}, and \ref{sec:robustness}. We demonstrate the ability to learn to use other tools with the same framework in Section~\ref{sec:other_tools}. Lastly in Section~\ref{sec:comparison} we conduct ablation studies and comparisons to validate each component of our framework. We encourage readers to watch our supplementary videos to better comprehend the quality of our learned chopsticks skills.}

\subsection{Diverse Chopsticks Gripping Styles}
\label{sec:BO_results}
\begin{figure}[t]
  \centering
      \includegraphics[width=0.8\linewidth]{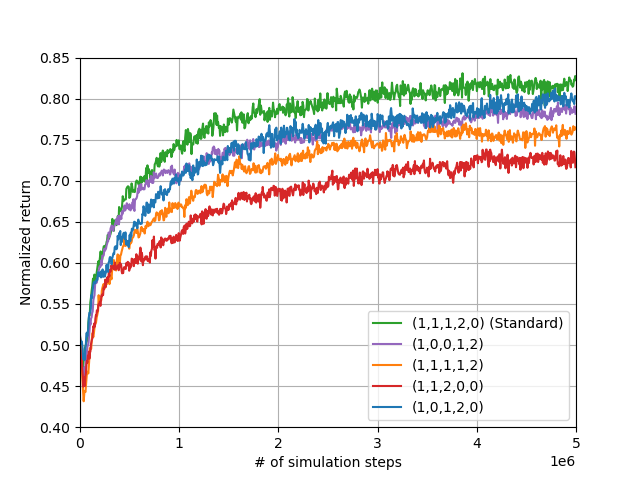}
  \caption{{Training curves of learning basic chopsticks maneuvers using different gripping styles. The standard style is indeed the most efficient way of using chopsticks.}}
  \label{fig:learning_curve}
\end{figure}

{The Bayesian Optimization takes roughly six days in total to optimize gripping poses for the seventeen valid gripping styles. In Figure~\ref{fig:pose_BO} we show the most distinctive five styles: $(1,1,1,2,0)$, $(1,1,1,1,2)$, $(1,1,2,0,0)$, $(1,0,1,2,0)$ and $(1,0,0,1,2)$. Four out of the five styles are documented in \cite{macro2021}, following which we name the gripping styles. The standard style with $c=(1,1,1,2,0)$ is the most common way to use chopsticks, and also considered the best way by common belief. Our BO results verify that it is indeed the most efficient way to use chopsticks, as policy learning for the standard style converges the fastest with the highest normalized return, as shown in Figure \ref{fig:learning_curve}. The optimized gripping poses for the remaining twelve styles are shown in Figure~\ref{fig:other12styles} in Appendix~\ref{sec:app:poses}.}

\input{fig/skills.tex}

\subsection{Chopsticks Skills for Object Relocation}
\label{sec:skill_learned}
{We evaluate our framework with object relocation tasks where the simulated hand uses the chopsticks to grasp objects of various shapes and sizes, and then move or throw them to some desired target locations.}

\subsubsection{Grasp and Move}
{We train hand control policies using sequential grasp-and-move object relocation tasks. Figure~\ref{fig:grasping_training} shows the learned chopsticks skills using different gripping styles. We also test the generalization ability of the learned controllers with a more challenging stacking task, as shown in the teaser Figure~\ref{fig:stack}. This task is not included during policy training, yet the learned controllers can directly finish the stacking task without any fine tuning.}

\subsubsection{Grasp and Throw}
{Throwing is an efficient means to relocate objects when the target position is out of reach by the arm and hand. Our motion planner generates kinematic chopsticks trajectories for throwing tasks following the method described in \cite{zeng2020tossingbot}. More specifically, we first estimate the needed position and velocity of the chopsticks at the end of the relocating phase, given the final target position of the object. Then we use trajectory optimization for the relocating phase to first move the chopsticks to the estimated releasing position and then rotate them to achieve the estimated releasing velocity. Trajectory generation for the approaching and releasing phases are the same as described in Section~\ref{sec:trajectoryGeneration}. The trajectory of the object after being thrown is assumed to be projectile, and aerodynamic drags are ignored. We train the hand control policy by tracking the above planned trajectories, the same as for the grasp-and-move task. Figure~\ref{fig:throwing} shows one of our throwing results where a box is grasped and thrown to hit another stack of boxes far away.}

\subsubsection{Parameterized Controllers for Different Holding Positions}
{Humans can hold chopsticks at different positions in similar poses. For example, children and beginners tend to hold chopsticks near the tips for easier control. Expert users hold chopsticks near the rear ends to increase the reachability and avoid potential harmful injury from hot food. Our system can learn such parameterized controllers with ease.}

\input{fig/threeHands.tex}

{We parameterize the holding position of the chopsticks by translating the chopsticks along the axial direction of the lower chopstick. Such translations do not change the relative rotation between the chopsticks and the palm. The holding position is also added into the input state representation of the hand control policy to enable learning of parameterized controllers. Figure~\ref{fig:parameterized_control} shows that the learned controllers can hold the chopsticks at multiple positions to relocate objects. We allow a range of $\pm5$ cm for the chopsticks translation in our tests. Note that policies using some holding positions are harder to train than others, which leads to biased sampling during DRL training. We adopt the adaptive sampling strategy similar to \cite{won2019learning} to increase samples for harder tasks. }

\subsection{Diverse Hand Morphologies}
\label{sec:morphology}
{Our framework can learn chopstick skills for drastically different hand morphologies. In Figure~\ref{fig:retarget_model} we visualize three additional hand models that we have tested: a hand with long fingers, a large hand, and a tri-finger hand. The fingers of the long hand double the lengths of the fingers of the standard hand. The large hand doubles the size of the standard hand in all dimensions for both the fingers and the palm. The tri-finger hand is designed by ourselves following claw toy grabbers commonly seen in arcade machines. We show some of the learned skills for each hand in the teaser Figures~\ref{fig:teaserLargehand}, \ref{fig:teaserTrifinger}, and Figure~\ref{fig:retarget_skill}. We are happy to find that the tri-finger hand can learn to use chopsticks successfully as well. Note that we still use the same arm model for these different hands, although we scale the radius of the arm to match the dimension of the hands for better visualization. Changing the length of the arm model mainly affects the reachability of the hand and does not affect the ability to use chopsticks by hand.}

\begin{figure}[t]
  \begin{subfigure}[b]{0.99\linewidth}
         \centering
         \includegraphics[width=\linewidth]{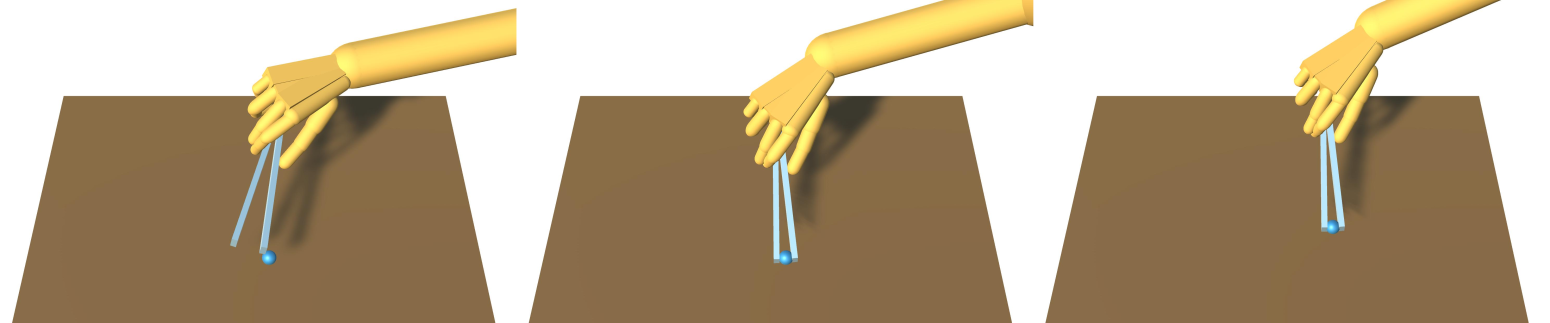}
         \label{fig:tong}
     \end{subfigure}\\
     \begin{subfigure}[b]{0.99\linewidth}
         \centering
         \includegraphics[width=\linewidth]{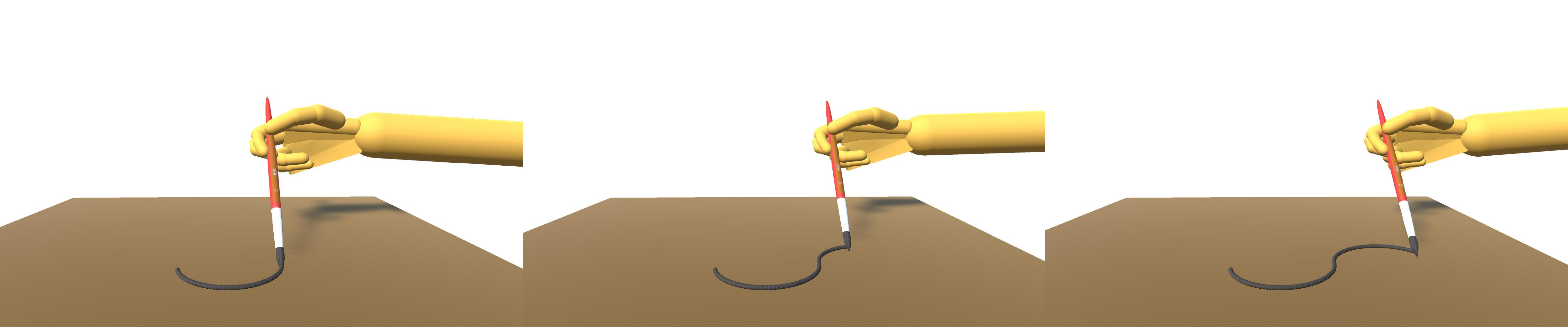}
         \label{fig:pen}
    \end{subfigure}\\
  \caption{Top: relocating a ball using a pair of tongs. Bottom: tracing a curve using a brush.}
  \label{fig:othertools}
\end{figure}

\subsection{Using Other Tools}
\label{sec:other_tools}
{Our framework can be applied to learn physics-based hand controllers to use other types of tools. In Figure~\ref{fig:othertools} we demonstrate relocating a ball with a pair of tongs, and tracing a curve using a brush. The tongs are modeled as a $7$-DoF gripper, with its two bars modeled as  $26cm\times0.8cm\times2cm$ cuboids. We model the pivot of the tongs with an angular spring to provide restoring torques. More specifically, the torques are computed as $-k (\theta - \theta_{\textrm{rest}})$, where $k=20 Nm$ is the stiffness parameter, and $\theta_{\textrm{rest}}=0.2 rad$ is the resting angle between the two bars. The brush is modeled as a long capsule of $26cm$ in length and $0.4cm$ in radius.}

{We follow the same pipeline designed to learn chopsticks skills to learn tongs skills. The only difference is that we only allow the thumb, index, and middle fingers to contact the tongs. For curve tracing with the brush, we use a much simplified version of the motion planner. As there is no tool-object interactions involved, the grasping model to predict a configuration for the tool to grasp the object is not needed. Trajectory generation for the tool is also much easier, by simply tracing the curves using the tool tip, and the hand and arm trajectories can be solved directly from IK.}

\input{fig/curves.tex}

\subsection{Robustness}
\label{sec:robustness}
{Our learned physics-based hand controllers can handle noises, perturbations, and uncertainties to certain extent. Figure~\ref{fig:grasping_dynamic} shows such an example. The target object rotates due to unexpected collisions when the chopsticks approach the object. The controller can still grasp the rotated box successfully according to the plan computed before the collision. Figure~\ref{fig:replanning} shows another example of relocating two stacked boxes together. The chopsticks only grasp the bottom box during the first move and the top box starts to fall when the first move ends. The controller can grasp the top box during falling according to the plan made at the beginning of the second move.}

{Generally speaking, the success rate of chopsticks-based grasps depends on many factors, such as the weight of the objects, the shape of the chopsticks, the hand gripping styles etc. Here we report the grasping success rates with respect to the object-chopsticks friction coefficients for consecutive relocation tasks. More specifically, we use our motion planner to generate $50$ sequential grasp-and-move tasks, each containing eight random objects to be relocated. The friction coefficients are set to one of three values around the MuJoCo default $1.0$. The success rates are plotted in Figure~\ref{fig:successrate}. Interestingly enough, we find our hand controllers always succeed for the first few grasps. Failures only occur at the latter stage. Objects with smaller friction coefficients are indeed more difficult to be grasped. Another contributing factor to the failures is likely to be the soft contact model implemented by MuJoCo as discussed in Section~\ref{sec:contactDynamics}, as we observe failure cases where the chopsticks gradually slip out of the hand after a few moves.}

{To handle failure cases and bigger perturbations, replanning by the motion planner according to the latest scene configuration is needed. We note that human-level performance is not 100\% for chopsticks-based grasping tasks either, especially for tiny or slippery objects, so replanning is also seen in real-life scenarios. We can simulate such behavior by deliberately adding Gaussian noises $\mathcal{N}(0, \sigma^2 I)$ to $\bm{p}_{\text{chop}}$ of the chopsticks configurations computed by the grasping model. $\sigma=3 mm$ in our experiments. The hand controller tracks the noisy motion plans, which eventually leads to failures in sequential relocation tasks. Upon such failures, we replan again with no added noise. Then the object can be grasped successfully. We encourage the readers to refer to the ``planning with noise'' examples in the supplementary video.}

\subsection{Ablation and Comparison}
\label{sec:comparison}
{We conduct ablation and comparative studies to justify and validate two major components of our control and learning framework: the hierarchical control structure, and the gripping pose optimization module. Additionally, we report the effect of related MuJoCo contact dynamics parameters.}

\subsubsection{Hierarchical Control}
{Our high-level motion planner computes kinematic motion trajectories for the low-level hand controllers to track, which greatly simplifies the DRL reward design and improves the overall motion quality without using pre-captured example data. For system ablation without using the motion planner, we use a sparse reward to directly measure the success of a task at the very end of the task. More specifically, we return a reward $1.0$ if the object can reach the target position within $0.5cm$ in distance, and a reward $0.0$ otherwise. We use exactly the same PPO algorithm with the same parameter and hyperparameter settings. As shown in Figure~\ref{fig:comparison}, this scheme does not learn at all, with close to zero returns after $100$ million simulation steps. The learned hand controller just randomly moves the fingers and cannot even hold the chopsticks firmly, let alone using chopsticks to grasp objects.}

\begin{figure}[t]
     \begin{subfigure}[b]{0.48\linewidth}
         \centering
         \includegraphics[width=\linewidth]{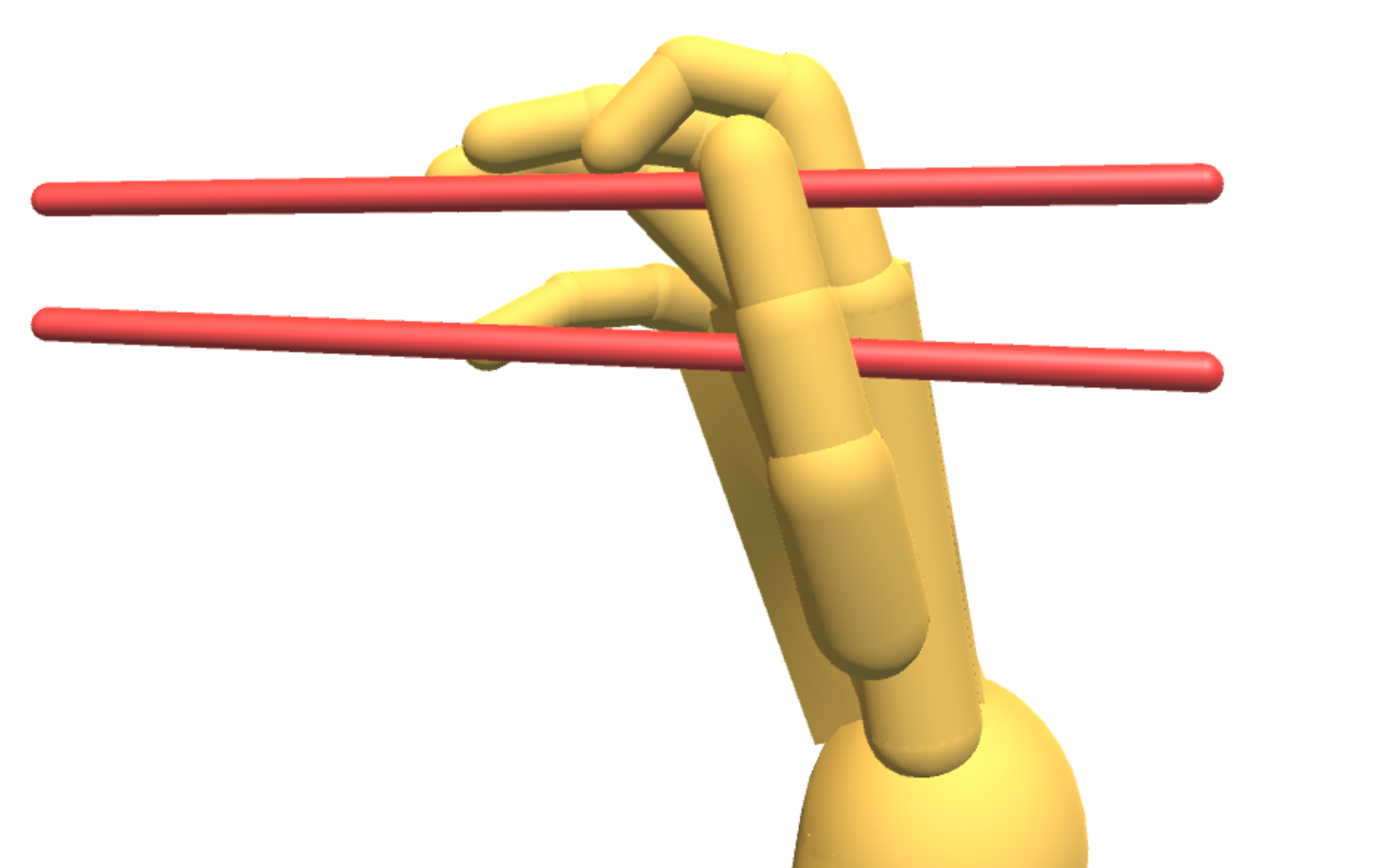}
         \label{fig:hand craft}
     \end{subfigure}
     \begin{subfigure}[b]{0.48\linewidth}
         \centering
         \includegraphics[width=\linewidth]{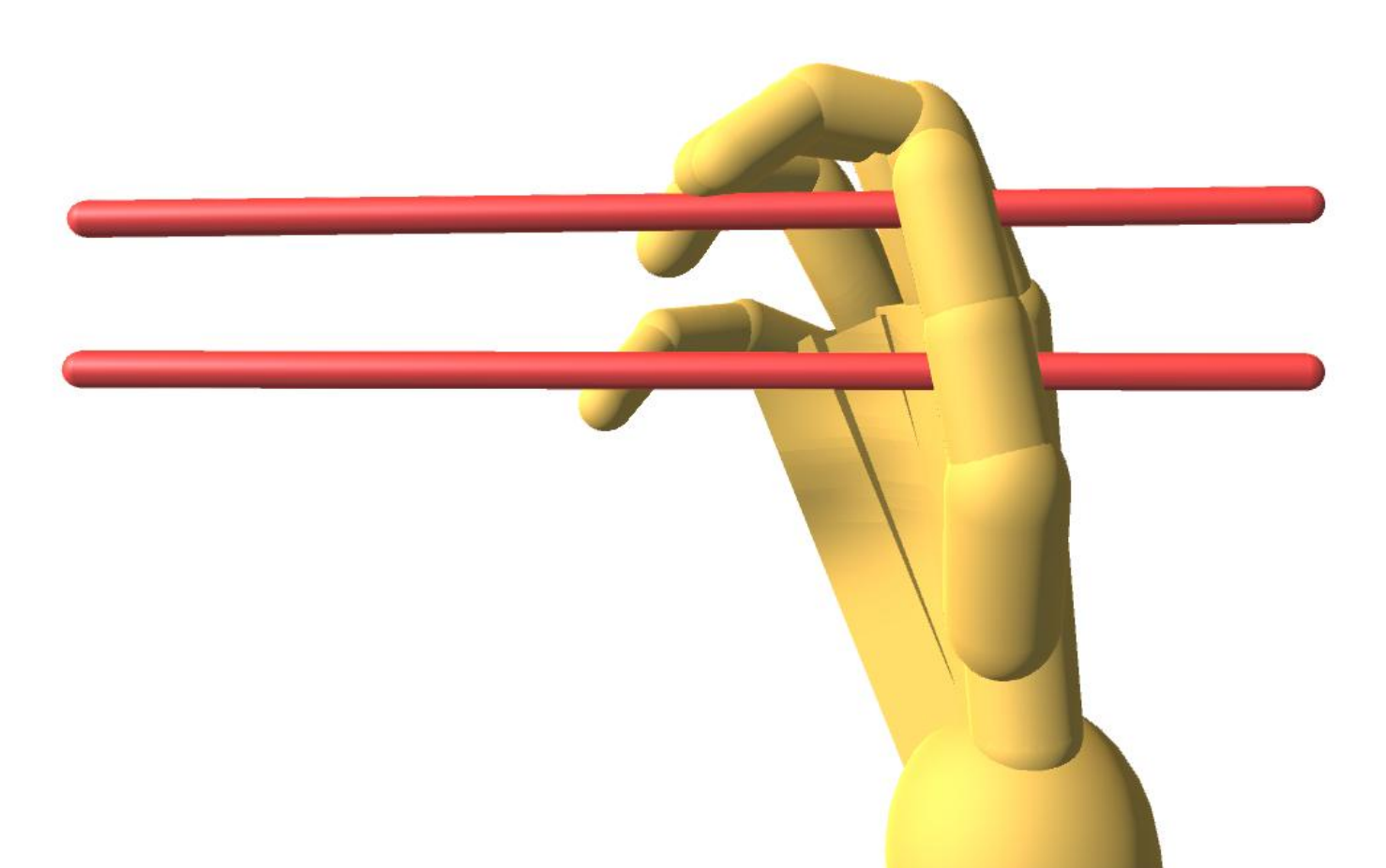}
         \label{fig:bighand_model}
     \end{subfigure}
  \caption{A handcrafted pose (left) vs. our optimized gripping pose (right) for the standard gripping style.}
  \label{fig:handcraft_pose}
\end{figure}

\subsubsection{Gripping Pose Optimization}
{We ablate the gripping pose optimization with two alternatives: policy learning with the default T-pose and a handcrafted gripping pose. The T-pose has no preferred finger-chopsticks contact relationships, so the reward term $r_{\textrm{contact}}$ is not used in DRL training. The handcrafted gripping pose, as shown in Figure~\ref{fig:handcraft_pose} left, is tuned to follow reference pictures in \cite{macro2021}. As shown in Figure~\ref{fig:comparison}, policy training with T-pose progresses poorly. The final policy cannot even hold the chopsticks steadily. Policy learning with the handcrafted pose shows better performance, but still converges slower and has inferior final policy, compared to learning with our BO optimized pose as shown in Figure~\ref{fig:handcraft_pose} right.}

\begin{figure}[t]
    \begin{subfigure}[b]{\linewidth}
         \centering
         \includegraphics[width=0.8\linewidth]{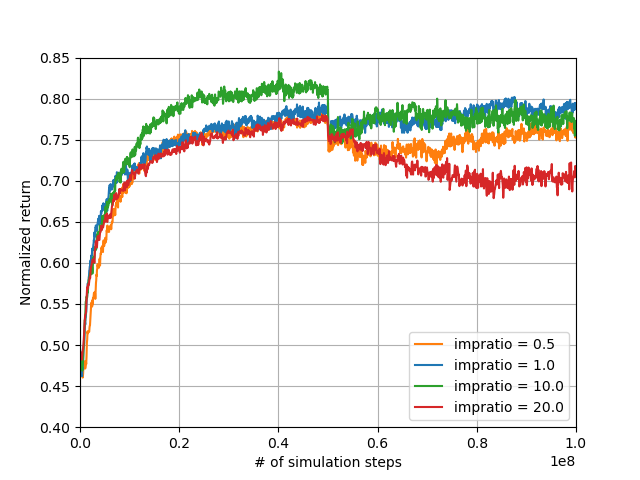}
          \caption{\emph{solimp} controls the contact constraint impedance. The default value performs the best in terms of policy learning.}
         \label{fig:solimp}         
    \end{subfigure}\\  \begin{subfigure}[b]{\linewidth}
         \centering
         \includegraphics[width=0.8\linewidth]{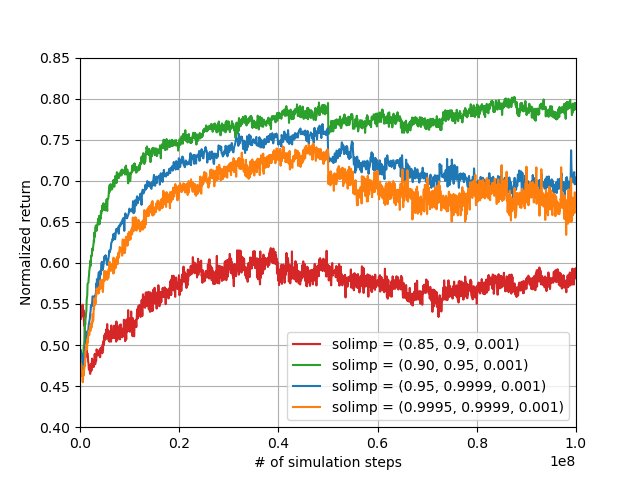}
         \caption{\emph{impratio} controls the frictional-to-normal constraint impedance. Mid-range values perform the best in terms of policy learning.}
         \label{fig:impratio}
    \end{subfigure}\\

     \caption{Policy training curves for different values of contact dynamics parameters in MuJoCo.}
\end{figure}

\subsubsection{Contact Dynamics}
\label{sec:contactDynamics}
{Chopsticks skills are contact-rich control problems for which contact dynamics plays an important role in the success, robustness, and quality of the learned controllers. We train our hand controllers and synthesize the demo animations using the MuJoCo default simulation parameters. However, the synthesized motions contain visible artifacts including penetrations of the chopsticks into the fingers, and objects being moved and stacked looking too soft or sticky. We therefore explore more settings for a few contact dynamics parameters in MuJoCo.}

{Contacts in MuJoCo are inherently soft and modeled as a convex optimization rather than the conventional LCP (Linear Complementarity Problem) \cite{todorov2014convex}. The default parameters are also set to prefer soft contacts to encourage stable simulations. In particular, the parameter \emph{solimp} parameterizes the impedance, which controls how ``hard'' the contact constraints are. The default setting of \emph{solimp} is $(0.90,0.95,0.001)$. The valid range of the first two values of \emph{solimp} is $[0.0001, 0.9999]$. The larger they are, the harder the contact constraints will be. In Figure~\ref{fig:solimp} we show the policy learning curves for different settings of \emph{solimp}. Policy training with the default setting does perform the best, although at the cost of visible penetrations. However, it does not mean that the more penetrations allowed, the better the policy will learn. Policy learning with (0.85,0.9,0.001) actually fails due to large penetrations between the fingers and chopsticks.}

{A related issue caused by MuJoCo's soft contact model is that gradual contact slip cannot be avoided, even with large friction coefficients. There is a parameter \emph{impratio} that determines the ratio of frictional-to-normal constraint impedance for friction cones. Settings larger than 1 cause friction forces to be “harder” than normal forces, having the general effect of reducing slip. We train the low-level control policy for the standard gripping style with different \emph{impratio} values centered around its default value $1.0$. As shown in Figure~\ref{fig:impratio}, policy learning with mid-range values have similar final performance, while policy learning with too small or too large values shows worse performance.}

%% file: fig/fiveStyles.tex
\begin{figure}[t]
  \centering
    \begin{subfigure}[b]{0.99\linewidth}
         \centering
         \includegraphics[width=\linewidth]{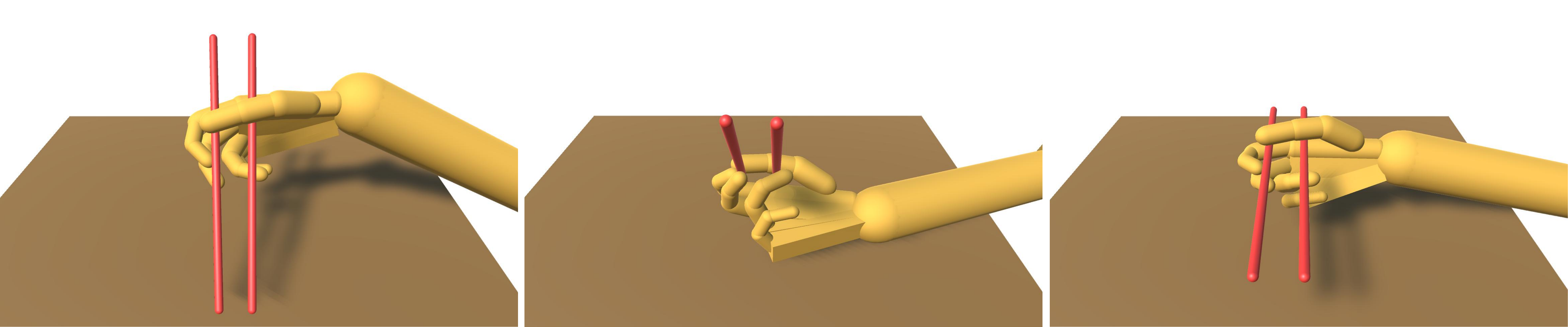}
         \caption{$\bm{c}=(1,1,1,2,0)$ Standard style}
         \label{fig:1120}
    \end{subfigure}\\
    \begin{subfigure}[b]{0.99\linewidth}
         \centering
         \includegraphics[width=\linewidth]{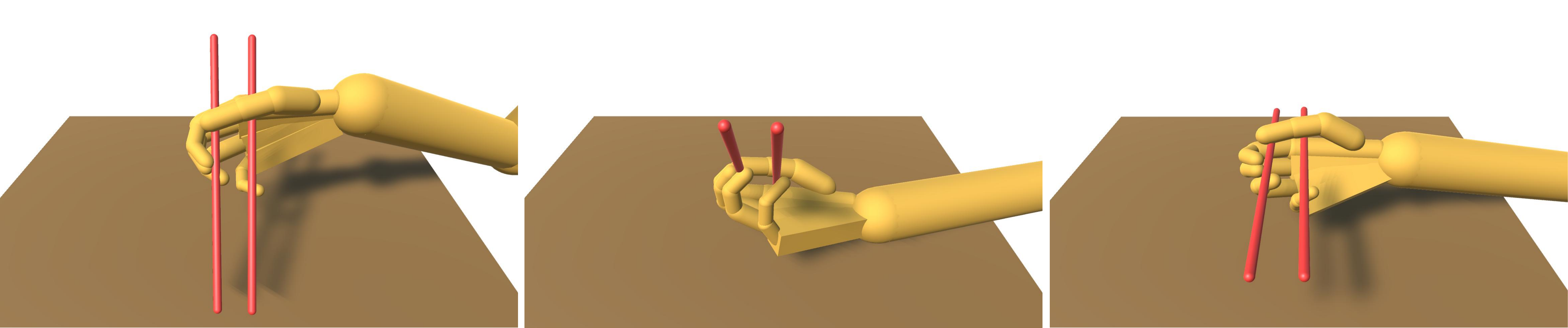}
         \caption{$\bm{c}=(1,1,1,1,2)$ Forsaken pinky style}
         \label{fig:1112}         
    \end{subfigure}\\
    \begin{subfigure}[b]{0.99\linewidth}
         \centering
         \includegraphics[width=\linewidth]{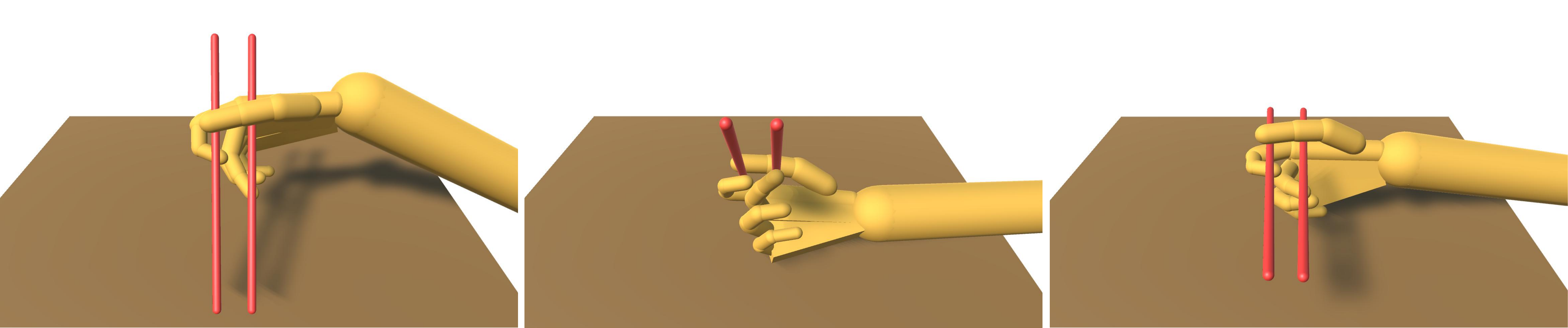}
         \caption{$\bm{c}=(1,1,2,0,0)$ Right-hand rule style}
         \label{fig:1200}
    \end{subfigure}\\
    \begin{subfigure}[b]{0.99\linewidth}
         \centering
         \includegraphics[width=\linewidth]{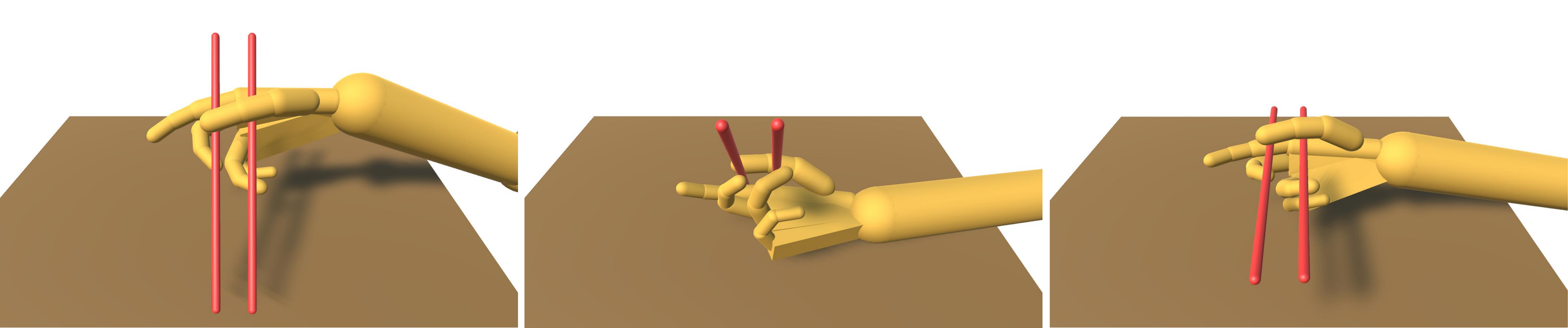}
         \caption{$\bm{c}=(1,0,1,2,0)$ Dino claws style}
         \label{fig:0120}         
    \end{subfigure}\\
    \begin{subfigure}[b]{0.99\linewidth}
         \centering
         \includegraphics[width=\linewidth]{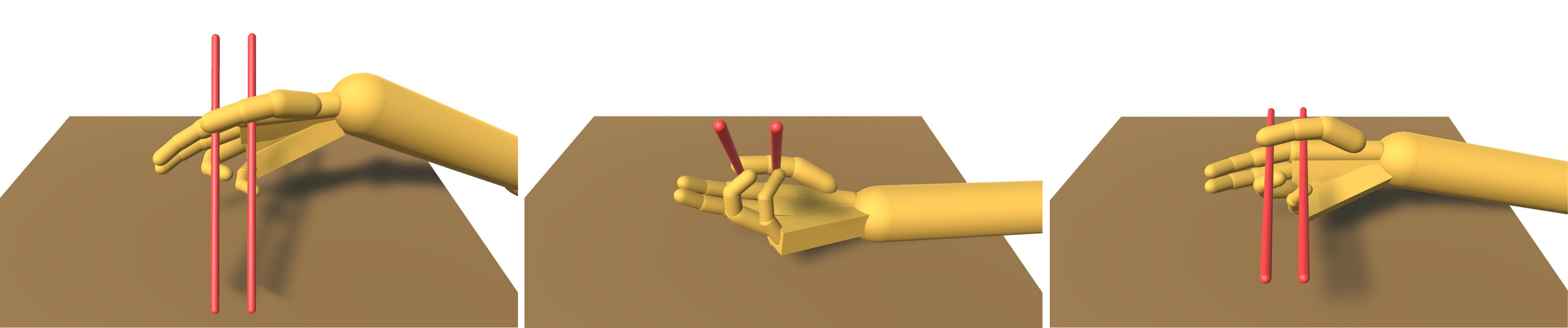}
         \caption{$\bm{c}=(1,0,0,1,2)$ style with no convention name}
         \label{fig:0012}
     \end{subfigure}
  \caption{{Visualization of the optimized chopsticks gripping poses in five styles, performing the basic open-and-close chopsticks maneuver.}}
  \label{fig:pose_BO}
\end{figure}

\begin{figure}[t]
  \centering
    \begin{subfigure}[b]{0.99\linewidth}
         \centering
         \includegraphics[width=\linewidth]{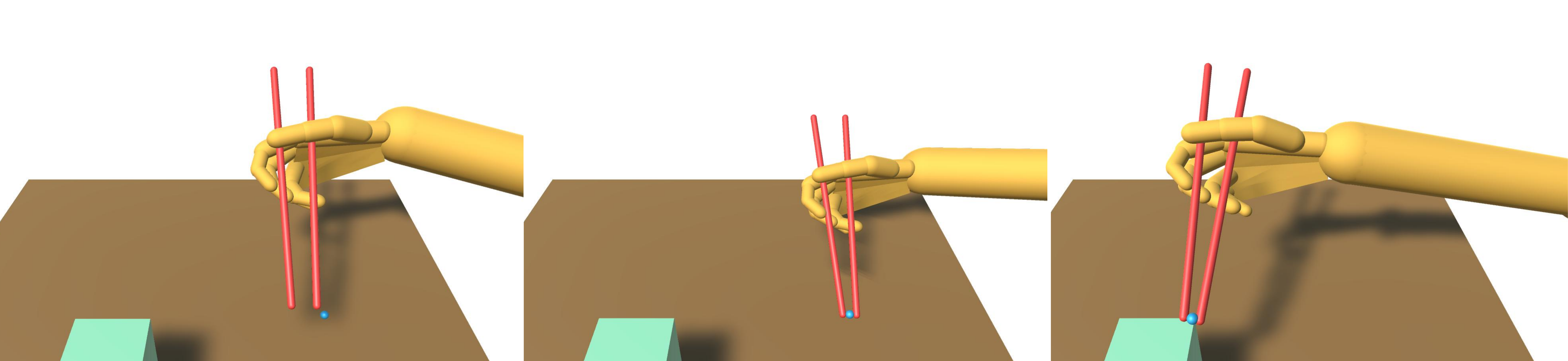}
         \caption{$\bm{c}=(1,1,1,2,0)$ Standard style}
         \label{fig:1120_seq}
    \end{subfigure}\\
    \begin{subfigure}[b]{0.99\linewidth}
         \centering
         \includegraphics[width=\linewidth]{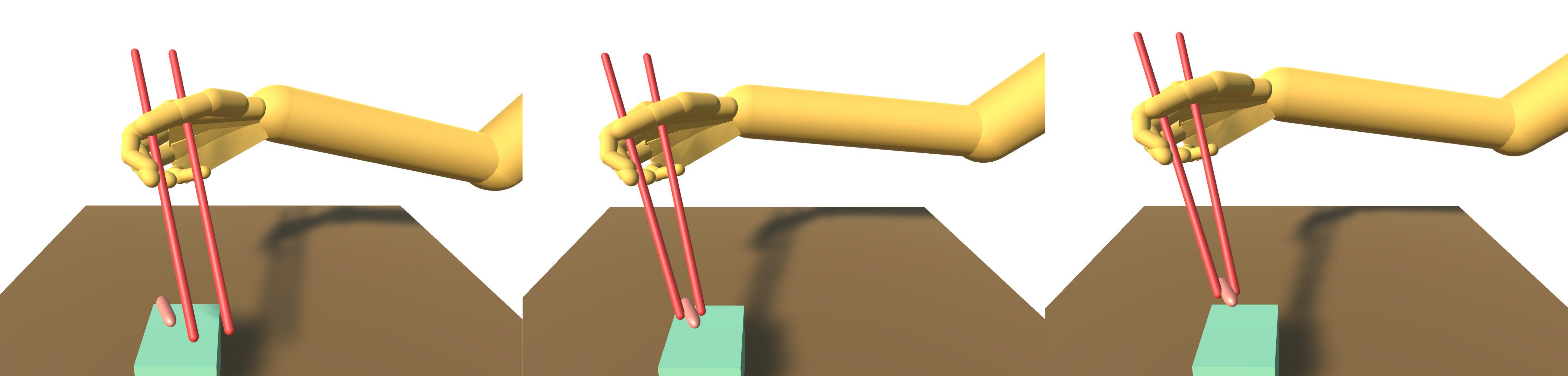}
         \caption{$\bm{c}=(1,1,1,1,2)$ Forsaken pinky style}
         \label{fig:1112_seq}         
    \end{subfigure}\\
    \begin{subfigure}[b]{0.99\linewidth}
         \centering
         \includegraphics[width=\linewidth]{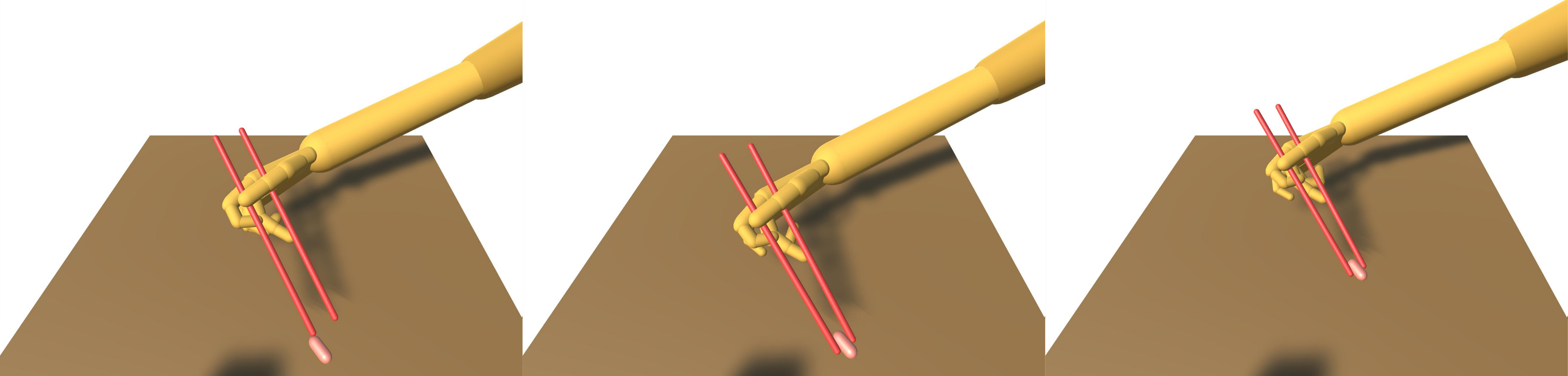}
         \caption{$\bm{c}=(1,1,2,0,0)$ Right-hand rule style}
         \label{fig:1200_seq}
    \end{subfigure}\\
    \begin{subfigure}[b]{0.99\linewidth}
         \centering
         \includegraphics[width=\linewidth]{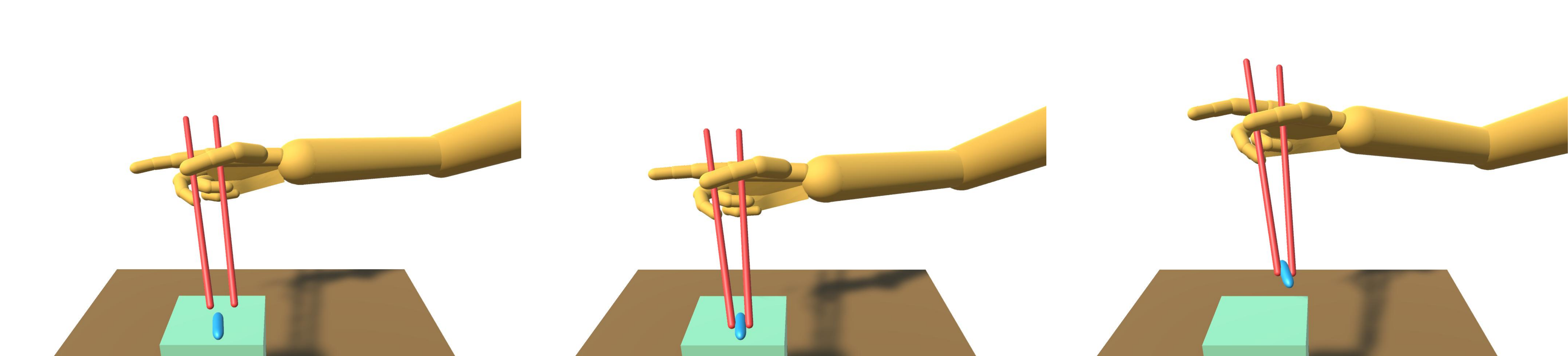}
         \caption{$\bm{c}=(1,0,1,2,0)$ Dino claws style}
         \label{fig:0120_seq}         
    \end{subfigure}\\
    \begin{subfigure}[b]{0.99\linewidth}
         \centering
         \includegraphics[width=\linewidth]{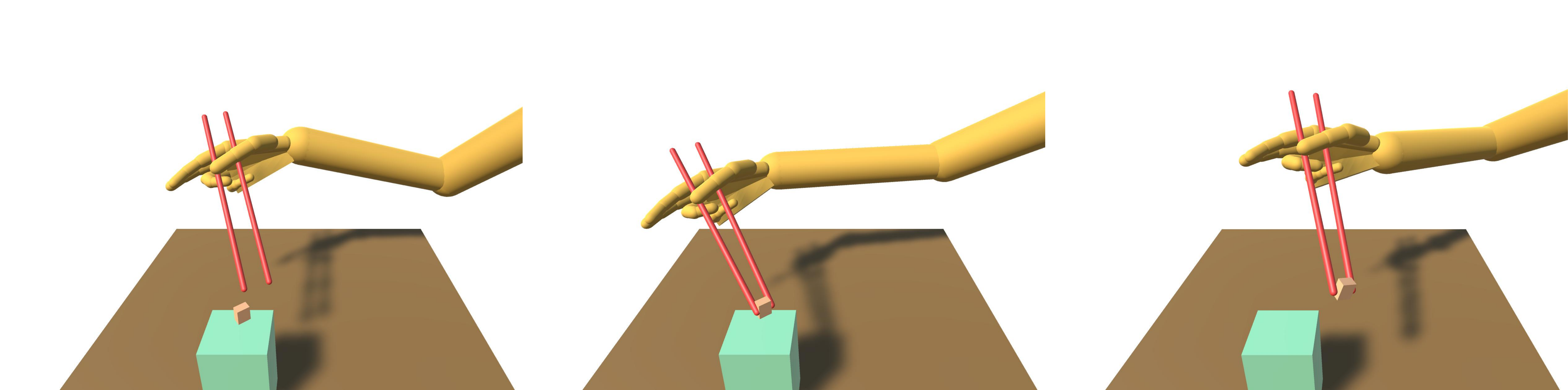}
         \caption{$\bm{c}=(1,0,0,1,2)$ style with no convention name}
         \label{fig:0012_seq}
     \end{subfigure}
  \caption{The hand controls chopsticks to grasp and move various objects with different gripping poses.}
  \label{fig:grasping_training}
\end{figure}

%% file: fig/skills.tex
\begin{figure*}[t]
  \centering
  \includegraphics[width=\linewidth]{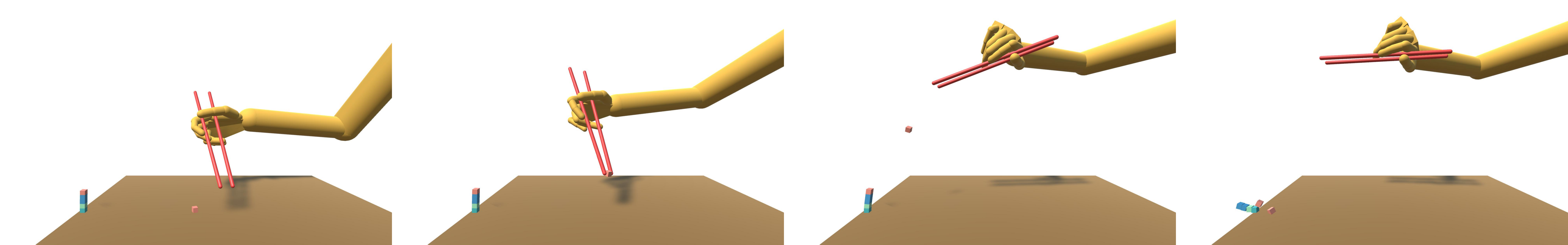}
  \caption{{Grasping and throwing a box to hit another stack of boxes.}}
  \label{fig:throwing}
\end{figure*}

\begin{figure*}[t]
  \centering
       \begin{subfigure}[b]{0.48\linewidth}
         \centering
         \includegraphics[width=\linewidth]{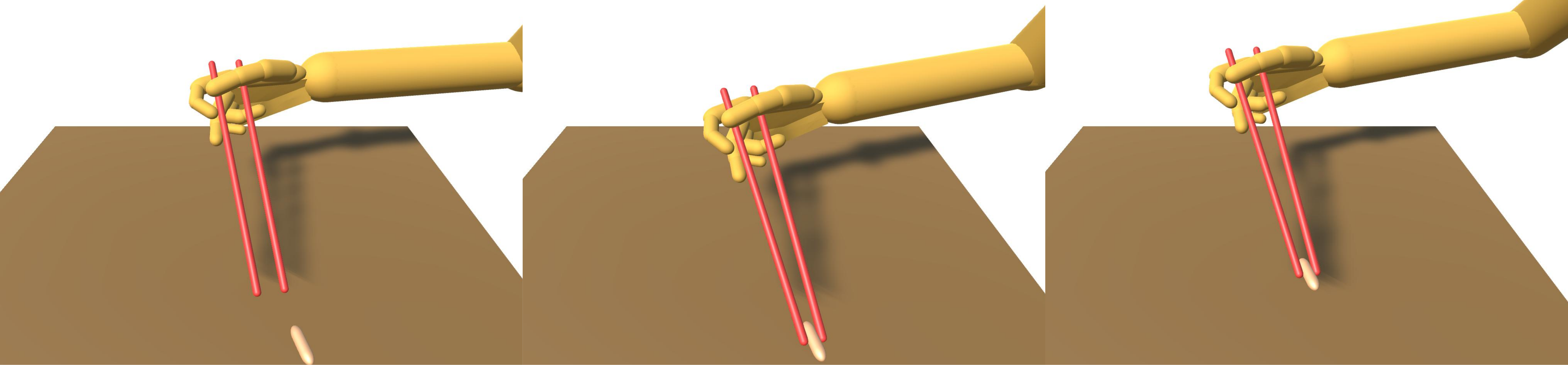}
         \caption{}
         \label{fig:hold_high}
         
    \end{subfigure}
          \begin{subfigure}[b]{0.48\linewidth}
         \centering
         \includegraphics[width=\linewidth]{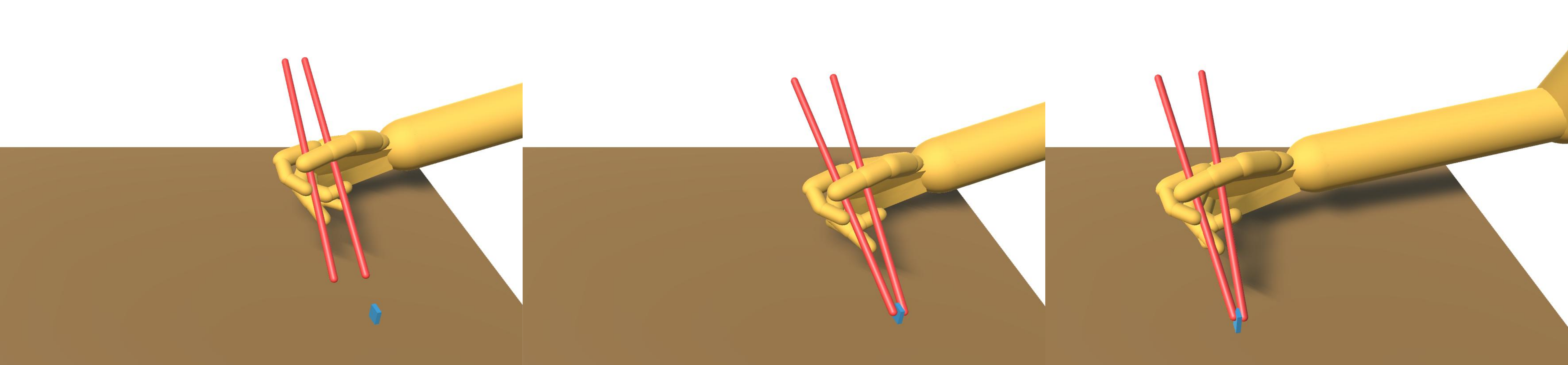}
         \caption{}
         \label{fig:hold_low}
     \end{subfigure}
  \caption{{Our controllers are parameterized so that the hand can hold the chopsticks high or low.}}
  \label{fig:parameterized_control}
\end{figure*}

\begin{figure*}[t]
  \centering
  \includegraphics[width=\linewidth]{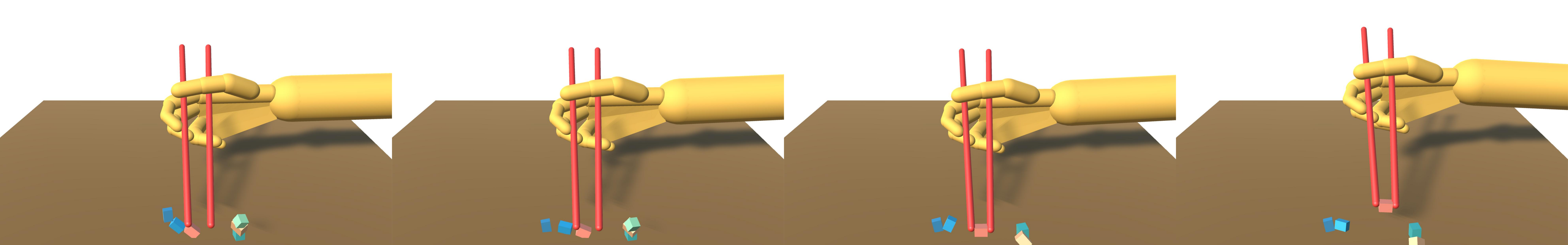}
  \caption{{Grasping a moving object without replanning.}}
  \label{fig:grasping_dynamic}
\end{figure*}

\begin{figure*}[t]
  \centering
  \includegraphics[width=\linewidth]{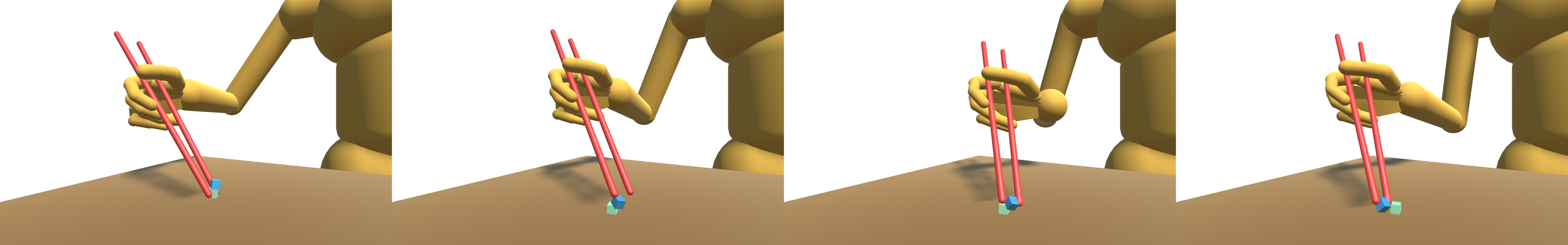}
  \caption{{Relocating two boxes together. The top blue box falls down at the end of the first move. With replanning, the chopsticks are able to grasp the top box during falling to continue the second move.}}
  \label{fig:replanning}
\end{figure*}

%% file: fig/threeHands.tex
\begin{figure}[t]
       \begin{subfigure}[b]{0.24\linewidth}
         \centering
         \includegraphics[width=\linewidth]{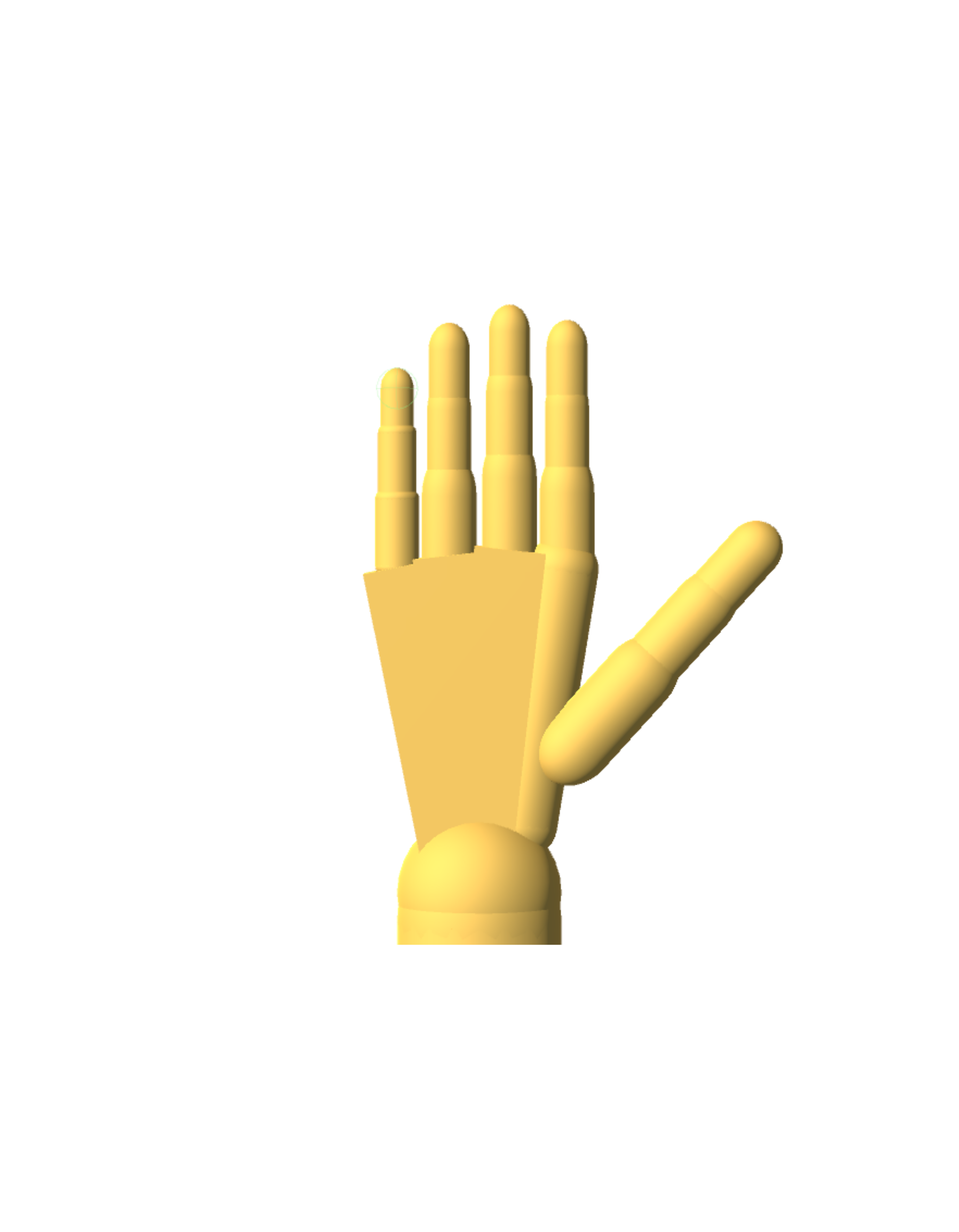}
         \caption{Standard hand}
         \label{fig:defaulthand_model}
     \end{subfigure}
    \begin{subfigure}[b]{0.24\linewidth}
         \centering
         \includegraphics[width=\linewidth]{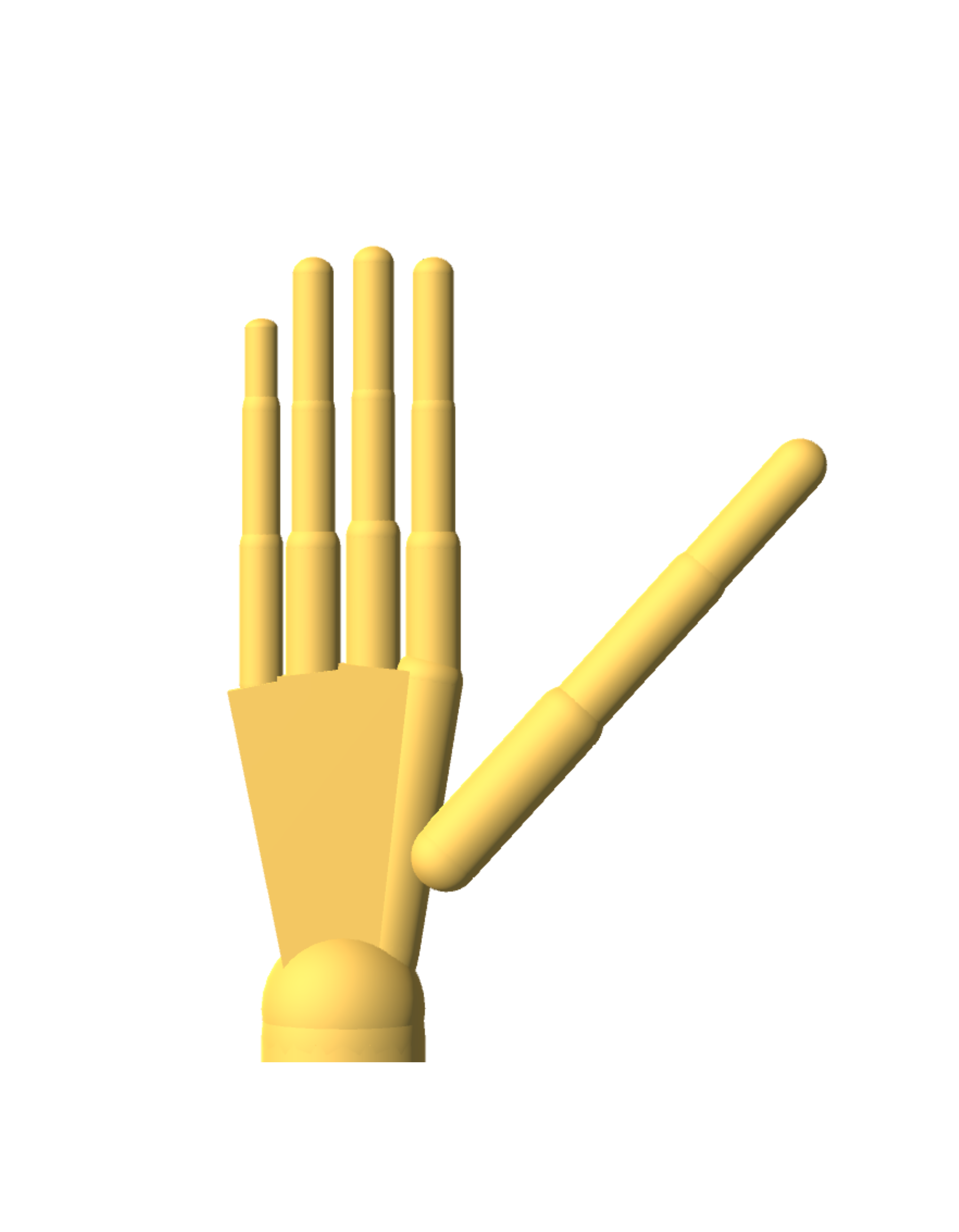}
         \caption{Long hand}
         \label{fig:longhand_model}
     \end{subfigure}
     \begin{subfigure}[b]{0.24\linewidth}
         \centering
         \includegraphics[width=\linewidth]{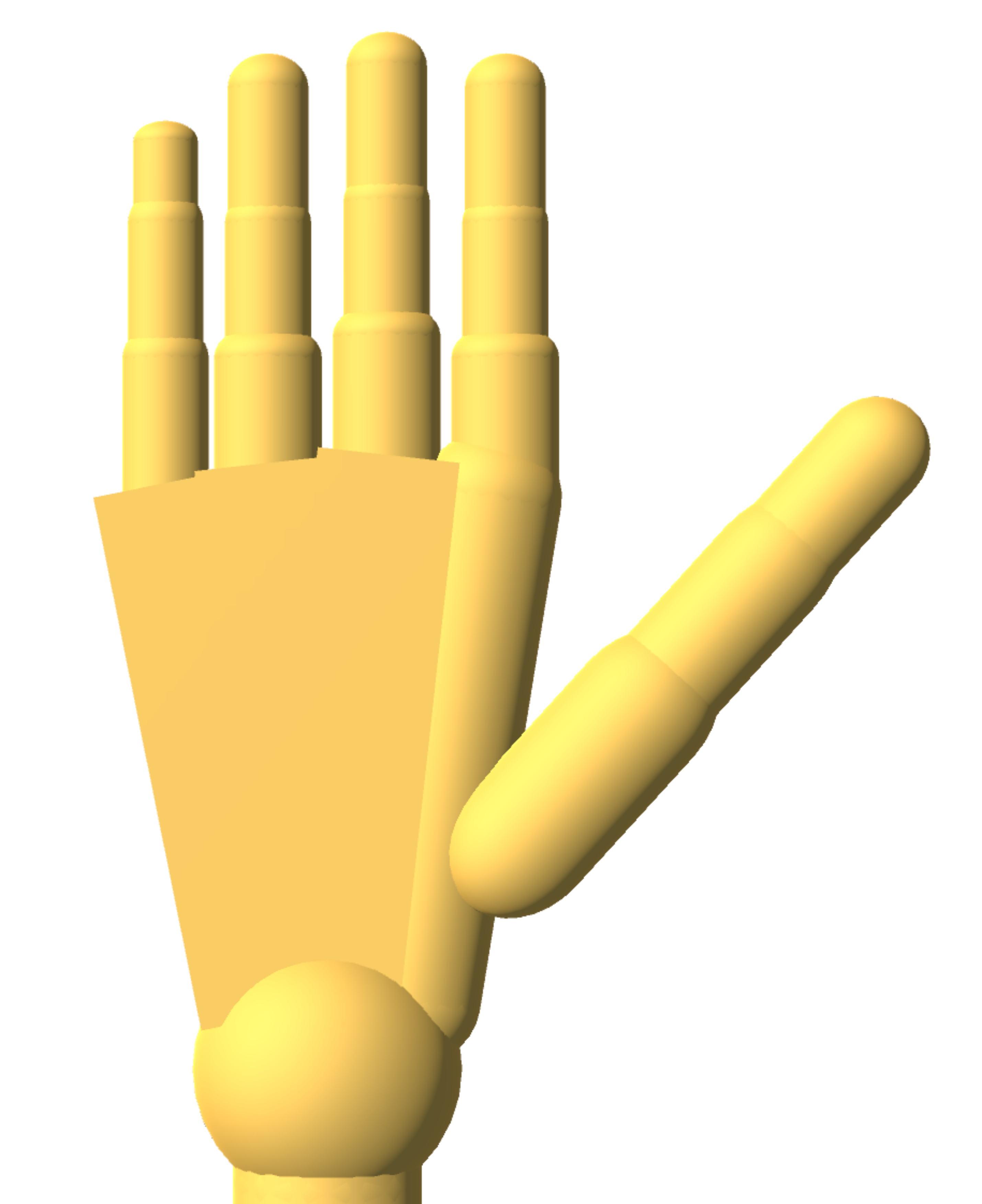}
         \caption{Large hand}
         \label{fig:bighand_model}
     \end{subfigure}
     \begin{subfigure}[b]{0.24\linewidth}
         \centering
         \includegraphics[width=\linewidth]{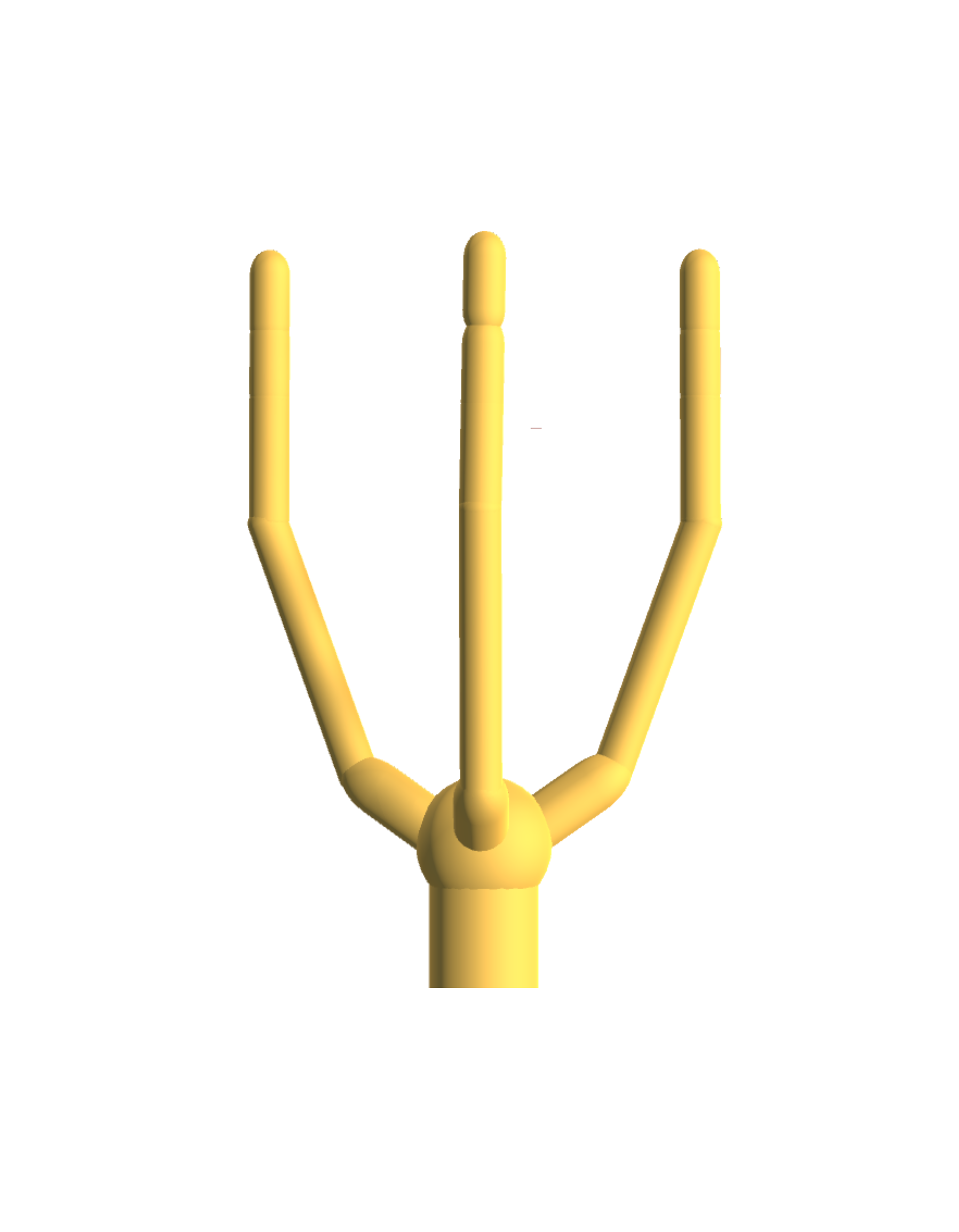}
         \caption{Tri-finger hand}
         \label{fig:trifinger_model}
     \end{subfigure}
  \caption{Our framework works well for drastically different hand morphologies.}
  \label{fig:retarget_model}
\end{figure}

\begin{figure}[t]
  \centering
  \begin{subfigure}[b]{0.99\linewidth}
         \centering
         \includegraphics[width=\linewidth]{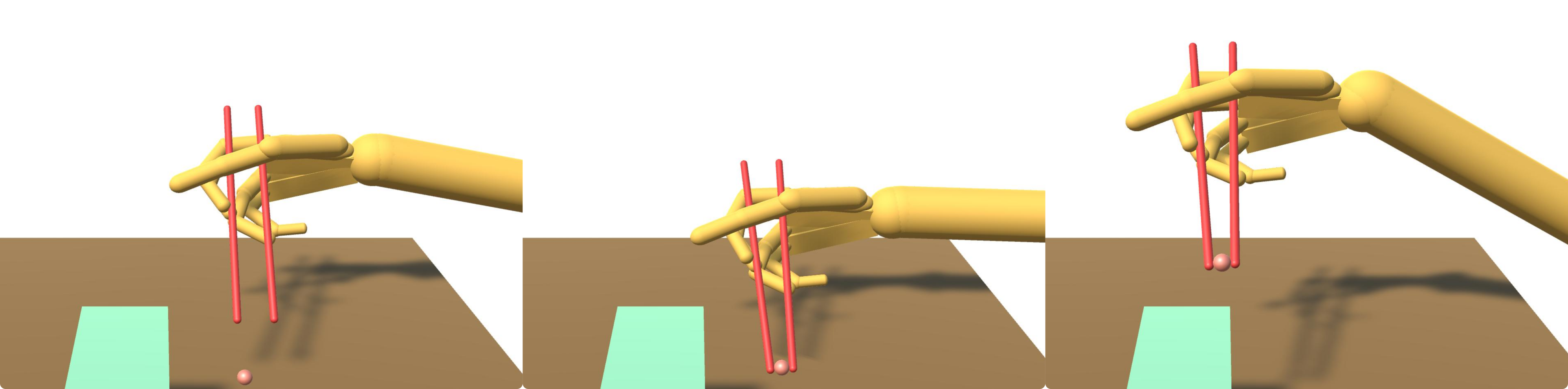}
         \label{fig:long_seq}
     \end{subfigure}\\
     \begin{subfigure}[b]{0.99\linewidth}
         \centering
         \includegraphics[width=\linewidth]{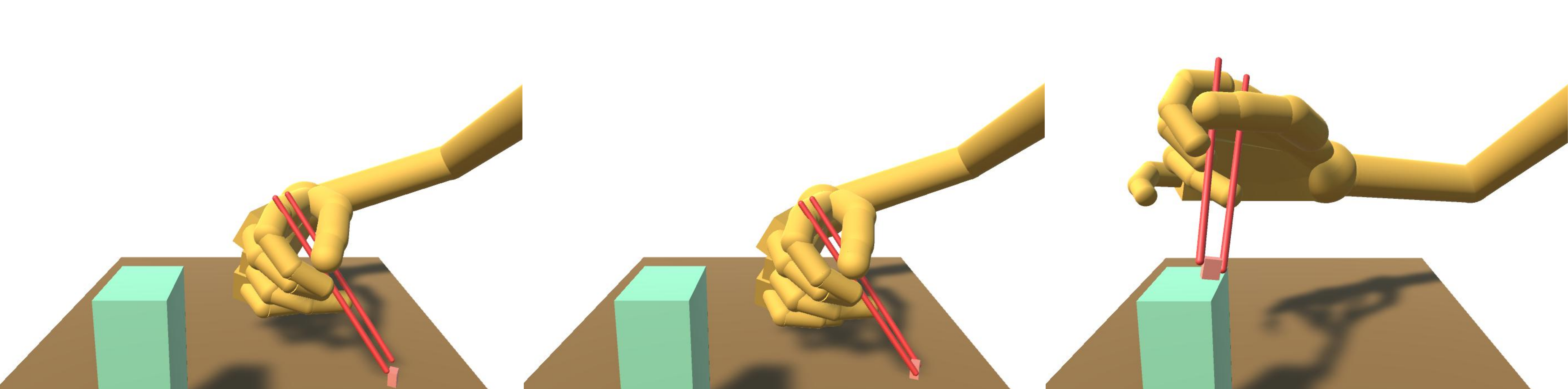}
         \label{fig:big_seq}
    \end{subfigure}\\
    \begin{subfigure}[b]{0.99\linewidth}
         \centering
         \includegraphics[width=\linewidth]{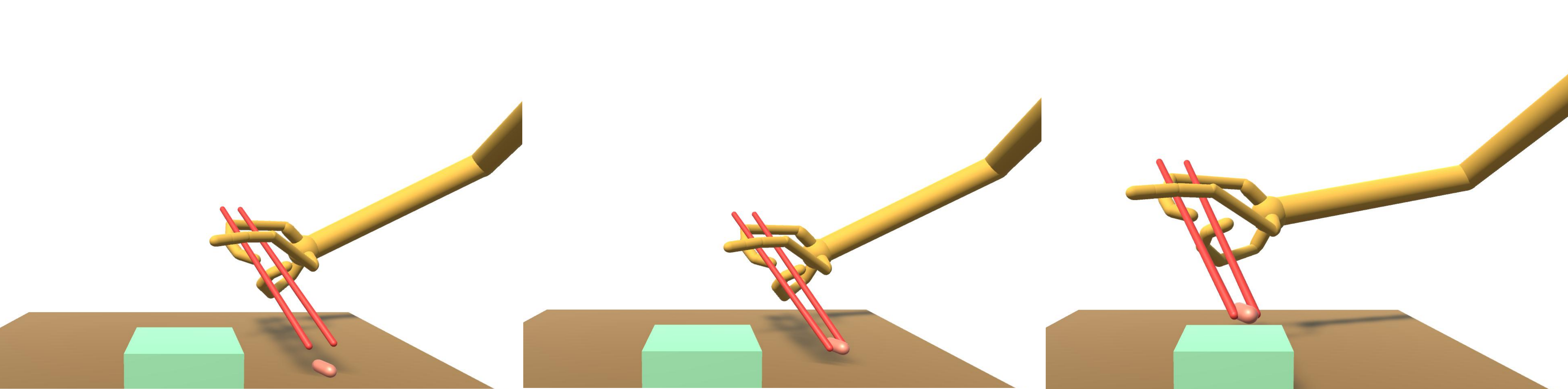}
         \label{fig:tri_seq}
     \end{subfigure}
  \caption{{Chopsticks skills learned for hands of different morphologies.}}
  \label{fig:retarget_skill}
\end{figure}

%% file: fig/curves.tex
\begin{figure}[t]
  \centering
      \includegraphics[width=0.8\linewidth]{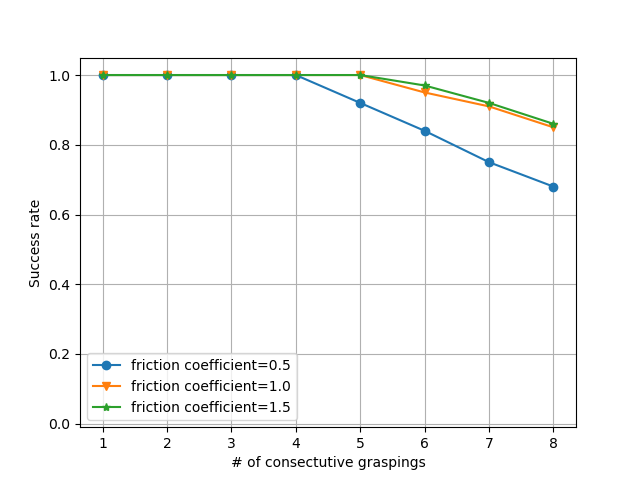}
  \caption{{The task success rate with respect to the number of consecutive object relocations performed. We test three object-chopsticks friction coefficients, around the default MuJoCo friction coefficient $1.0$.}}
  \label{fig:successrate}
\end{figure}

\begin{figure}[t]
  \centering
      \includegraphics[width=0.8\linewidth]{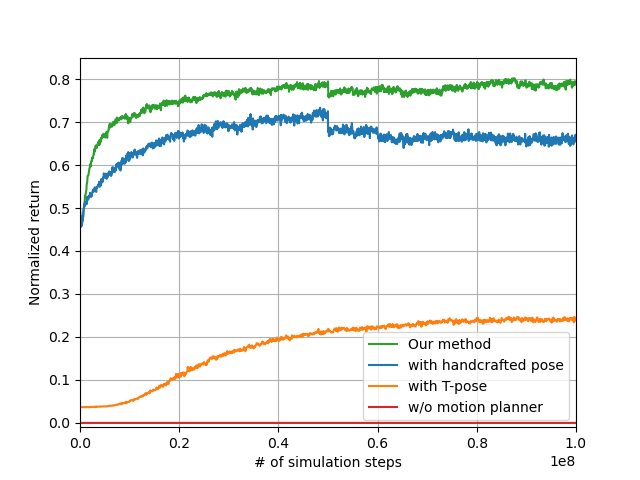}
  \caption{Learning curves for policies trained with different ablations. Without the high-level motion planner, the policy does not learn at all. Using the default T-pose shown in Figure~\ref{fig:pose_pregrasping}, the policy learns very poorly. Our system can also learn using a handcrafted gripping pose, but using the BO-optimized pose performs better. Note that the performance drop around the middle of the training is due to lengthening the training episodes gradually from then on.}
  \label{fig:comparison}
\end{figure}


%% file: sec/8_conclusions.tex
\section{Conclusion and Discussion}

{We have presented a physics-based learning and control system for object manipulation using chopsticks. This is a challenging task that involves complex interactions between the hand, chopsticks, and objects. Our key insight is to tackle the problem by solving two sub-problems: finding good grips to hold the chopsticks stably first, and then using them to grasp and relocate objects. More specifically, the gripping pose is a strong determinant of successful chopstick use later, and we use Bayesian optimization to efficiently explore the gripping pose space. Candidate gripping styles are evaluated through deep reinforcement learning on basic chopsticks maneuvers. After good gripping poses are found through BO and DRL, the actual object relocation using chopsticks is learned in two stages. A high-level motion planner first generates kinematic trajectories for the chopsticks and hand to satisfy task requirements. Then low-level hand control policies are trained to track the generated motion plans using a chosen gripping pose through DRL again. Whenever possible, we adopt common design choices, such as PPO for DRL learning, and MuJoCo with default parameters for simulation.}

{We have demonstrated physics-based object relocation using chopsticks in diverse gripping styles for multiple hand morphologies, for the first time in the literature. Our framework does not need any motion capture data or human demonstrations, and is easily applicable to virtual creatures and robotic manipulators. The framework is also transferable to learning other tools such as tongs, tweezers, pens and brushes, with minor tool-specific changes for style pruning and trajectory generation. Manipulating scissors without actually cutting things is also doable, although real cutting with scissors requires additional components such as a thin-shell simulator and a deformable object motion planner.}

{We have focused on learning visually robust policies with a limited number of quantitative tests. We hope our paper will stimulate future work on more systematic investigations on factors and variations that can affect the success rates of chopsticks-based object manipulations tasks, such as object shapes and sizes, object weights, material and friction properties of the chopsticks and objects, chopsticks geometries, and gripping styles. Different robustness metrics other than the task success rates are also worth exploring.}

{There are many limitations of our current solution that we would like to further investigate in the near future. We use a $7$-DoF parallel gripper to simplify the kinematic planning of chopsticks movements. However, some gripping styles allow chopsticks to move more freely, such as the ``dangling stick'' and the ``Italian'' style shown in Figure~\ref{fig:pose_name}. To reproduce such styles, we need to model the chopsticks as a more versatile gripper with higher DoFs in the motion planner. Another possible solution is to learn a low-dimensional chopsticks motion manifold, similar to the approach described in \cite{holden2015learning} for learning a full-body human motion manifold, and train the grasping model in the latent space. Such a data-driven approach may produce more natural and diverse chopsticks manipulations, at the cost of pre-capturing all the chopsticks skills beforehand.}

{Our learned controllers cannot use chopsticks to successfully manipulate objects for an infinitely long time. The chopsticks gradually slip in hand after relocating objects repeatedly, and this destroys the good grips. Most likely this is caused by the MuJoCo contact dynamics models. As far as we know, no rigid body contact dynamics solver can completely avoid unstable contacts or contact slips. Modelling fingers and palms as truly soft bodies as in~\cite{jain2011controlling} may lead to more robust skills. In the long run, we would like to learn to adjust chopsticks grips with dexterous finger gating maneuvers, such as picking up chopsticks from a table top, changing grips, and moving chopsticks up and down in hand.}

{Currently our system runs at interactive rates but not realtime, bottlenecked at the motion planner. An idea for improvement is to replace the trajectory optimization component with a neural network model, which may be trained jointly with the low-level control policies, similar to the ideas of \cite{peng2017deeploco} and \cite{zeng2019end}. We would also like to explore the possibility of combining gripping pose optimization and control policy learning into one iterative learning pipeline, which alternates the training between optimizing gripping poses and improving low-level control policies, evaluated on the same set of object relocation tasks. Such a scheme may potentially achieve better learning efficiency and control performance.}

{We have only touched the tip of the iceberg for physics-based chopsticks skills. How to eat and cut noodles with chopsticks? How to flip a piece of meat using chopsticks? How to beat an egg with chopsticks? Is it easier to use Chinese, Japanese, or Korean chopsticks for a certain type of food? Modeling and planning with soft or amorphous materials as in \cite{zhang2021manipnet} is a must. Last but not least, we would like to transfer our chopsticks controllers to real anthropomorphic robot hands using ``Sim2Real'' techniques such as domain randomization and adaptation similar to \cite{peng2018sim,peng2020learning}.}

%% file: sec/x_ack.tex
\begin{acks}
    We would like to thank the anonymous reviewers for their constructive suggestions and feedback. We also thank Zhiqi Yin, Yujie Wang, and Michiel van de Panne for various discussions and help. Yin is partially supported by NSERC Discovery Grants Program RGPIN-06797 and RGPAS-522723.
\end{acks}

%% file: sec/x_appendix.tex
\appendix

\section{Input State of the Hand Controllers}
\label{sec:app:states}

{We denote the input state of our low-level hand controllers as a vector $\bm{s}$. It consists of multiple components as listed in Table~\ref{tab:control_state}. $\handstate$, $\chopstate$, and $\objstate$ are the simulation states of the hand and arm, the chopsticks, and the object, respectively. $\bm{d}_{\text{hand}}$ measures the distances between the fingertips and their desired contact locations on the chopsticks as specified in the gripping pose. $\bm{f}_{\text{hand}}$ and $\bm{f}_{\text{chop}}$ are the magnitude of the normal forces between the fingertips and the chopsticks, and between the chopsticks and the object, respectively. }


{Quantities with a tilde on top represent desired states in the planned trajectory to track. In particular, $\kinechopstate=(\bm{q}_{\text{chop}},\dot{\bm{q}}_{\text{chop}})$, where $\bm{q}_{\text{chop}}$ is parameterized as the 7-DoF parallel gripper as described in Section~\ref{sec:graspingModel}. To encourage smooth tracking, we include six frames of these desired states sampled every $0.05s$ for the next $0.3s$.} 




\begin{table}[hbp]
    \caption{Components of the state $s$ of our hand controllers.}
    \begin{tabular}{rl}
    \toprule
    symbol &  description \\
    \hline
    $\handstate$ & simulation state of the hand and arm\\ 
    $\chopstate$ &  simulation state of the chopsticks\\
    $\objstate$ & simulation state of the object \\
    $\bm{t}_{\text{obj}}$ &  shape parameters of the object \\
    $\bm{d}_{\text{hand}}$ & \begin{tabular}[c]{@{}l@{}}distance between the fingertips and their \\ desired contact locations on the chopsticks\end{tabular}\\
    $\bm{f}_{\text{hand}}$ &   {magnitude of contact forces on fingertips}\\
    $\bm{f}_{\text{chop}}$ &    {magnitude of contact forces on  chopsticks tips}\\
    $\bm{\tilde{s}}_{\textrm{hand}} \times 6$ &  planned states of the hand and arm\\
    $\bm{\tilde{s}}_{\textrm{chop}} \times 6$ & planned states of the chopsticks\\
    $\bm{\tilde{s}}_{\textrm{obj}} \times 6$ &  planned states of the object\\
    \bottomrule
    \end{tabular}
    \label{tab:control_state}
\end{table}

\section{Additional Optimized Gripping Poses}
\label{sec:app:poses}
{Optimized gripping poses are found by Bayesian Optimization and DRL for seventeen valid gripping styles. In the main text we show the five most distinctive styles, and here we show the remaining twelve styles in Figure~\ref{fig:other12styles}.}

\input{fig/twelvePoses.tex}


%% file: fig/twelvePoses.tex
\begin{figure}[hb]
  \centering
    \begin{subfigure}[b]{0.3\linewidth}
         \centering
         \includegraphics[width=\linewidth]{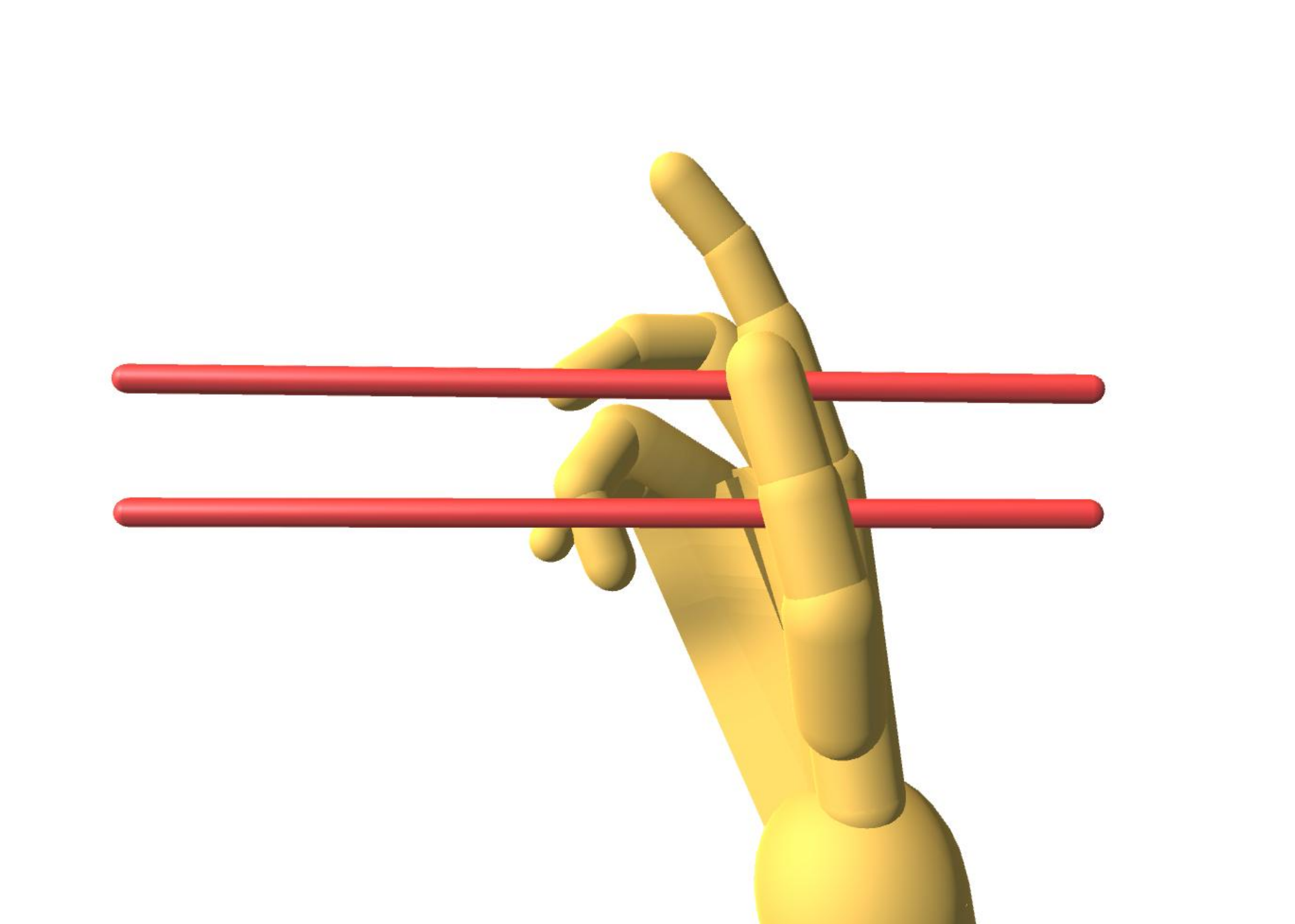}
         \caption{$\bm{c}=(1,0,1,0,2)$}
         \label{fig:10102}
    \end{subfigure}
    \begin{subfigure}[b]{0.3\linewidth}
         \centering
         \includegraphics[width=\linewidth]{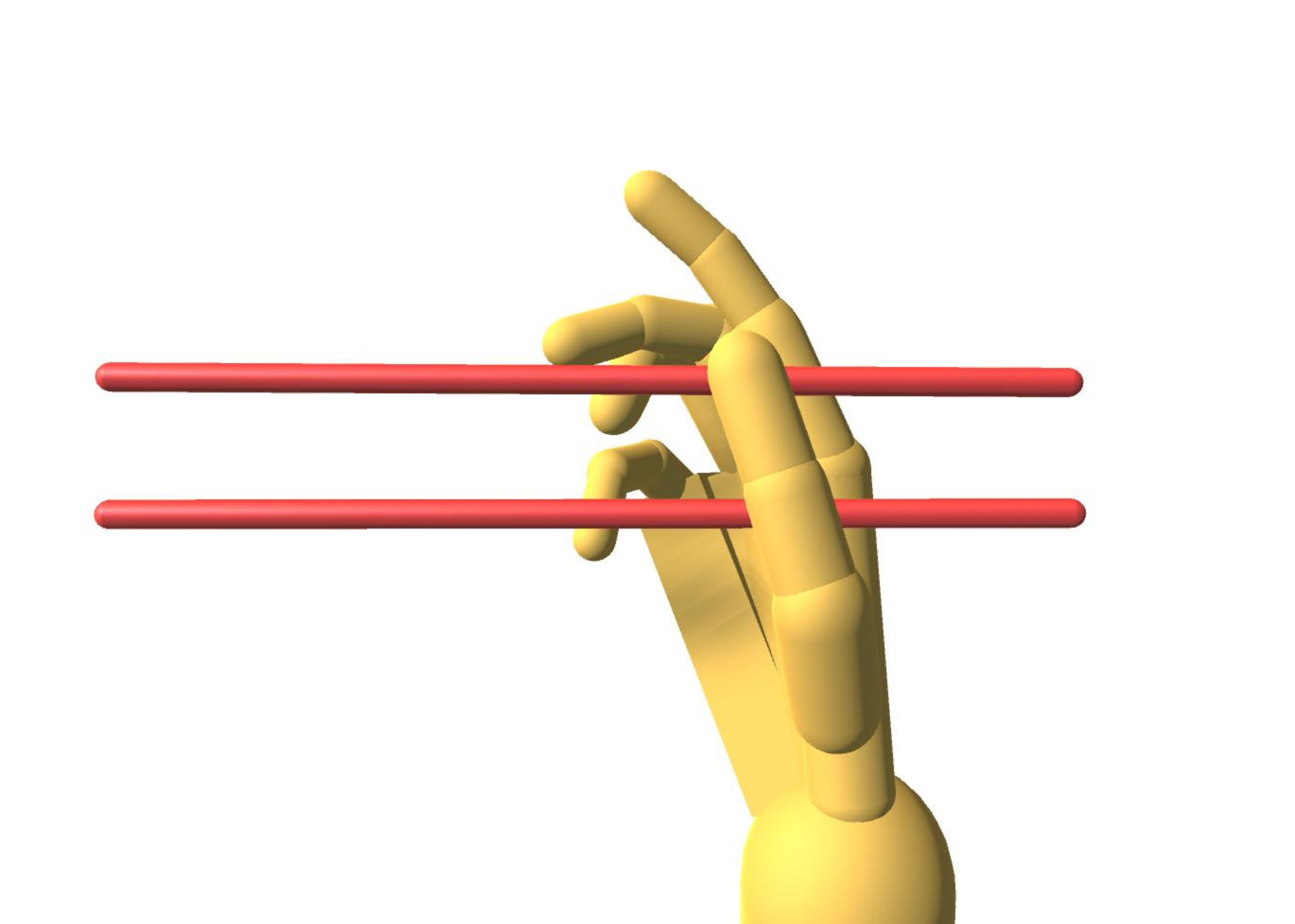}
         \caption{$\bm{c}=(1,0,1,1,2)$}
         \label{fig:10112}         
    \end{subfigure}
    \begin{subfigure}[b]{0.3\linewidth}
         \centering
         \includegraphics[width=\linewidth]{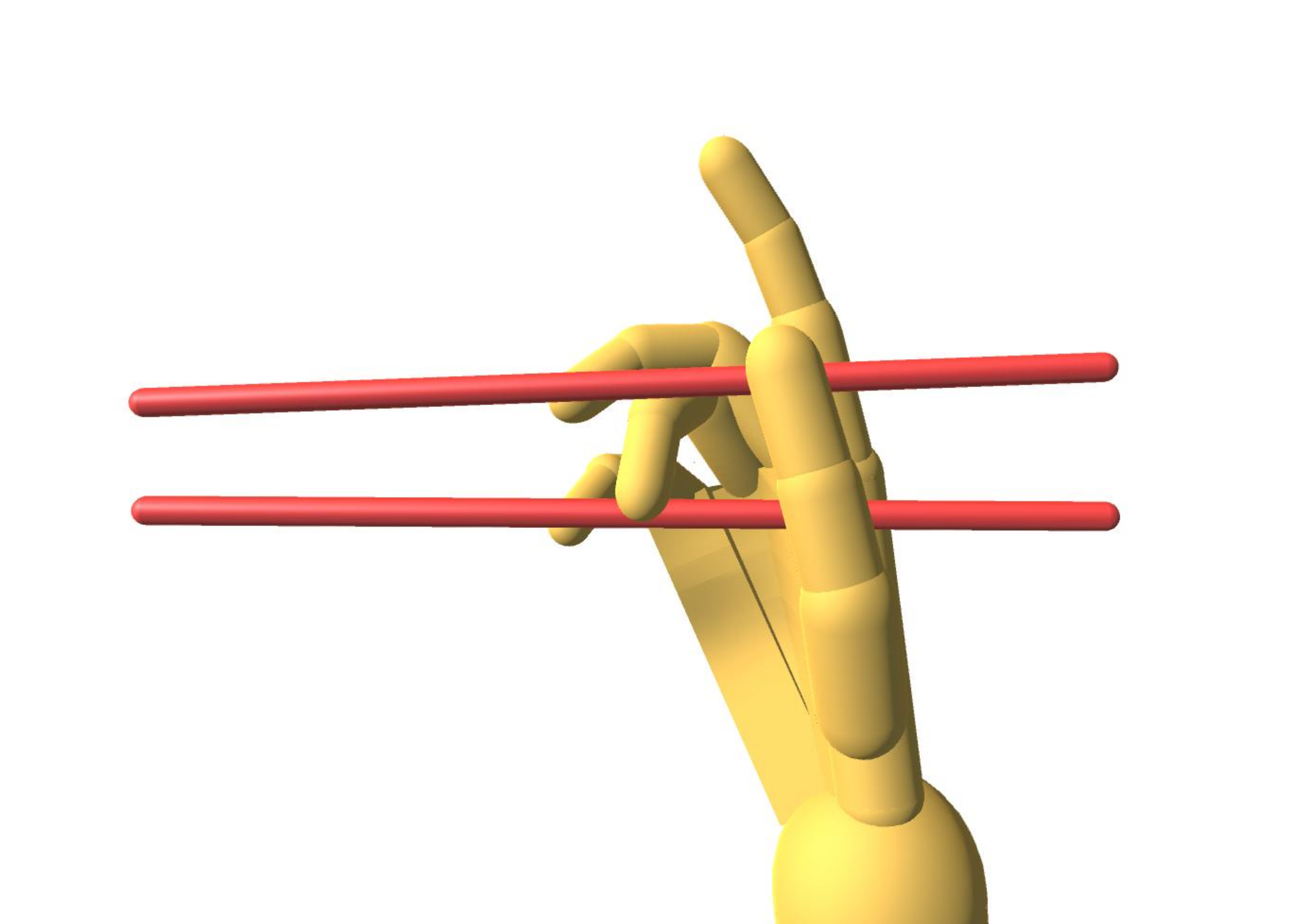}
         \caption{$\bm{c}=(1,0,1,2,2)$}
         \label{fig:10122}
    \end{subfigure}\\
    \begin{subfigure}[b]{0.3\linewidth}
         \centering
         \includegraphics[width=\linewidth]{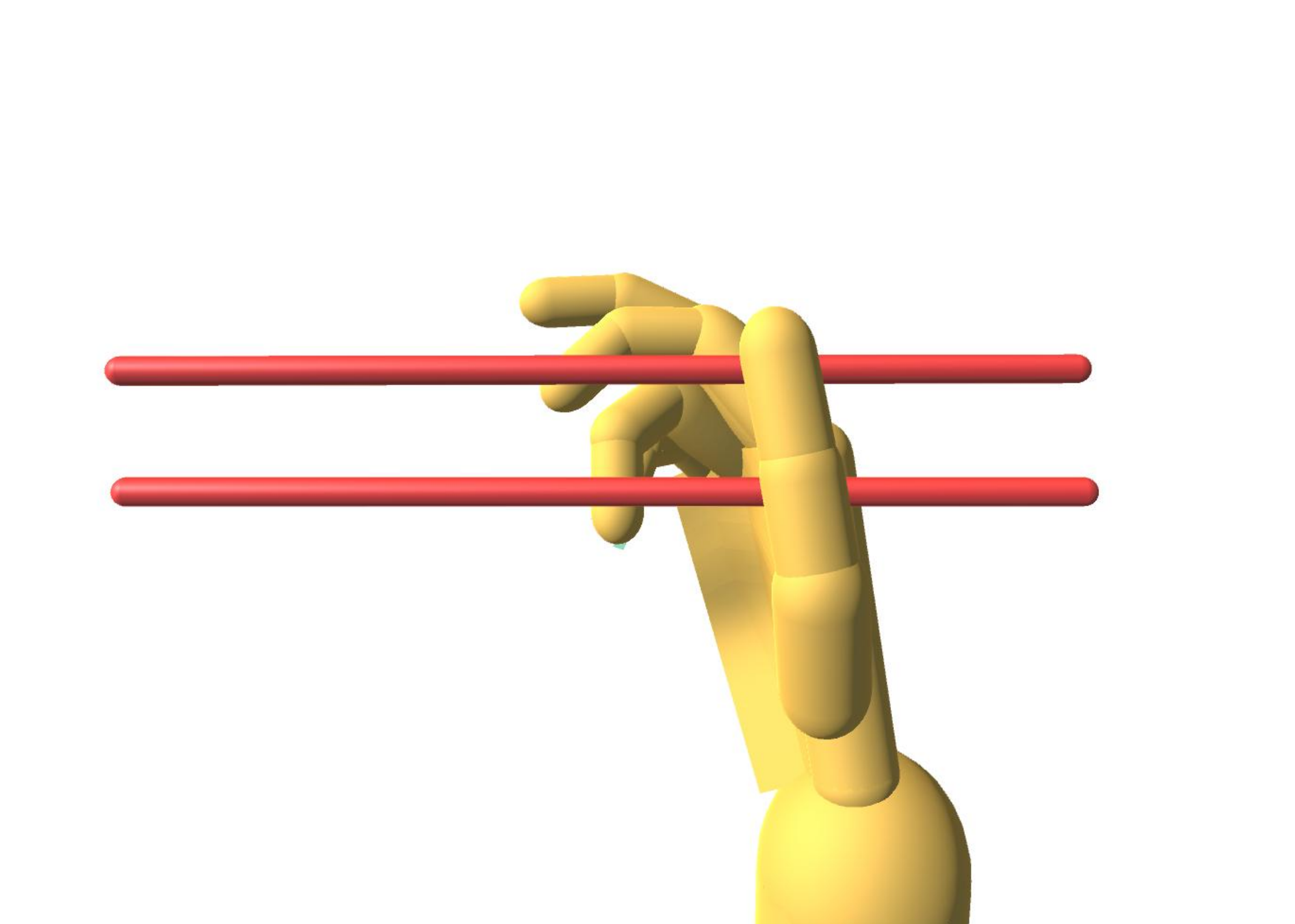}
         \caption{$\bm{c}=(1,1,0,0,2)$}
         \label{fig:11002}         
    \end{subfigure}
    \begin{subfigure}[b]{0.3\linewidth}
         \centering
         \includegraphics[width=\linewidth]{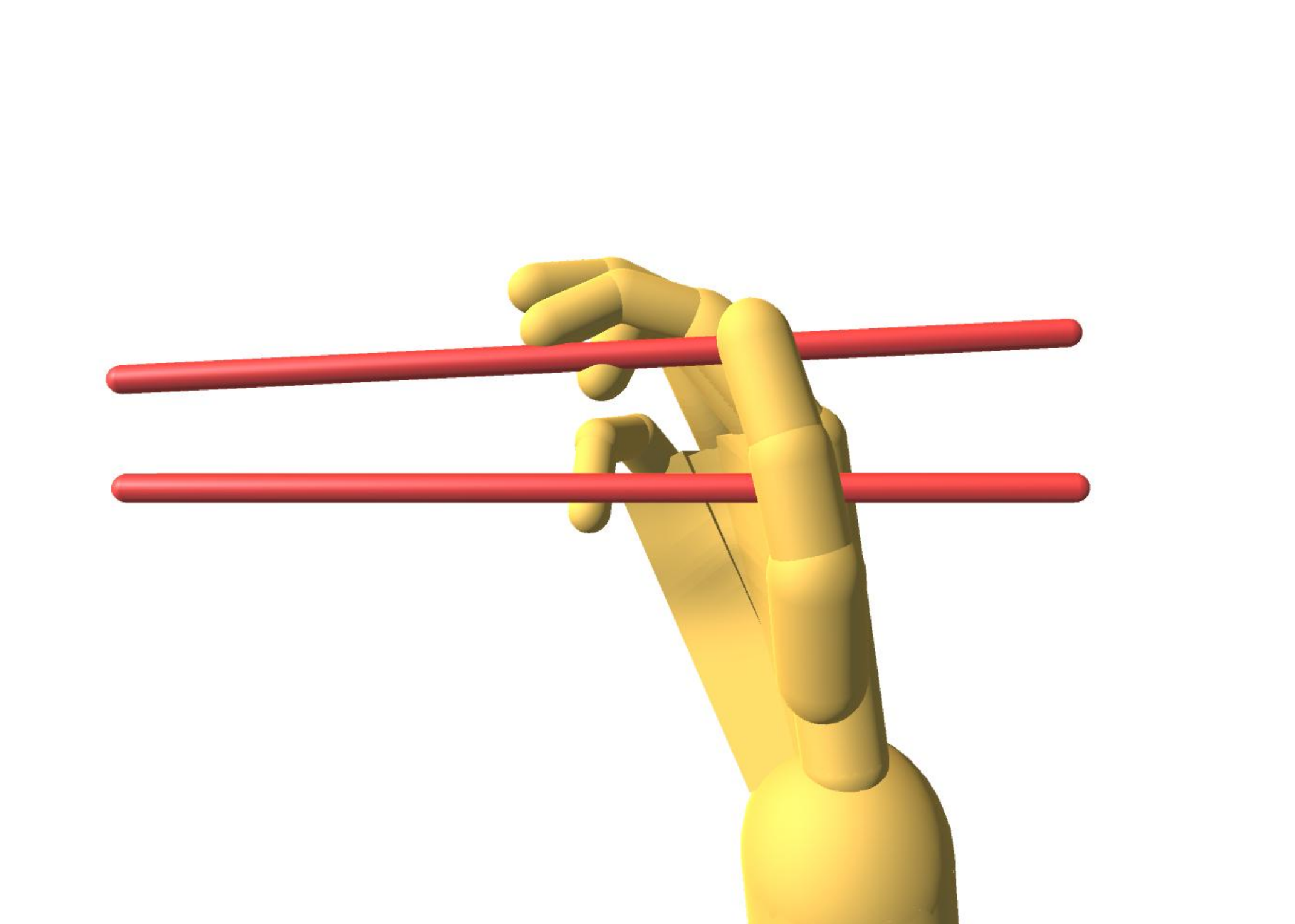}
         \caption{$\bm{c}=(1,1,0,1,2)$}
         \label{fig:11012}
     \end{subfigure}
      \begin{subfigure}[b]{0.3\linewidth}
         \centering
         \includegraphics[width=\linewidth]{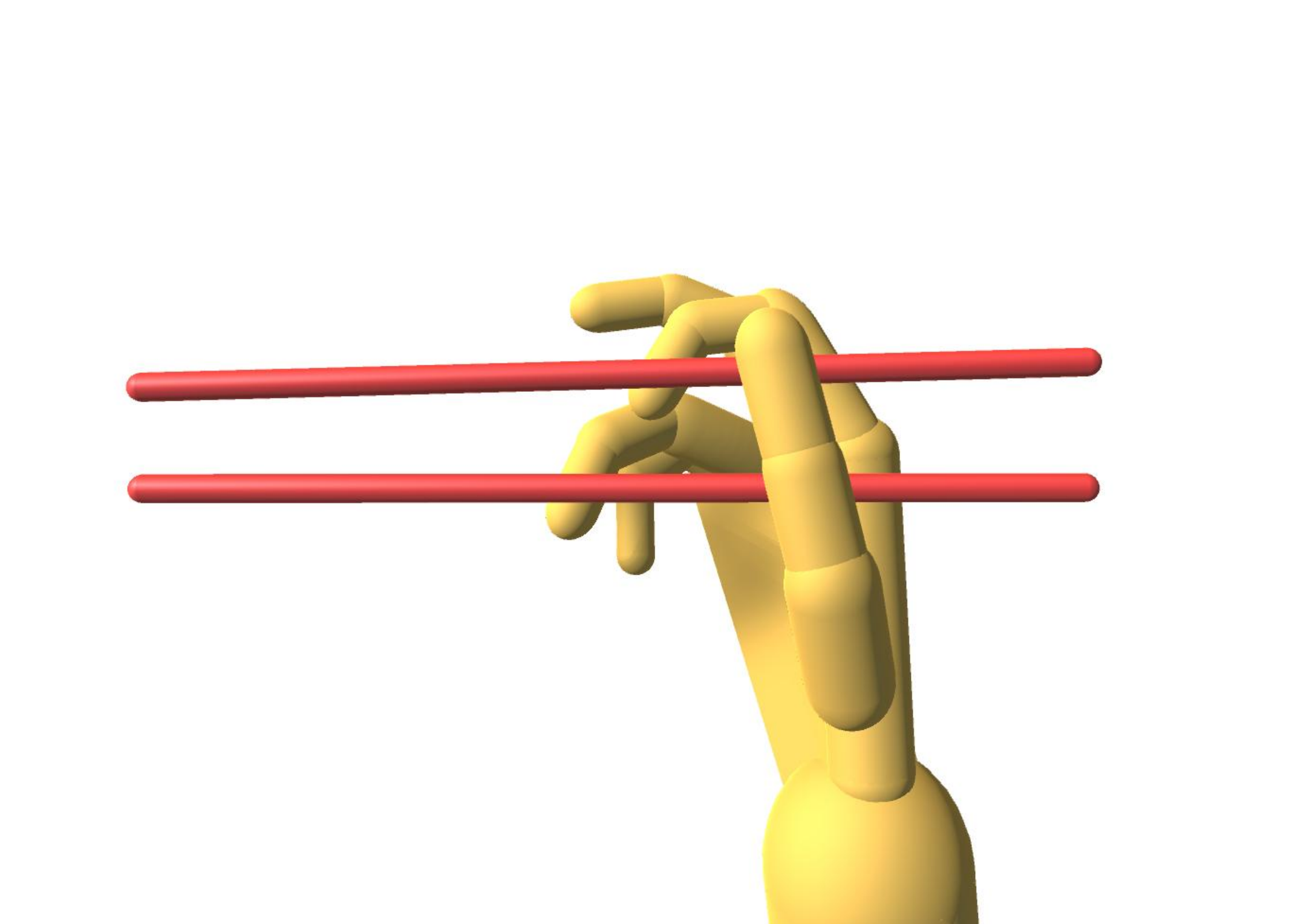}
         \caption{$\bm{c}=(1,1,0,2,0)$}
         \label{fig:11020}
    \end{subfigure}\\
      \begin{subfigure}[b]{0.3\linewidth}
         \centering
         \includegraphics[width=\linewidth]{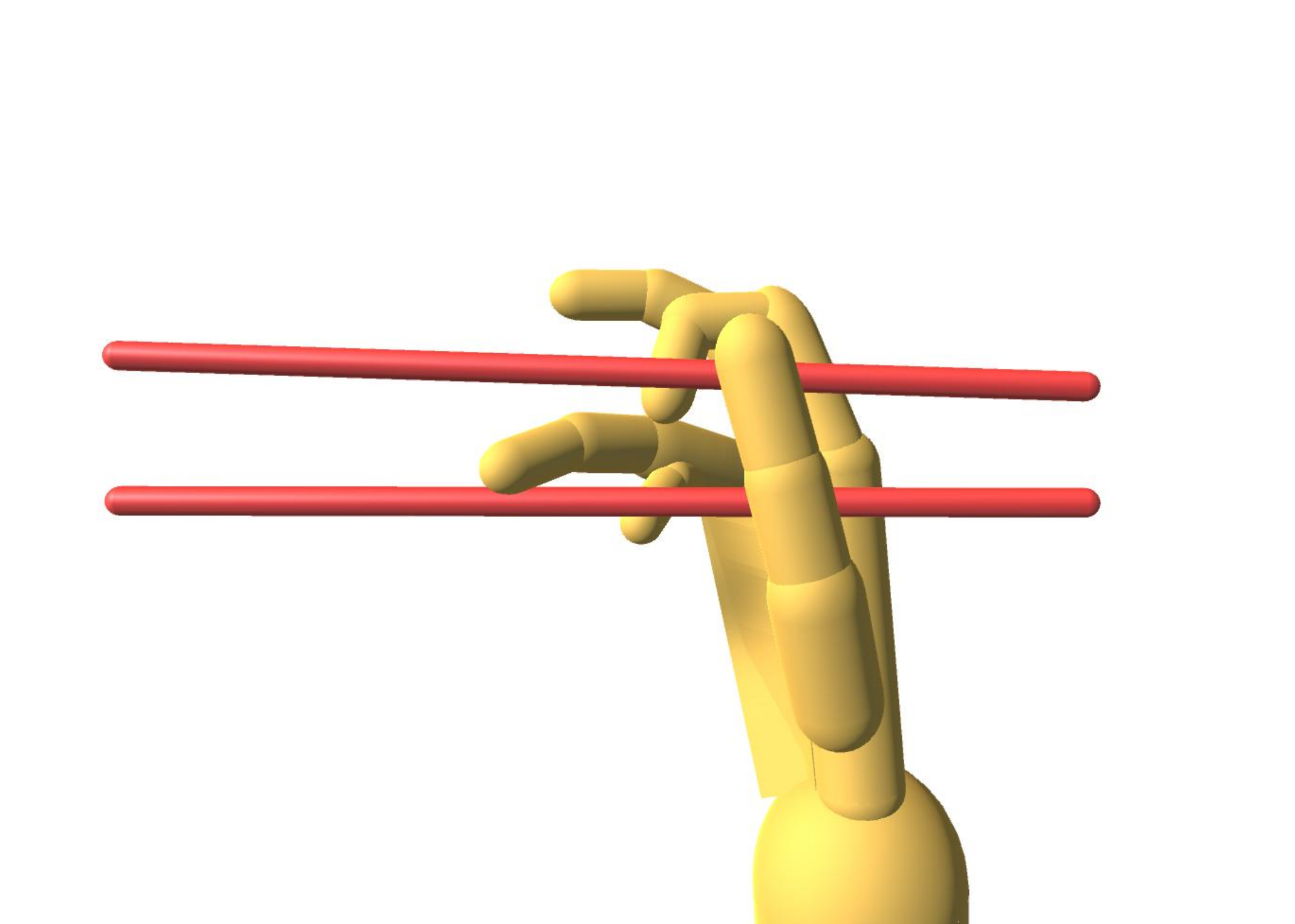}
         \caption{$\bm{c}=(1,1,0,2,2)$}
         \label{fig:11022}
    \end{subfigure}
    \begin{subfigure}[b]{0.3\linewidth}
         \centering
         \includegraphics[width=\linewidth]{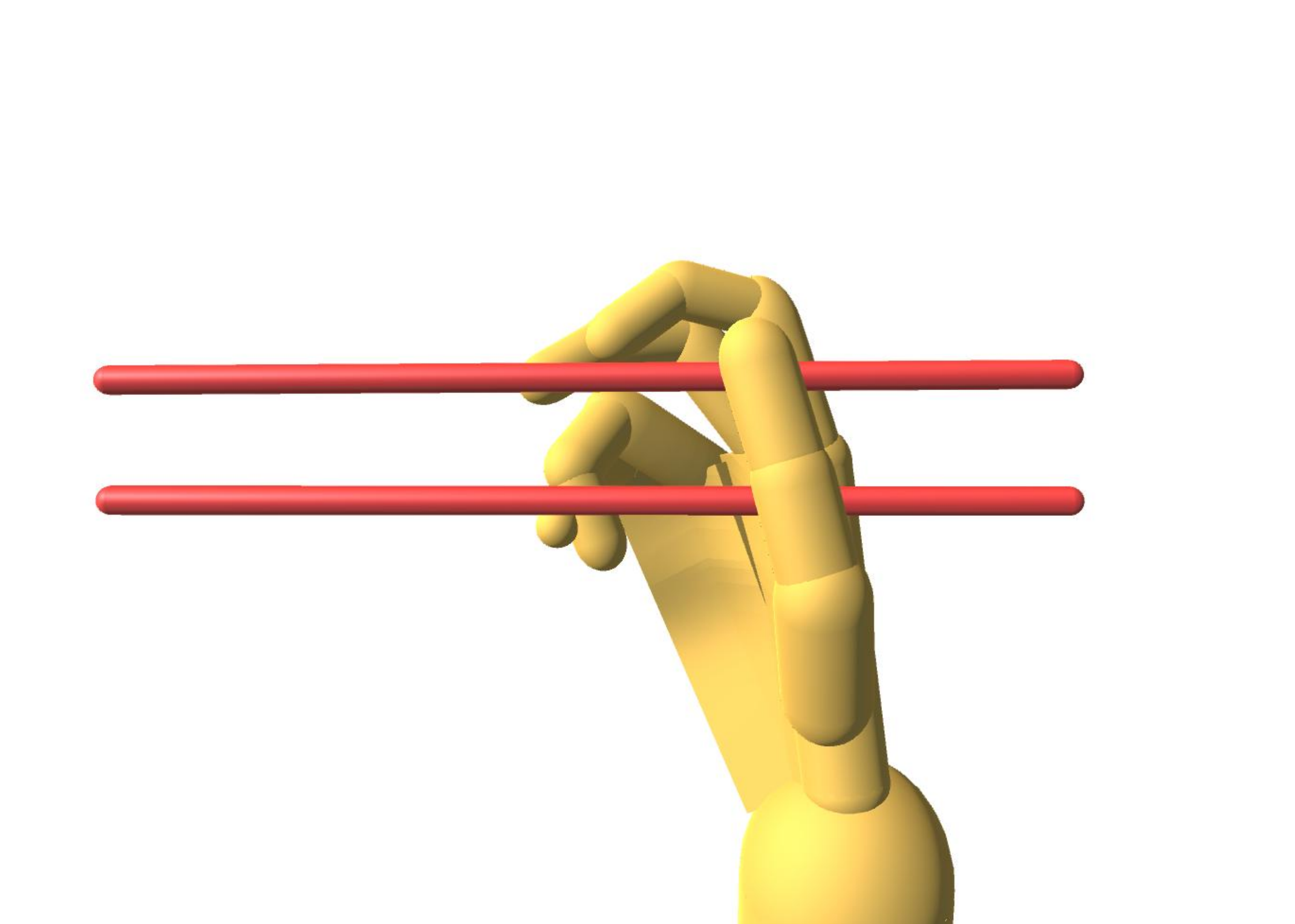}
         \caption{$\bm{c}=(1,1,1,0,2)$}
         \label{fig:111102}         
    \end{subfigure}
    \begin{subfigure}[b]{0.3\linewidth}
         \centering
         \includegraphics[width=\linewidth]{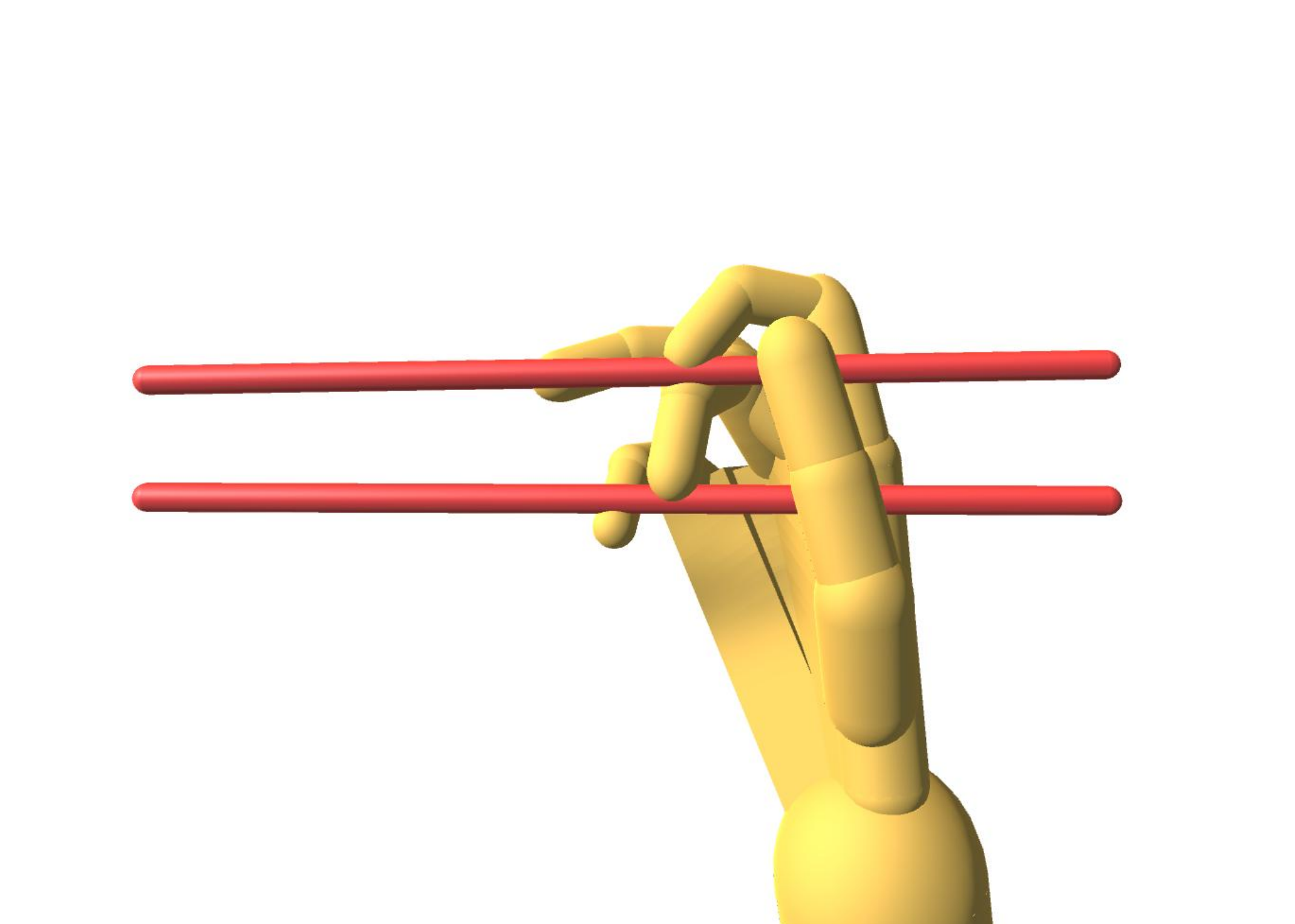}
         \caption{$\bm{c}=(1,1,1,2,2)$}
         \label{fig:11122}
    \end{subfigure}\\
    \begin{subfigure}[b]{0.3\linewidth}
         \centering
         \includegraphics[width=\linewidth]{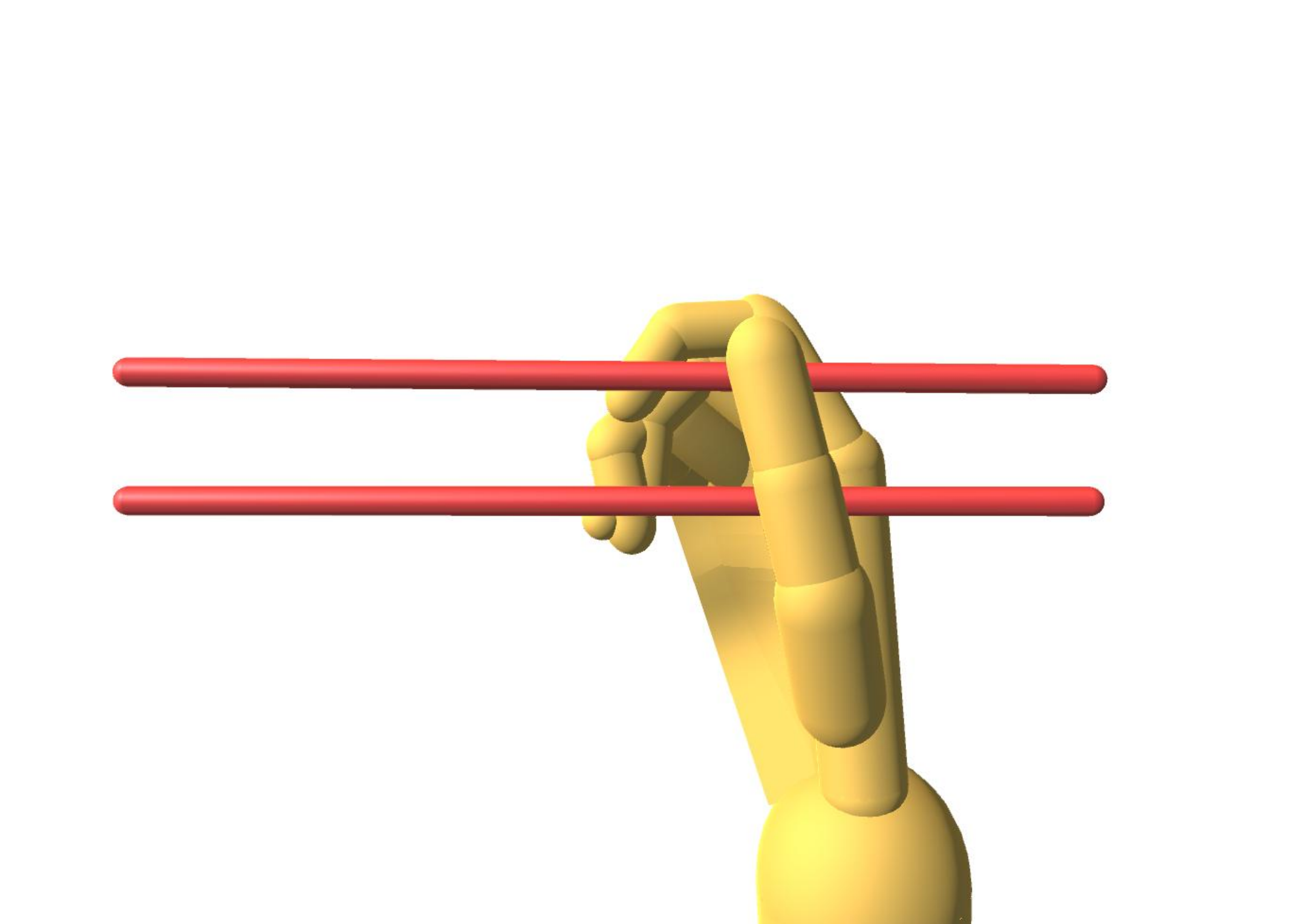}
         \caption{$\bm{c}=(1,1,2,0,2)$}
         \label{fig:11202}         
    \end{subfigure}
    \begin{subfigure}[b]{0.3\linewidth}
         \centering
         \includegraphics[width=\linewidth]{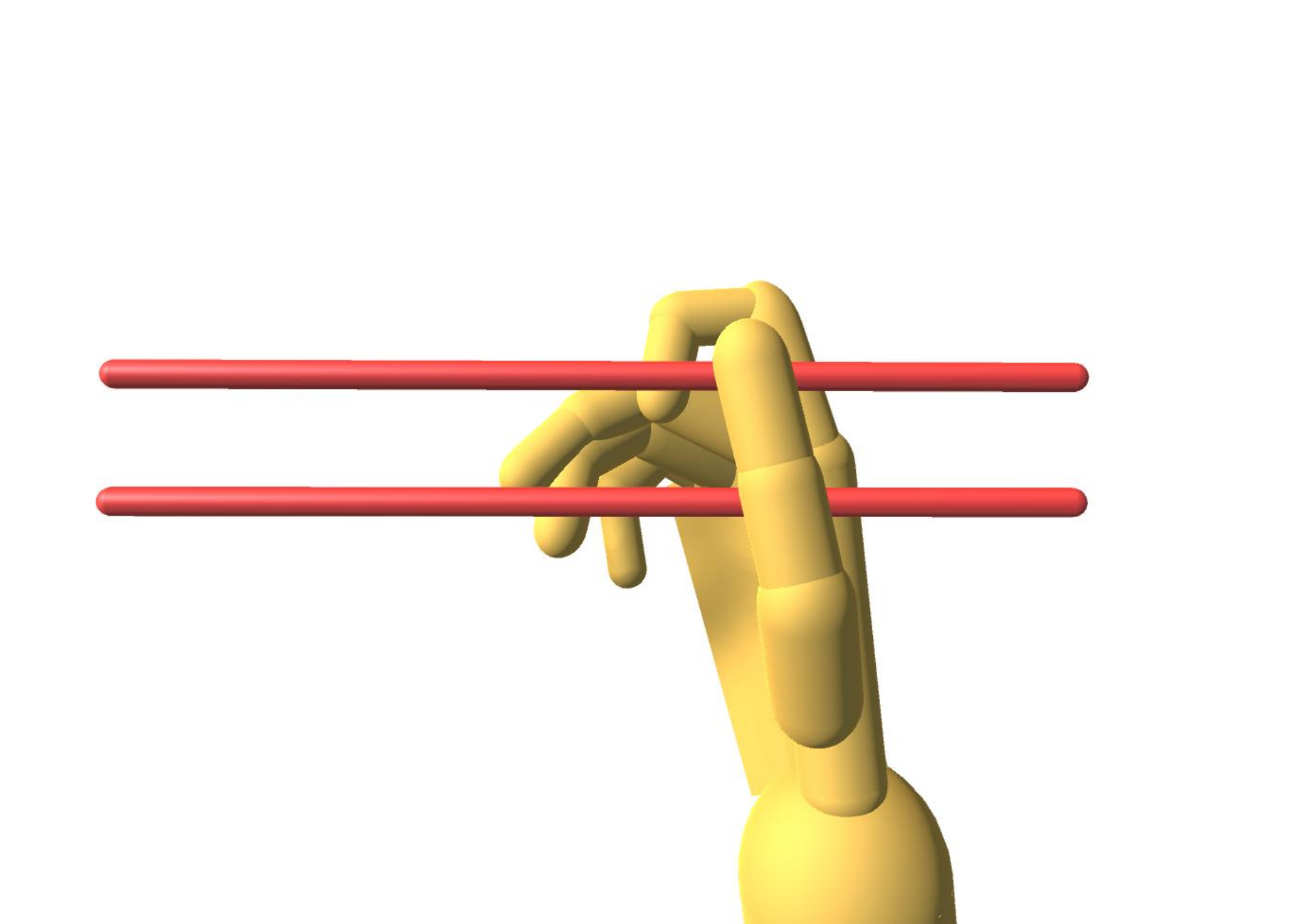}
         \caption{$\bm{c}=(1,1,2,2,0)$}
         \label{fig:11220}
     \end{subfigure}
      \begin{subfigure}[b]{0.3\linewidth}
         \centering
         \includegraphics[width=\linewidth]{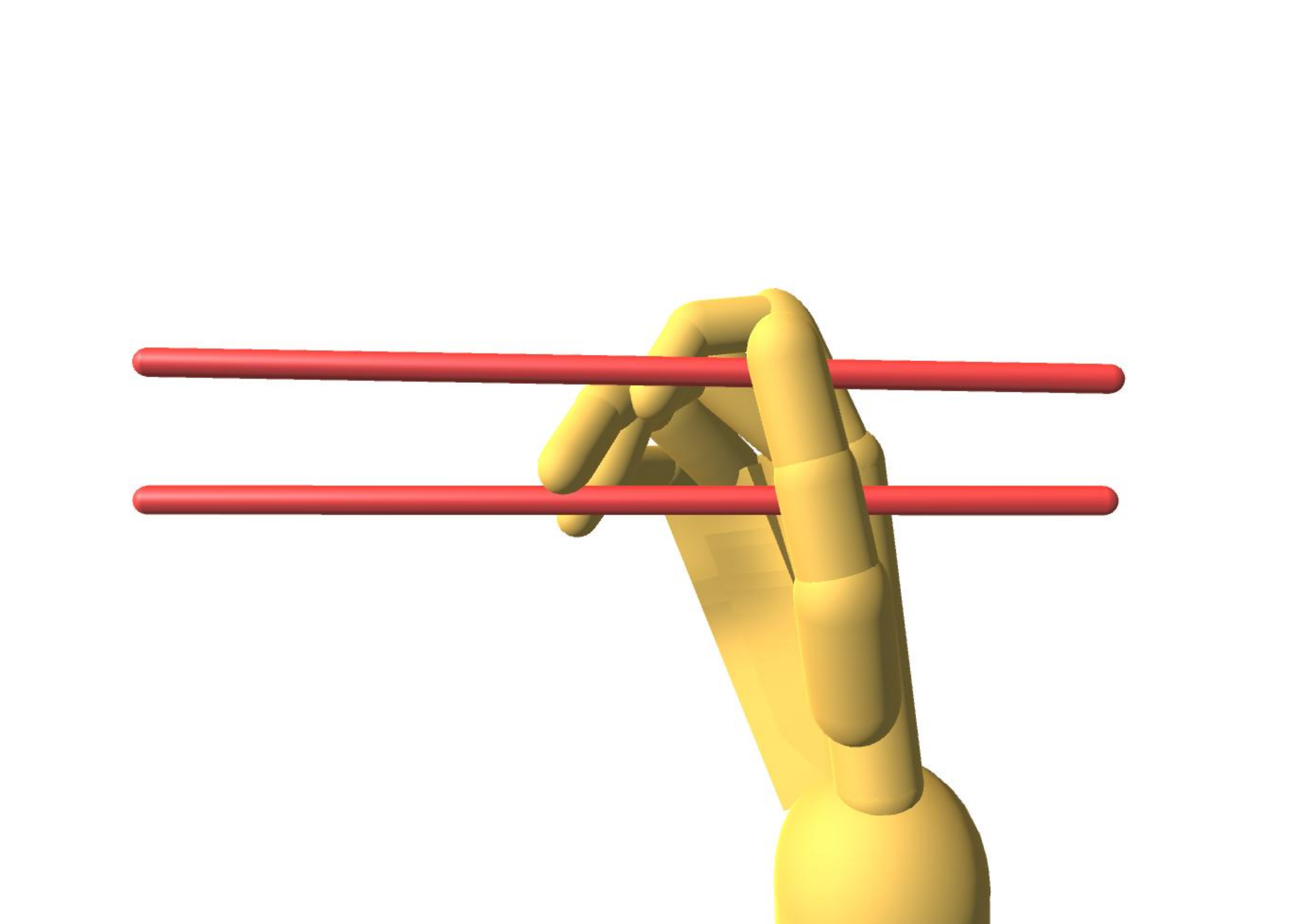}
         \caption{$\bm{c}=(1,1,2,2,2)$}
         \label{fig:11222}
    \end{subfigure}
  \caption{{Visualization of twelve additional optimized gripping poses.}}
  \label{fig:other12styles}
\end{figure}